\documentclass[11pt]{article}
\pdfoutput=1

\usepackage{mathrsfs}
\usepackage{amsmath,amssymb}
\usepackage{bm}
\usepackage{longtable}
\usepackage{natbib}
\usepackage[usenames]{color}
\usepackage{amsthm}
\usepackage{algorithm}
\usepackage{algorithmic}
\usepackage{multirow} 
\usepackage{caption}
\usepackage{subcaption}
\usepackage[shortlabels]{enumitem}
\usepackage{dsfont}
\usepackage{mathtools}
\usepackage{caption}
\usepackage{booktabs}

\usepackage[colorlinks,
linkcolor=red,
anchorcolor=blue,
citecolor=blue
]{hyperref}

\usepackage{float}  
\usepackage{setspace}
\usepackage[left=1in, right=1in, top=1in, bottom=1in]{geometry}

\setlist[itemize]{itemsep=0pt, topsep=2pt}
\setlist[enumerate]{itemsep=-2pt, topsep=2pt}

\ifx\proof\undefined
\newenvironment{proof}{\par\noindent{\bf Proof\ }}{ \\[2mm]}
\fi

\usepackage{xcolor}

\ifdefined\final
\usepackage[disable]{todonotes}
\else
\usepackage[textsize=tiny]{todonotes}
\fi
\allowdisplaybreaks

\title{Inference for Heterogeneous Treatment Effects with Deep Survival Learners}

\author{
Sattwik Ghosal\textsuperscript{1},
Xuran Meng\textsuperscript{1},
Yuming Sun\textsuperscript{2},
and Yi Li\textsuperscript{1}\\[0.5em]
{\small \textsuperscript{1}Department of Biostatistics,
University of Michigan}\\
{\small \textsuperscript{2}Department of Mathematics,
William \& Mary}\\[0.3em]
{\small
\texttt{\{ghosal,xuranm,yili\}@umich.edu},
\texttt{ysun30@wm.edu}
}
}
\date{}

\ifx\proof\undefined
\newenvironment{proof}{\par\noindent{\bf Proof\ }}{ \\[2mm]}
\fi

\ifx\theorem\undefined
\newtheorem{theorem}{Theorem}
\fi
\ifx\question\undefined
\newtheorem{question}[theorem]{Question}
\fi
\ifx\example\undefined
\newtheorem{example}[theorem]{Example}
\fi
\ifx\property\undefined
\newtheorem{property}{Property}
\fi
\ifx\lemma\undefined
\newtheorem{lemma}[theorem]{Lemma}
\fi
\ifx\proposition\undefined
\newtheorem{proposition}[theorem]{Proposition}
\fi
\ifx\remark\undefined
\newtheorem{remark}[theorem]{Remark}
\fi
\ifx\target\undefined
\newtheorem{target}[theorem]{Target}
\fi
\ifx\corollary\undefined
\newtheorem{corollary}[theorem]{Corollary}
\fi
\ifx\definition\undefined
\newtheorem{definition}[theorem]{Definition}
\fi
\ifx\conjecture\undefined
\newtheorem{conjecture}[theorem]{Conjecture}
\fi
\ifx\fact\undefined
\newtheorem{fact}[theorem]{Fact}
\fi
\ifx\claim\undefined
\newtheorem{claim}[theorem]{Claim}
\fi
\ifx\assumption\undefined
\newtheorem{assumption}[theorem]{Assumption}
\fi
\numberwithin{equation}{section}
\numberwithin{theorem}{section}

\begin{document}

\maketitle

 \begin{abstract}

Conditional average treatment effects (CATEs) provide
a principled framework for assessing treatment-effect
heterogeneity and identifying patients who may benefit
 from an intervention. For survival outcomes, CATEs may also evolve over  time and pose challenges for
estimation and inference. Existing methods can accommodate
complex covariate heterogeneity but often fail to exploit
shared temporal structure or provide valid inference for
time-varying treatment effects and their contrasts.
We propose a deep survival learning framework that
combines neural networks for nonparametric covariate effects
with a shared spline representation of temporal variation,
enabling joint estimation of CATE trajectories.
Using doubly robust pseudo-outcomes and sample splitting,
we establish nonasymptotic oracle bounds accounting for
network and temporal approximation, nuisance estimation,
and optimization errors. Our theory shows that exploiting
shared temporal structure yields sharper error bounds for
temporal contrasts than those derived from separate
time-specific estimators. For inference, we develop a subsampled
ensemble procedure and establish asymptotic normality for
pointwise CATEs, temporal contrasts, and covariate contrasts,
including temporal contrasts over shrinking time intervals.
Under suitable regularity and bias-control conditions,
consistent infinitesimal jackknife variance estimators
yield asymptotically valid confidence intervals.
Simulations evaluate estimation accuracy, interval coverage,
and sensitivity to temporal dimension and basis mismatch.
An application to lung cancer survival data illustrates
the framework's ability to characterize heterogeneous
treatment benefits over time.

\end{abstract}

\section{Introduction}
\label{sec:intro}

For time-to-event outcomes, understanding treatment-effect heterogeneity across patients and over time is essential for precision medicine and individualized decision making. Within the
potential-outcomes framework \citep{imbens2015causal},
such heterogeneity is commonly quantified by the
conditional average treatment effect (CATE), which
contrasts potential outcomes under treatment and
control conditional on patient characteristics.
Beyond the challenges posed by censoring,
characterizing time-varying CATEs involves two
types of contrasts: temporal contrasts, which
quantify changes in treatment benefit over time for a given
patient profile, and covariate contrasts, which
compare treatment benefits across patient profiles
at a given time. Inference for these quantities
is essential for distinguishing genuine differences
in treatment benefit from sampling variation.

Temporal contrasts help assess the timing and
durability of treatment benefits and inform
clinical reassessment. Our motivating data from
the Boston Lung Cancer Study 
\citep{alvarez2026effects} illustrate these
challenges in evaluating adjuvant chemotherapy
among patients with stage II or III lung cancer.
Our
analysis estimates three-year survival benefits
of 20 and 14.9 percentage points for stage II
and III patients, respectively. Estimated benefits
increase over the first three to four years
following diagnosis before leveling off, and
vary further with age, body mass index, and
smoking history. These findings motivate
inference for both temporal and covariate
contrasts. More broadly, covariate heterogeneity
may involve complex nonlinearities and
high-order interactions, for which deep neural
networks offer flexible approximation with
convergence rates governed by intrinsic rather
than ambient dimension under suitable
compositional assumptions
\citep{schmidt2020deep}. In contrast, time is
one-dimensional, and treatment effects may
evolve smoothly.

Progress has been made in flexible CATE
estimation through meta-learners
\citep{kunzel2019metalearners,NieWager2021,
curth2021nonparametric}
and doubly robust methods that mitigate the
effects of nuisance estimation errors
\citep{kennedy2020towards}. Extensions to
right-censored outcomes include censoring-adjusted
pseudo-outcomes \citep{yang2025doubly,bo2025evaluating},
causal survival forests \citep{cui2023estimating},
and neural-network models for treatment-specific
survival or hazard functions \citep{curth2021survite}.
 In particular,
\citet{cui2023estimating} establish asymptotic
normality and pointwise inference for causal
survival forests, but do not explicitly exploit
a shared low-dimensional temporal representation
to control estimation errors in local CATE changes. Neural-network survival
learners such as \citet{curth2021survite}
emphasize flexible estimation of treatment-specific
outcome functions rather than distributional
inference for correlated CATE contrasts.
\citet{westling2024inference} develop doubly robust,
cross-fitted estimators of treatment-specific
survival curves and establish pointwise and
uniform-in-time inference. Closely related to our
work, \citet{Sun2026-wa} develop a deep survival
learner that combines doubly robust pseudo-outcomes
with a multi-output neural network for joint
estimation of time-varying  CATEs.
Their theoretical analysis focuses on estimation
over fixed evaluation grids under assumed
network approximation and empirical convergence
rates.
 
 However, two important
 gaps remain. First, existing theory
for flexible survival CATE estimation primarily
addresses accuracy at individual or fixed
evaluation times, without characterizing how
estimation errors propagate to temporal contrasts
as the evaluation grid becomes finer. Accurate
CATE estimation at individual times does not
necessarily ensure accurate estimation of local
treatment-effect changes. 
Second, valid inference for nonlinear survival
CATE estimators and their correlated temporal
and covariate contrasts remains underdeveloped.
Although subsampling and ensemble-based inference
have been established using incomplete
$U$-statistics and infinitesimal jackknife
variance estimation
\citep{wager2014confidence,mentch2016quantifying},
these results do not directly accommodate
jointly modeled survival CATEs with estimated
nuisance functions. Related inference for
log-hazard ratios in nonparametric Cox regression
\citep{Ghosal2026} does not address causal
treatment effects or their temporal contrasts.
Inference for temporal contrasts over shrinking time intervals is  challenging because the contrast variance may vanish, requiring precise control of nonlinear approximation errors and covariance estimation relative to the shrinking variance scale.

We develop a deep survival learning framework that
combines flexible neural-network modeling of covariate
heterogeneity with a smooth temporal structure for the
true CATE trajectory.  
For estimation, we approximate the CATE trajectory
using a finite-dimensional spline basis and jointly
learn its covariate-dependent coefficients through
a multi-output neural network. We construct doubly
robust pseudo-outcomes using sample splitting to
account for censoring and nuisance estimation. 
By exploiting shared temporal structure, the proposed
approach improves the theoretical error bounds for
temporal contrasts, particularly at nearby  times.

Our main contributions are fourfold. First, we establish
nonasymptotic oracle bounds for CATE levels and temporal
contrasts, accounting for nuisance estimation, approximation,
and optimization errors. Bounded temporal derivatives and
Gram-matrix stability yield time-normalized contrast risk
bounds that remain stable under grid refinement, unlike
contrast bounds derived from separate time-specific estimators.
Second, we establish asymptotic normality for ensemble-based
CATE estimators using incomplete $U$-statistics and H\'ajek
projections, with localization and undersmoothing controlling
pointwise estimation bias. Third, we develop inference for
temporal contrasts over fixed or shrinking intervals and
covariate contrasts between patient profiles. A gap-uniform
nonlinear expansion, normalized bias control, and a
contrast-specific eigenvalue lower bound of the order of
the squared time gap ensure valid inference as temporal
intervals shrink. Finally, we establish consistency of a
finite-ensemble-adjusted infinitesimal jackknife covariance
estimator, including relative variance consistency for
shrinking temporal contrasts. These results yield
asymptotically valid confidence intervals for pointwise
CATEs and their temporal and covariate contrasts.

In the following, Section~\ref{sec:method} introduces the data structure,
target estimands, and function class.
Section~\ref{sec:estimation} develops the estimation
procedure, and Section~\ref{sec:cross_fitting}
presents ensemble-based inference for CATE.
Section~\ref{sec:theory} establishes the theoretical
results. Sections~\ref{sec:sim} and~\ref{sec:realdata}
present simulation studies and an application to the
Boston Lung Cancer Study, respectively.
Proofs and auxiliary results are provided in the Appendix.

\noindent {\it Notation.} 
We use boldface for vectors and matrices, and bold
lowercase letters also denote vector-valued functions.  
Let $\mathbf{X}$ be a $d$-dimensional covariate
vector with distribution $P_{\mathbf{X}}$
supported on $\mathcal{X}\subseteq\mathbb{R}^d$.
We use $\mathbf{x}_*$ to denote a fixed covariate
value and $\mathbf{x}$ to denote a generic element
of $\mathcal{X}$. 
Expectation subscripts indicate the random quantities
over which integration is taken, e.g.,
\(
\mathbb{E}_{\mathbf{X}}\{g(\mathbf{X})\}
=\int_{\mathcal{X}}g(\mathbf{x})\,dP_{\mathbf{X}}(\mathbf{x}).
\)
Let $\mathbb{I}(\cdot)$ denote the indicator function. For $\mathbf{x}=(x_1,\ldots,x_d)^\top$, define
$|\mathbf{x}|_p
=\bigg(\sum_{k=1}^d|x_k|^p\bigg)^{1/p}, 1\le p<\infty,
|\mathbf{x}|_\infty=\max_k|x_k|$, $|\mathbf{x}|_0=\sum_{k=1}^d\mathbb{I}(x_k\neq0).$
For a matrix $\mathbf{A}=(a_{kl})$, let $\|\mathbf{A}\|_0
=\sum_{k,l}\mathbb{I}(a_{kl}\neq0),
\|\mathbf{A}\|_\infty=\max_{k,l}|a_{kl}|.$ 
For $f:\mathcal{X}\to\mathbb{R}$, define $\|f\|_\infty
=\sup_{\mathbf{x}\in\mathcal{X}}|f(\mathbf{x})|,
\|f\|_{L_2(P_{\mathbf{X}})}
=\big\{\mathbb{E}_{\mathbf{X}}
[f^2(\mathbf{X})]\big\}^{1/2}.$  
  For deterministic sequences $a_n$ and $b_n>0$, we write
$a_n=O(b_n)$ if $|a_n/b_n|\le C$ for some constant
$C>0$ and all sufficiently large $n$, and
$a_n=o(b_n)$ if $a_n/b_n\to0$.
For random variables $X_n$ and a deterministic sequence
$a_n>0$, we write $X_n=O_p(a_n)$ if, for every
$\epsilon>0$, there exists $C>0$ such that
$\mathbb{P}(|X_n/a_n|>C)\le\epsilon$ for all sufficiently
large $n$, and $X_n=o_p(a_n)$ if
$X_n/a_n\overset{p}{\to}0$.
For nonnegative sequences, we write
$a_n\lesssim b_n$ if $a_n\le Cb_n$ for some constant
$C>0$ independent of $n$, and $a_n\asymp b_n$ if both
$a_n\lesssim b_n$ and $b_n\lesssim a_n$.
For $a\in\mathbb R$, let $\lceil a\rceil$ and
$\lfloor a\rfloor$ denote the smallest integer greater
than or equal to $a$ and the largest integer less than
or equal to $a$, respectively.
Finally, $a\wedge b=\min(a,b)$ and
$a\vee b=\max(a,b)$.

\section{Methodological Framework}\label{sec:method}

\subsection{Causal Identification via Pseudo-Outcomes}
\label{sec:causal_setup}

We consider right-censored survival data under the potential
outcomes framework \citep{imbens2015causal}.
For $i=1,\ldots,n$, let $T_i$ and $C_i$ denote the event
and censoring times, with observed time
$U_i=\min(T_i,C_i)$ and event indicator
$\Delta_i=\mathbb I(T_i\le C_i)$.
We observe i.i.d. records
$D_i=(U_i,\Delta_i,\mathbf X_i,W_i)$,
where $\mathbf X_i\in\mathcal X\subset\mathbb R^d$
are pre-treatment covariates with distribution $P_{\mathbf X}$
and compact support $\mathcal X$, and
$W_i\in\{0,1\}$ is the treatment indicator.
Write $\mathcal D_n=\{D_i\}_{i=1}^n$.
Let $T_i^w$ denote the potential survival time under
treatment $w\in\{0,1\}$, with consistency
$T_i=T_i^{W_i}$.
We assume unconfoundedness and treatment positivity
\citep{rosenbaum1983propensity}, and 
independent censoring  conditional on covariates.

 We evaluate treatment effects on a fixed compact time
window $\mathcal{T}=[t_{\min},t_{\max}]$, on which the conditional
censoring survival probabilities are bounded away from zero.
The formal identification and positivity conditions are stated
in Assumption~\ref{assum:causal}. For any fixed covariate profile
$\mathbf{x}_*\in\mathcal{X}$, the estimand is the
conditional average treatment effect (CATE):
\[
\tau_0(\mathbf{x}_*;t)
=
\mathbb{P}(T_i^1>t\mid\mathbf{X}_i=\mathbf{x}_*)
-
\mathbb{P}(T_i^0>t\mid\mathbf{X}_i=\mathbf{x}_*),
\qquad t\in\mathcal{T}.
\]
 
We estimate the CATE trajectory over a prespecified
deterministic grid
$t_{\min}=t_1<\cdots<t_J=t_{\max}$,
enabling joint estimation of treatment effects
at clinically relevant times and their
temporal changes.
The number of grid points $J$ may increase with $n$
to provide finer temporal resolution over a compact $\mathcal{T}.$

Because only one potential outcome can be observed per
individual and is subject to right censoring, we construct
an augmented inverse probability weighted pseudo-outcome
whose conditional expectation equals the CATE 
\citep{robins1994estimation}.  
Let $\eta=\{S^w,G^w,e^w;w=0,1\}$ denote the true
nuisance functions, where $S^w(\mathbf{x};t)
=\mathbb{P}(T>t\mid\mathbf{X}=\mathbf{x},W=w),\,
G^w(\mathbf{x};t)
=\mathbb{P}(C>t\mid\mathbf{X}=\mathbf{x},W=w),\,$ and $
e^w(\mathbf{x})=
\mathbb{P}(W=w\mid\mathbf{X}=\mathbf{x}).$
For $w\in\{0,1\}$ and $t\in\mathcal{T}$, define
\[
\pi_i^w(t; \mathbf{X}_i,\eta)
=
S^w(\mathbf{X}_i;t)
+
\frac{\mathbb{I}(W_i=w)}{e^w(\mathbf{X}_i)}
\bigg\{
\frac{Y_i(t)}{G^w(\mathbf{X}_i;t)}
-
S^w(\mathbf{X}_i;t)
\bigg\},
\]
where $Y_i(t)=\mathbb{I}(U_i>t)$.
Under conditional independent censoring,
\[
\mathbb{E}\{Y_i(t)\mid
\mathbf{X}_i=\mathbf{x},W_i=w\}
=
S^w(\mathbf{x};t)G^w(\mathbf{x};t).
\]
Consequently, the temporal pseudo-outcome
$\varphi_i(t)=\pi_i^1(t; \mathbf{X}_i,\eta)-\pi_i^0(t;\mathbf{X}_i,\eta)$
satisfies $\mathbb{E}\{\varphi_i(t)\mid\mathbf{X}_i=\mathbf{x}_*\}
=
\tau_0(\mathbf{x}_*;t)$, 
 reducing CATE estimation to conditional mean regression. 
In practice, we replace $\eta$ with estimates
$\widehat{\eta}$ fitted on an independent
subsample $\mathcal{D}_{\mathrm{nuis}}$, using the sample
split described in Section~\ref{sec:estimation}, and
construct a pseudo-outcome for each individual at each
evaluation time point:
\[
\widehat{\varphi}_i(t_j)
=
\pi_i^1(t_j;\mathbf{X}_i,\widehat{\eta})
-
\pi_i^0(t_j;\mathbf{X}_i,\widehat{\eta}).
\]
The construction is doubly robust with respect to $(e,S)$
when $G$ is consistently estimated: the conditional bias
vanishes if either $e$ or $S$ is consistently estimated,
under the stated boundedness and positivity conditions
\citep{Sun2026-wa}. We use these pseudo-outcomes to
construct a DNN estimator of the CATE trajectory.

\subsection{True Functional Class of CATE}
\label{sec:function_class}

We impose structural conditions on the CATE with
respect to both covariates and time: at each 
$t\in\mathcal{T}$, the function
$\mathbf{x}\mapsto\tau_0(\mathbf{x};t)$ belongs to a
compositional H\"older class, allowing nonlinearity governed by intrinsic dimensions
\citep{schmidt2020deep};  
for each fixed $\mathbf{x}\in\mathcal{X}$, the true
CATE trajectory $t\mapsto\tau_0(\mathbf{x};t)$
is assumed to be smooth on
$\mathcal{T}=[t_{\min},t_{\max}]$.

\begin{definition}[H\"older Class]
\label{def:standard_holder}
Let $\mathcal{D}\subset\mathbb{R}^p$ be a compact rectangle.  For $\beta>0$,  define $s_\beta=\lceil\beta\rceil-1,\,\,
\alpha_\beta=\beta-s_\beta\in(0,1].$
The H\"older class
$\mathcal{H}_p^\beta(\mathcal{D},\mathcal{K})$
consists of functions $f:\mathcal{D}\to\mathbb{R}$ satisfying
\[
\sum_{|\mathbf{k}|_1\le s_\beta}
\|\partial^{\mathbf{k}}f\|_\infty
+
\sum_{|\mathbf{k}|_1=s_\beta}
\sup_{\substack{\mathbf{u},\mathbf{v}\in\mathcal{D}\\
                 \mathbf{u}\ne\mathbf{v}}}
\frac{
|\partial^{\mathbf{k}}f(\mathbf{u})
-\partial^{\mathbf{k}}f(\mathbf{v})|
}{
|\mathbf{u}-\mathbf{v}|_\infty^{\alpha_\beta}
}
\le\mathcal{K}.
\]
Here, $\mathbf{k}\in\mathbb{N}_0^p$ is a multi-index,
$|\mathbf{k}|_1=\sum_{\ell=1}^p k_\ell$, and
$\partial_{\mathbf{0}}f=f$.  
\end{definition}

\begin{definition}[Composite H\"older Class]
\label{def:composite_holder}
For depth $q\ge0$, denote by 
$\mathbf{d}=(d_0,\ldots,d_{q+1})$ with
$d_0=d$ and $d_{q+1}=1$, and let
$\mathbf{d}^*=(d_0^*,\ldots,d_q^*)$ satisfy
$1\le d_i^*\le d_i$  and
$\boldsymbol{\gamma}=(\gamma_0,\ldots,\gamma_q)$
have positive entries. The class
$\mathcal{G}(q,\mathbf{d},\mathbf{d}^*,
\boldsymbol{\gamma},\mathcal{K})$
consists of scalar functions $
f=\mathbf{g}^{(q)}\circ\cdots\circ\mathbf{g}^{(0)},$
where
\[
\mathbf{g}^{(i)}:
[a_i,b_i]^{d_i}
\longrightarrow
[a_{i+1},b_{i+1}]^{d_{i+1}},
\qquad i=0,\ldots,q,
\]
with $\mathcal{X}\subseteq[a_0,b_0]^d$ and
$\max_i\{|a_i|,|b_i|\}\le\mathcal{K}$.
Each component $g_j^{(i)}$ depends on at most $d_i^*$ input
coordinates and, as a function of those coordinates with unused
arguments added if necessary, belongs to $ \mathcal{H}_{d_i^*}^{\gamma_i}
([a_i,b_i]^{d_i^*},\mathcal{K}).$
%Here, $q$ describes the target function and is distinct from the depth of the fitted neural network.
\end{definition}

As $\tau_0 \in (-1, 1)$,  we define its latent representation  as
\begin{equation}
\label{eq:psi_true}
\psi_0(\mathbf x;t)
=\operatorname{arctanh}\{\tau_0(\mathbf x;t)\} \in \mathbb R,
\end{equation}
where $\operatorname{arctanh}(z)
=\frac12\log\{(1+z)/(1-z)\}$ for $z\in(-1,1)$.
We specify the CATE function class through
the covariate and temporal regularity of  $\psi_0$ as follows.

\begin{definition}[Composite--Temporal Class]
\label{def:joint_composite}
Let $\mathcal G=
\mathcal G(q,\mathbf d,\mathbf d^*,
\boldsymbol\gamma,\mathcal K)$.
For constants $K_t>0$ and $\beta\in(0,1]$,
define
\[
\begin{aligned}
\mathcal G_{\mathrm{joint}}^*
(\mathcal G,K_t,\beta)
=\bigg\{
\tau=\tanh\circ\psi:\;&
\psi(\cdot;t)\in\mathcal G
\quad\text{for all }t\in\mathcal T,\\
&
\sup_{\mathbf x\in\mathcal X,\,
t\in\mathcal T}
\big\{
|\psi(\mathbf x;t)|
+|\partial_t\psi(\mathbf x;t)|
\big\}\le K_t,\\
&
\sup_{\mathbf x\in\mathcal X}
|\partial_t\psi(\mathbf x;t)
-\partial_t\psi(\mathbf x;t')|
\le K_t|t-t'|^\beta,
\quad t,t'\in\mathcal T
\bigg\}.
\end{aligned}
\]
\end{definition}

We assume
$\tau_0\in\mathcal G_{\mathrm{joint}}^*
(\mathcal G,K_t,\beta)$.
Thus, $\psi_0(\cdot;t)$ belongs to the
compositional H\"older class $\mathcal G$
at each fixed $t$, while its temporal derivative
is uniformly bounded and H\"older continuous. Following \citet{schmidt2020deep}, we define the effective
smoothness indices
\(
\gamma_i^* =
\gamma_i \prod_{\ell=i+1}^{q}(\gamma_\ell \wedge 1),
\quad i=0,\ldots,q.
\)
The corresponding intrinsic exponent and squared-error rate
sequence are
\begin{align}\label{eq:c_eff&phi_n}
c_{\mathrm{eff}}
=
\min_{0\le i\le q}
\frac{2\gamma_i^*}{2\gamma_i^*+d_i^*},
\qquad
\phi_n
=
\max_{0\le i\le q}
n^{-\frac{2\gamma_i^*}{2\gamma_i^*+d_i^*}}
=
n^{-c_{\mathrm{eff}}}.
\end{align}
%Define the effective index $i_{\mathrm{eff}}\in
%\operatorname*{arg\,min}_{0\le i\le q} \frac{2\gamma_i^*}{2\gamma_i^*+d_i^*}$. 
The sequence $\phi_n$ determines the intrinsic covariate
rate, with temporal approximation, nuisance estimation,
and optimization errors treated separately. 
%We nextintroduce the working temporal basis and DNN architecture, distinguishing them from the unknown true representation.

% ----------------Sec2.3------------

\subsection{Spline-Projected Multi-Output Deep Learning Architecture}
\label{sec:dnn_architecture}

We propose a multi-output neural network to estimate
covariate-dependent coefficients in the temporal
representation defined below.

 %We refer to the resulting estimator as \textbf{SPINE}, for \textbf{Sp}line-projected \textbf{I}nference for \textbf{N}eural \textbf{E}ffects.
\paragraph{Temporal representation.}
For each $m$, we use $m$ B-spline basis functions of fixed order
on $\mathcal T$, with prespecified quasi-uniform knots
\citep{deboor2001practical,schumaker2007spline}.
Define the working temporal space and evaluation matrix by
\begin{equation}
\label{eq:readout}
\mathcal B_m=\operatorname{span}\{B_{1,m},\ldots,B_{m,m}\},
\qquad
\mathbf B=\bigl[B_{\ell,m}(t_j)\bigr]_{j,\ell}
\in\mathbb R^{J\times m},
\end{equation}
such that for every $t\in\mathcal T$, the basis functions satisfy
\(
B_{\ell,m}(t)\ge0,\) for \(\ell=1,\ldots,m,\) and \(
\sum_{\ell=1}^m B_{\ell,m}(t)=1.\) Here, $m\le J$ and is allowed to grow with $n$ and assume that $\mathbf B$
has full column rank. Section~\ref{sec:sim} examines sensitivity
to the working dimension and basis.

\paragraph{Coefficient network.}  
Let $L\ge1$ denote the number of hidden layers and
$\mathbf p=(p_0,\ldots,p_{L+1})$ the layer widths,
where $p_0=d$ and $p_{L+1}=m$.
Write $\sigma(z)=\max\{z,0\}$ for the componentwise
ReLU activation and
$\sigma_{\mathbf v}(\mathbf z)=\sigma(\mathbf z-\mathbf v)$.
The coefficient network is
\begin{equation}
\label{eq:DNN}
\mathbf f
=(\mathcal W_L\circ\sigma_{\mathbf v_L})
\circ\cdots\circ
(\mathcal W_1\circ\sigma_{\mathbf v_1})
\circ\mathcal W_0,
\end{equation}
where $\mathcal W_l\in\mathbb R^{p_{l+1}\times p_l}$
for $l=0,\ldots,L$ and
$\mathbf v_l\in\mathbb R^{p_l}$ for $l=1,\ldots,L$. We use ReLU networks for their established sparse approximation guarantees for compositional H\"older functions  \citep{schmidt2020deep}.

For each $\ell=1,\ldots,m$, consider the subnetwork
consisting of the $\ell$-th output and all hidden
units that contribute to it through directed paths
of nonzero weights. Denote its weight matrices
and shifts by $\mathcal W_l^{(\ell)}$ and
$\mathbf v_l^{(\ell)}$.
For $\mathbf s=(s_1,\ldots,s_m)$, define
\begin{equation}
\label{dnn-class}
\begin{aligned}
\mathcal F_{\mathrm{joint}}(L,\mathbf p,\mathbf s,F)
=\Big\{&\mathbf f\text{ of the form \eqref{eq:DNN}}:\;
\max_{0\le l\le L}\|\mathcal W_l\|_\infty\le1,\quad
\max_{1\le l\le L}\|\mathbf v_l\|_\infty\le1,\\
&\|f_\ell\|_\infty\le F,\quad
\sum_{l=0}^{L}\|\mathcal W_l^{(\ell)}\|_0
+\sum_{l=1}^{L}\|\mathbf v_l^{(\ell)}\|_0
\le s_\ell,\quad \ell=1,\ldots,m
\Big\}.
\end{aligned}
\end{equation} 
Hidden parameters may be shared across outputs and are
counted separately toward each output's sparsity budget.
When $m=1$, the network has a scalar output.
Choosing $B_{1,1}(t)=1$ gives a time-invariant CATE model,
whereas $B_{1,1}(t)=\mathbb I(t=t_j)$ gives a fixed-time
DNN estimator that can be fitted separately without
requiring temporal smoothness.

\paragraph{CATE output.}
For $\mathbf f\in\mathcal F_{\mathrm{joint}}(L,\mathbf p,\mathbf s,F)$,
define
\[
\psi_{\mathbf f}(\mathbf x;t)
=\sum_{\ell=1}^m f_\ell(\mathbf x)B_{\ell,m}(t),
\qquad
\tau_{\mathbf f}=\tanh\circ\psi_{\mathbf f}.
\]
Define the class of CATE functions by
\(
\mathcal T_m
=\big\{\tau_{\mathbf f}:
\mathbf f\in\mathcal F_{\mathrm{joint}}(L,\mathbf p,\mathbf s,F)\big\}.
\)
On the evaluation grid, this implies $\big(\tau_{\mathbf f}(\mathbf x;t_1),\ldots,
     \tau_{\mathbf f}(\mathbf x;t_J)\big)^\top
=\tanh\{\mathbf B\mathbf f(\mathbf x)\}$,
where $\tanh$ acts componentwise.

\section{Estimation}
\label{sec:estimation}

We randomly partition  $\mathcal D_n$
into nuisance and training samples,
$\mathcal D_{\mathrm{nuis}}$ and
$\mathcal D_{\mathrm{train}}$ with $n_1=|\mathcal D_{\mathrm{train}}|$.
We estimate $\eta$ from Section~\ref{sec:causal_setup}
using $\mathcal D_{\mathrm{nuis}}$ and construct
$\widehat\varphi_i(t_j)$ for each
$i\in\mathcal D_{\mathrm{train}}$ using $\widehat\eta$. 
 {
Let $\mathbb P$ denote the joint probability measure
governing $\mathcal D_n$, random splitting and subsampling,
and any randomness arising from DNN fitting.
For any integrable random variable $U$ measurable with
respect to these sources of randomness, define
\(
\mathbb E(U)=\int U\,d\mathbb P.
\)
Let $\mathcal S_{\mathrm{nuis}}$ denote the sigma-field
generated by $\mathcal D_{\mathrm{nuis}}$.
For this $U$ and an event $A$, define \(
\mathbb E_{\mathrm{train}}(U)
=\mathbb E(U\mid\mathcal S_{\mathrm{nuis}}),
\,\,
\mathbb P_{\mathrm{train}}(A)
=\mathbb P(A\mid\mathcal S_{\mathrm{nuis}}).
\)
 Denote by $\mathbb P_{\mathrm{nuis}}$ the probability measure
governing $\mathcal D_{\mathrm{nuis}}$ and define
\(
\mathbb E_{\mathrm{nuis}}(U)
=\int U\,d\mathbb P_{\mathrm{nuis}}
\)
for any integrable $\mathcal S_{\mathrm{nuis}}$-measurable
 $U$.}  For ease of reference, a notation dictionary for
the theoretical development is provided in
Table~\ref{tab:proof_notation} in the Appendix.

%A subscript $*$ on an expectation, variance, or covariance conditions additionally on $\mathcal D_{\mathrm{train}}$, so only subsample selection and network fitting remain random.}
%The $L^2(P_{\mathbf{X}})$ norm is taken with
%respect to an independent covariate vector $\mathbf{X}\sim P_{\mathbf{X}}$.
 
We fit the coefficient network by minimizing the
time-averaged empirical squared loss
\begin{equation}
L_{n,J}(\mathbf{f})
=
\frac{1}{n_1J}
\sum_{i\in\mathcal{D}_{\mathrm{train}}}
\sum_{j=1}^J
\big\{
\widehat{\varphi}_i(t_j)
-\tau_{\mathbf{f}}(\mathbf{X}_i;t_j)
\big\}^2.
\label{eq:empirical_loss}
\end{equation}
Optimization is over
$\mathbf{f}\in\mathcal{F}_{\mathrm{joint}}(L,\mathbf{p},s,F)$,
with the temporal basis fixed. An approximate minimizer
$\widehat{\mathbf{f}}$ yields $
\widehat\tau(\mathbf{x};t)
=
\tanh\!\big\{
\sum_{\ell=1}^m
\widehat f_\ell(\mathbf{x})B_{\ell,m}(t)
\big\}.$
We quantify optimization error by  
\begin{equation}
\mathcal{E}_{\mathrm{opt},n,J}
=
\mathbb{E}_{\mathrm{train}}
\left[
L_{n,J}(\widehat{\mathbf{f}})
-
\inf_{\mathbf{f}\in\mathcal{F}_{\mathrm{joint}}
(L,\mathbf{p},s,F)}
L_{n,J}(\mathbf{f})
\right].
\label{eq:optimization_error}
\end{equation}

%%%%%%%%%%%%%%%%%%%%%%%%%%%%%%%%%%%%%%%%%%

We assess estimation accuracy through two complementary
risks. The average CATE estimation risk is defined to be
\begin{equation}\label{eq:joint_risk}
R_J(\widehat\tau,\tau_0)
=
\frac{1}{J}
\mathbb E_{\mathrm{train}}
\sum_{j=1}^J
\|\widehat\tau(\cdot;t_j)-\tau_0(\cdot;t_j)\|_{L^2(P_{\mathbf X})}^2,
\end{equation}
while the time-normalized temporal contrast risk is defined to be
\begin{equation}\label{eq:contrast_risk}
R_{\Delta,J}(\widehat\tau,\tau_0)
=
\frac{1}{J-1}
\sum_{j=1}^{J-1}
\mathbb E_{\mathrm{train}}
\left\|
\frac{\widehat\tau(\cdot;t_{j+1})-\widehat\tau(\cdot;t_j)}
{t_{j+1}-t_j}
-
\frac{\tau_0(\cdot;t_{j+1})-\tau_0(\cdot;t_j)}
{t_{j+1}-t_j}
\right\|_{L^2(P_{\mathbf X})}^2.
\end{equation}
The time normalization facilitates comparisons across
grid resolutions. 
Defining
$w_J(t)=\{(J-1)(t_{j+1}-t_j)\}^{-1}$
for $t\in[t_j,t_{j+1})$, as
$\max_j(t_{j+1}-t_j)\to0$, it follows
\[
R_{\Delta,J}
=
\mathbb E_{\mathrm{train}}
\int_{\mathcal T}w_J(t)
\|\partial_t(\widehat\tau-\tau_0)(\cdot;t)\|_{L^2(P_{\mathbf X})}^2\,dt
+o_{\mathbb P_{\mathrm{nuis}}}(1).
\]
 Thus, $R_{\Delta,J}$ measures accuracy in  temporal
changes.

%For an equallyspaced grid, $w_J(t)=1/(t_{\max}-t_{\min})$, yielding the average integrated derivative error; for unequal spacing, the weights reflect the distribution of evaluation times.

%%%%%%%%%%%%%%%%%%%%%%%%%%%%%%%%%%%%%%%%%%

The proposed estimator yields estimates of
$\tau_0(\mathbf{x}_*;t)$ over the evaluation grid,
from which two types of contrasts can be constructed:
\begin{equation}\label{eq:cate_contrasts}
\begin{aligned}
\theta^{\mathrm{tm}}(\mathbf{x}_*;t,t')
&=\tau_0(\mathbf{x}_*;t')
-\tau_0(\mathbf{x}_*;t), \quad t<t',\\
\theta^{\mathrm{pt}}(\mathbf{x}_*,\mathbf{x}'_*;t)
&=\tau_0(\mathbf{x}_*;t)
-\tau_0(\mathbf{x}'_*;t).
\end{aligned}
\end{equation}
These  quantify temporal changes in treatment
benefit and treatment-effect heterogeneity between
covariate profiles, respectively. Their plug-in
estimators replace $\tau_0$ with $\widehat\tau$.

\section{Inference for CATE}
\label{sec:cross_fitting}

\begin{algorithm}[t]
\caption{Ensemble Subsampling for CATE Inference (Single Split)}
\label{algo:single_split_esm}
\begin{algorithmic}[1]
\REQUIRE Data $\mathcal D_n=\{(U_i,\Delta_i,\mathbf X_i,W_i)\}_{i=1}^n$;
training grid $t_1,\ldots,t_J$ and working spline basis;
subsample size $1\le r<n_1$; number of fits $B\ge2$;
evaluation profiles and times; significance level $\varsigma$.
\STATE Split $\mathcal D_n$ into $\mathcal D_{\mathrm{nuis}}$ and
$\mathcal D_{\mathrm{train}}$ as in Section~\ref{sec:estimation}.
%Index the $n_1$ training subjects by $i=1,\ldots,n_1$.
\STATE Fit the nuisance functions on $\mathcal D_{\mathrm{nuis}}$
and construct $\widehat\varphi_i(t_j)$ for each training subject
and grid time as in Section~\ref{sec:causal_setup}.
\FOR{$b=1$ \TO $B$}  
  \STATE 
Sample
$\mathcal I^b$ uniformly from all size-$r$
subsets of the indices of $\mathcal D_{\mathrm{train}}$.
 Record
  $I_{b,i}=\mathbb I(i\in\mathcal I^b)$. \label{step4}
  \STATE Approximately minimize
  \[
  \frac{1}{rJ}\sum_{i\in\mathcal I^b}\sum_{j=1}^J
  \{\widehat\varphi_i(t_j)-\tau_{\mathbf f}(\mathbf X_i;t_j)\}^2
  \]
  over $\mathbf f\in\mathcal F_{\mathrm{joint}}
  (L,\mathbf p,\mathbf s,F)$ in \eqref{dnn-class} and obtain $\widehat f_\ell^b(\cdot), l=1, \ldots,m$.
 % For temporal inference under Theorem~\ref{thm:contrast_clt}(ii),impose its slope bound on these fits.
  \STATE At each requested profile $\mathbf x_*$ and time $t$, compute
  $\widehat\psi^b(\mathbf x_*;t)=
   \sum_{\ell=1}^{m}\widehat f_\ell^b(\mathbf x_*)B_{\ell,m}(t)$.
\ENDFOR
\STATE Compute
$\bar\psi_B=B^{-1}\sum_{b=1}^B\widehat\psi^b$,
set $\bar\tau_B=\tanh(\bar\psi_B)$, and form
the requested temporal and covariate contrast estimates defined in (\ref{eq:cate_contrast_estimates}).
\STATE Compute the adjusted IJ covariance using \eqref{eq:ij_cov}
and the contrast standard errors using \eqref{eq:var_temporal}.
\ENSURE Pointwise CATE intervals \eqref{eq:CATE_CI} and contrast
intervals \eqref{eq:contrast_CI}.
\end{algorithmic}
\end{algorithm}

We construct pointwise and contrast estimators
from the ensemble of $B$ subsample fits in Algorithm~\ref{algo:single_split_esm}.
We average the fitted latent
functions before  the $\tanh$ transformation:
\[
\bar\psi_B(\mathbf x_*;t)
=\frac1B\sum_{b=1}^B\widehat\psi^b(\mathbf x_*;t),
\qquad
\bar\tau_B(\mathbf x_*;t)
=\tanh\{\bar\psi_B(\mathbf x_*;t)\}.
\]
Then, we obtain the  covariate and temporal contrast
estimators of \eqref{eq:cate_contrasts} as
\begin{align}
\widehat\theta_B^{\mathrm{pt}}
(\mathbf x_*,\mathbf x'_*;t)
&=\bar\tau_B(\mathbf x_*;t)
-\bar\tau_B(\mathbf x'_*;t),
\nonumber\\
\widehat\theta_B^{\mathrm{tm}}(t,t';\mathbf x_*)
&=\bar\tau_B(\mathbf x_*;t')
-\bar\tau_B(\mathbf x_*;t),
\qquad t<t'.
\label{eq:cate_contrast_estimates}
\end{align}

Conditional on the estimated nuisance functions,
the order-invariant subsample fits yield an
incomplete U-statistic representation of
$\bar\psi_B$, motivating the IJ covariance estimator
\citep{mentch2016quantifying,Wager&Athey}.
Specifically, we  first estimate the first-order covariance of $\bar\psi_B$.  Let
$\bar I_{\cdot,i}=B^{-1}\sum_{b=1}^B I_{b,i}$
, where $ I_{b,i}$
is defined in Step \ref{step4} of Algorithm~\ref{algo:single_split_esm}.
For two evaluation points, the adjusted IJ estimator is
\begin{align}
\widehat\Gamma_\psi(\mathbf x_*,t;\mathbf x'_*,t')
={}&
\frac{n_1(n_1-1)}{(n_1-r)^2B(B-1)}
\sum_{b_1\ne b_2}
\left\{\sum_{i=1}^{n_1}
(I_{b_1,i}-\bar I_{\cdot,i})(I_{b_2,i}-\bar I_{\cdot,i})\right\}
\nonumber\\
&\quad\times
\{\widehat\psi^{b_1}(\mathbf x_*;t)-\bar\psi_B(\mathbf x_*;t)\}
\{\widehat\psi^{b_2}(\mathbf x'_*;t')-\bar\psi_B(\mathbf x'_*;t')\}.
\label{eq:ij_cov}
\end{align}
The sum over ordered pairs of distinct fits implements the
finite-$B$ correction to the usual IJ covariance
\citep{Ghosal2026}. Its diagonal gives
$\widehat\sigma_\psi^2(\mathbf x_*;t)
=\widehat\Gamma_\psi(\mathbf x_*,t;\mathbf x_*,t)$.

Since $\tanh$ is increasing, a pointwise $(1-\varsigma)$
confidence interval for $\tau_0(\mathbf x_*;t)$ is obtained by
transforming the two endpoints of the latent-scale interval:
\begin{equation}
\label{eq:CATE_CI}
\tanh\!\left\{\bar\psi_B(\mathbf x_*;t)
\pm z_{1-\varsigma/2}\widehat\sigma_\psi(\mathbf x_*;t)\right\},
\end{equation}
where $z_{1-\varsigma/2}$ is the standard normal quantile.

For both contrast estimates in \eqref{eq:cate_contrast_estimates},  
 the standard errors are estimated to be 
\begin{align}
&\widehat{\sigma}^{\,2}_{\mathrm{pt}}(\mathbf x_*,\mathbf x'_*;t)
={}
\{1-\bar\tau^2_B(\mathbf x_*;t)\}^2
   \widehat\sigma_\psi^2(\mathbf x_*;t)
+\{1-\bar\tau^2_B(\mathbf x'_*;t)\}^2
   \widehat\sigma_\psi^2(\mathbf x'_*;t)
\nonumber\\
&\hspace{30mm}-2\{1-\bar\tau^2_B(\mathbf x_*;t)\}
    \{1-\bar\tau^2_B(\mathbf x'_*;t)\}
    \widehat\Gamma_\psi(\mathbf x_*,t;\mathbf x'_*,t).\nonumber\\
&\widehat{\sigma}^{\,2}_{\mathrm{tm}}(\mathbf x_*;t,t')
={}
\{1-\bar\tau^2_B(\mathbf x_*;t)\}^2
   \widehat\sigma_\psi^2(\mathbf x_*;t)
+\{1-\bar\tau^2_B(\mathbf x_*;t')\}^2
   \widehat\sigma_\psi^2(\mathbf x_*;t')
\nonumber\\
&\hspace{30mm}-2\{1-\bar\tau^2_B(\mathbf x_*;t)\}
    \{1-\bar\tau^2_B(\mathbf x_*;t')\}
    \widehat\Gamma_\psi(\mathbf x_*,t;\mathbf x_*,t').
\label{eq:var_temporal}
\end{align}
The corresponding $(1-\varsigma)$ Wald intervals are
\begin{align}
&\widehat\theta_B^{\mathrm{pt}}
  (\mathbf x_*,\mathbf x'_*;t)
\ \pm\ z_{1-\varsigma/2}\,
\widehat{\sigma}_{\mathrm{pt}}(\mathbf x_*,\mathbf x'_*;t),
\nonumber\\
&\widehat\theta_B^{\mathrm{tm}}(t,t';\mathbf x_*)
\ \pm\ z_{1-\varsigma/2}\,
\widehat{\sigma}_{\mathrm{tm}}(\mathbf x_*;t,t').
\label{eq:contrast_CI}
\end{align}
%%%%%%%%%%%%%%%%%%%%%%%%%%%%
\section{Theoretical Properties}\label{sec:theory}

We establish oracle bounds and consistency for fixed-time and
joint CATE estimators, followed by asymptotic normality and
variance consistency for ensemble-based pointwise and contrast
estimators at prespecified evaluation points.
As in Algorithm~\ref{algo:single_split_esm},  recall
$n_1=|\mathcal D_{\mathrm{train}}|$. We assume 
$n_1/n$ approaches a constant in $(0,1)$ as $n\to \infty$.

\begin{assumption}[Causal Identification, Censoring, and Positivity]
\label{assum:causal}
With the notation of Section \ref{sec:causal_setup}, take $\mathcal X=[0,1]^d$ and
a fixed interval $\mathcal T=[t_{\min},t_{\max}]$, $t_{\max}<\infty$.
For all $i$, $w\in\{0,1\}$, and $\mathbf x\in\mathcal X$,
\begin{gather*}
T_i=T_i^{W_i},\qquad
(T_i^0,T_i^1)\perp\!\!\!\perp W_i\mid\mathbf X_i,\qquad
C_i\perp\!\!\!\perp T_i^w\mid(\mathbf X_i,W_i=w),\\
c\le e^w(\mathbf x)\le1-c,\qquad
G^w(\mathbf x;t_{\max})\ge c_G,
\end{gather*}
for fixed $c\in(0,1/2)$ and $c_G\in(0,1]$.
\end{assumption}

\begin{assumption}[Uniform Convergence of Nuisance Estimators]
\label{assum:nuisance}
Under the sample split of Section~\ref{sec:estimation}, for
$w\in\{0,1\}$ and deterministic $r_n^{(S)},r_n^{(G)},r_n^{(e)}\to0$, we assume
\begin{align*}
\|\widehat S^w-S^w\|_{\infty}
&=O_{\mathbb P_{\mathrm{nuis}}}(r_n^{(S)}),\\
\|\widehat G^w-G^w\|_{\infty}
&=O_{\mathbb P_{\mathrm{nuis}}}(r_n^{(G)}),\\
\|\widehat e^w-e^w\|_{\infty}
&=O_{\mathbb P_{\mathrm{nuis}}}(r_n^{(e)}),
\end{align*}
where the first two $\|\cdot\|_{\infty}$ are taken on $\mathcal X\times \mathcal T$ and the last on $\mathcal X$.
\end{assumption}

\begin{assumption}[Network Architecture]
\label{assum:dnn}
Fix any $\delta\in[0,c_{\mathrm{eff}})$ and $\mu>1$, with
$c_{\mathrm{eff}}$ and $\phi_{n_1}$ as in \eqref{eq:c_eff&phi_n}.
The parameters of the network class \eqref{dnn-class} satisfy
\begin{align*}
\textnormal{(i)}\hspace{1mm}&
\log_2 n_1 \lesssim L \lesssim (\log n_1)^\mu,\qquad
\textnormal{(ii)}\hspace{1mm} m n_1^{1+\delta}\phi_{n_1} \lesssim p_l \lesssim m n_1,
\qquad l=1,\ldots,L,\\
\textnormal{(iii)}\hspace{1mm}&
s_\ell \asymp n_1^{1+\delta}\phi_{n_1}\log n_1,
\quad \ell=1,\ldots,m,\quad
\textnormal{(iv)}\hspace{1mm}
F\ge1 \text{ is a sufficiently large fixed constant}.
\end{align*}
%The bounds hold uniformly over layers and outputs, with constants independent of $m$ and $n_1$.
\end{assumption}

Assumption~\ref{assum:causal} ensures causal identification
and bounded inverse probability weights
\citep{rosenbaum1983propensity,tsiatis2006semiparametric},
while Assumption~\ref{assum:nuisance} controls nuisance-estimation
errors and pseudo-outcome bias \citep{Sun2026-wa}.
Assumption~\ref{assum:dnn} extends the scalar architecture
of \citet{schmidt2020deep} to $m$ temporal coefficient functions,
with hidden-layer widths scaling with $m$ and each output
retaining the scalar sparsity budget.
Allowing $\delta>0$ permits undersmoothing for inference
\citep{Meng2025-zh}.
Additional conditions for negligible pointwise bias are
introduced later. Under these assumptions, we establish
oracle bounds for the proposed estimator.

 %The known $\tanh$ link does not increase approximation error or covering numbers because it is $1$-Lipschitz.
 %The fixed readout \eqref{eq:readout} contributes no free parameters and is therefore unconstrained, and the envelope bound in (iv) carries the basis overlap constant $C_B$ so that the induced class of surface estimators covers the radius of $\mathcal{G}^{*}_{\mathrm{joint}}$.

Fix $j\in\{1,\ldots,J\}$. By Definition~\ref{def:joint_composite},
$\psi_0(\cdot;t_j)\in\mathcal G$. Taking $m=p_{L+1}=1$ in
\eqref{eq:DNN} and $B_{1,1}(t)=\mathbb I(t=t_j)$ gives a separate
fit $\widehat\tau(\cdot;t_j)$ at this time. \begin{equation}
\label{eq:tau_fixed_apx}
\tau_{\mathrm{apx}}(\cdot;t_j)
\in\operatorname*{arg\,min}_{g\in
\{\tanh\circ f:\,f\in\mathcal F_{\mathrm{joint}}(L,\mathbf p,\mathbf s,F)\}}
\|g-\tau_0(\cdot;t_j)\|_\infty
\end{equation}
and the nuisance-estimation error
\begin{equation}
\label{nuis-error}
\widehat{\mathcal E}_n
=\sum_{w\in\{0,1\}}\left\{
\|\widehat S^w-S^w\|_\infty
+\|\widehat e^w-e^w\|_\infty
+\|\widehat G^w-G^w\|_\infty
\right\},
\end{equation}
where the suprema in \eqref{nuis-error} are over $\mathcal X$
for $e^w$ and over $\mathcal X\times\mathcal T$ for $S^w$ and $G^w$. The following theorem bounds the population risk
\(
R_1\big(\widehat\tau(\cdot;t_j),\tau_0(\cdot;t_j)\big)
=\mathbb E_{\mathrm{train}}\!\Big[
\|\widehat\tau(\cdot;t_j)-\tau_0(\cdot;t_j)\|_{L_2(P_{\mathbf X})}^2
\Big],
\)
the single-time case of \eqref{eq:joint_risk}.

%%%%%%%%%%Theorem1%%%%%%%%%%%%

\begin{theorem}[Fixed-Time Oracle Bound and Convergence Rate]
\label{prop:oracle_fixed_t}
Fix $j\in\{1,\ldots,J\}$ and suppose $\psi_0(\cdot;t_j)\in\mathcal G$
as in Definition~\ref{def:composite_holder}.
Under Assumptions~\ref{assum:causal}--\ref{assum:dnn} with
$m=p_{L+1}=1$ and $0\le\delta<c_{\mathrm{eff}}$, the following hold
with $\mathcal E_{\mathrm{opt},n,1}$, $\tau_{\mathrm{apx}}(\cdot;t_j)$, $\phi_{n_1}$, and $\widehat{\mathcal E}_n$ as in
\eqref{eq:optimization_error}, \eqref{eq:tau_fixed_apx},
\eqref{eq:c_eff&phi_n}, and \eqref{nuis-error}, respectively.

\begin{enumerate}[(i)]
\item There exist a constant $C>0$ and an $\mathcal S_{\mathrm{nuis}}$-measurable event
$\Omega_n$ with
$\mathbb P_{\mathrm{nuis}}(\Omega_n)\to1$ such that on $\Omega_n$, it holds that 
\[
\begin{aligned}
R_1\big(\widehat\tau(\cdot;t_j),\tau_0(\cdot;t_j)\big)
\le C\Big\{&
\mathcal E_{\mathrm{opt},n,1}
+n_1^\delta\phi_{n_1}L\log^2 n_1+\|\tau_{\mathrm{apx}}(\cdot;t_j)-\tau_0(\cdot;t_j)\|_\infty^2
+\widehat{\mathcal E}_n^2\Big\}.
\end{aligned}
\]

\item If $\mathcal E_{\mathrm{opt},n,1}+\widehat{\mathcal E}_n^2
=O_{\mathbb P_{\mathrm{nuis}}}(n_1^\delta\phi_{n_1}L\log^2 n_1)$, then
\[
\|\widehat\tau(\cdot;t_j)-\tau_0(\cdot;t_j)\|_{L_2(P_{\mathbf X})}
=O_{\mathbb P}\!\left(
[n_1^\delta\phi_{n_1}L\log^2 n_1]^{1/2}\right).
\]
\end{enumerate}
\end{theorem}

Theorem~\ref{prop:oracle_fixed_t}(i) decomposes the
conditional risk into optimization, stochastic,
approximation, and nuisance errors.
Since $\psi_0(\cdot;t_j)\in\mathcal G$ by
Definition~\ref{def:joint_composite}, the approximation
error depends only on covariate regularity.
Part~(ii) establishes the resulting
$L_2(P_{\mathbf X})$ convergence rate, which,
for $\delta=0$, matches the composite H\"older rate
up to logarithmic factors \citep{schmidt2020deep}.
Allowing $\delta>0$ accommodates undersmoothing
for inference.

For $1<m\le J$, joint estimation uses the working
temporal basis to estimate the CATE surface over
$\mathcal T$, including times between grid points.
Define the following approximations, with supremum
norms taken over $\mathcal X\times\mathcal T$:
\begin{align}
\psi_{m,\mathrm{apx}}
&\in\operatorname*{arg\,min}_{\psi:\,\tanh\circ\psi\in\mathcal T_m}
\|\psi-\psi_0\|_\infty,\qquad
\tau_{m,\mathrm{apx}}
=\tanh\circ\psi_{m,\mathrm{apx}}.
\label{eq:tau_m_apx}
\end{align}

\begin{theorem}[Joint Non-Asymptotic Risk Bounds]
\label{thm:joint_oracle}
 Suppose Assumptions~\ref{assum:causal}--\ref{assum:dnn} hold
with $\delta\in[0,c_{\mathrm{eff}})$ and
$\tau_0\in\mathcal G_{\mathrm{joint}}^*(\mathcal G,K_t,\beta)$.
Let $t_{\min}=t_1<\cdots<t_J=t_{\max}$ be a deterministic grid,
with $m\le J$, where both $m$ and $J$ may increase with $n$.
Assume that the evaluation matrix $\mathbf B$ in
\eqref{eq:readout} satisfies
\(
\frac{c_B}{m}I_m
\preceq \frac{\mathbf B^\top\mathbf B}{J}
\preceq \frac{C_B}{m}I_m,
\)
for constants $c_B,C_B>0$ independent of $m$ and $J$.
On the event $\Omega_n$ defined in
Theorem~\ref{prop:oracle_fixed_t}, with
$\mathbb P_{\mathrm{nuis}}(\Omega_n)\to1$,
the joint estimator $\widehat\tau\in\mathcal T_m$
satisfies the following bounds.
\begin{enumerate}[(i)]
\item The average population risk in \eqref{eq:joint_risk}
satisfies
\[
R_J(\widehat\tau,\tau_0)
\le C\Big\{
\mathcal E_{\mathrm{opt},n,J}
+n_1^\delta\phi_{n_1}L\log^2 n_1
+\widehat{\mathcal E}_n^2
+\|\tau_{m,\mathrm{apx}}-\tau_0\|_\infty^2
\Big\}.
\]

\item The time-normalized contrast risk in
\eqref{eq:contrast_risk} satisfies
\[
\begin{aligned}
R_{\Delta,J}(\widehat\tau,\tau_0)
\le{}& C_\Delta m^2\Big\{
\mathcal E_{\mathrm{opt},n,J}
+n_1^\delta\phi_{n_1}L\log^2 n_1
+\widehat{\mathcal E}_n^2
+\|\tau_{m,\mathrm{apx}}-\tau_0\|_\infty^2
\Big\}\\
&+C_\Delta
\|\partial_t(\tau_{m,\mathrm{apx}}-\tau_0)\|_\infty^2.
\end{aligned}
\]
\end{enumerate}
Here $\mathcal E_{\mathrm{opt},n,J}$, $\phi_{n_1}$,
$\widehat{\mathcal E}_n$, and $\tau_{m,\mathrm{apx}}$
are defined in \eqref{eq:optimization_error},
\eqref{eq:c_eff&phi_n}, \eqref{nuis-error}, and
\eqref{eq:tau_m_apx}, respectively.
The constants $C,C_\Delta>0$ are independent of
$n_1$, $m$, $J$, and the grid gaps.
\end{theorem}

The local support of the B-spline basis and the
per-output sparsity condition in
Assumption~\ref{assum:dnn} control the complexity
of the joint estimator. Together with the joint
optimization bound, these yield part~(i).
For part~(ii), the derivative bound on the spline
basis introduces an additional factor of $m^2$
in the squared temporal contrast risk.

%\xrcm{We will reconstruct all these paragraphs.}

%\xrcm{1. By the Lemma~??? the final rate of Theorem~XXX is XXXX. (Directly give the rate of $R_J$ and $R_{\Delta,J}$).}

%\xrcm{2. Talk about $m_{opt}$, then we can have best order.}

%\xrcm{3. Compared with the separation case, what are the conditions for the benefits of Joint? In one paragraph. Then talk about $J$ is in our hand.}

Suppose the optimization and squared nuisance errors are
$O_{\mathbb P_{\mathrm{nuis}}}
(n_1^\delta\phi_{n_1}L\log^2 n_1)$.
By Lemma~\ref{lem:apprx} in the Appendix, we have
\[
\begin{aligned}
\|\tau_{m,\mathrm{apx}}-\tau_0\|_\infty^2
\lesssim m^{-2(1+\beta)}
+\phi_{n_1}^{\,1+\delta/(1-c_{\mathrm{eff}})},\qquad\|\partial_t(\tau_{m,\mathrm{apx}}-\tau_0)\|_\infty^2
\lesssim m^{-2\beta}
+m^2\phi_{n_1}^{\,1+\delta/(1-c_{\mathrm{eff}})}.
\end{aligned}
\]
Since $0\le\delta<c_{\mathrm{eff}}<1$ and
$\phi_{n_1}\to0$, the terms involving
$\phi_{n_1}^{\,1+\delta/(1-c_{\mathrm{eff}})}$
are absorbed into the corresponding stochastic
errors. Substituting these bounds
in Theorem~\ref{thm:joint_oracle} yields
\[
\begin{aligned}
R_J(\widehat\tau,\tau_0)
&=O_{\mathbb P_{\mathrm{nuis}}}\!\left(
n_1^\delta\phi_{n_1}L\log^2 n_1
+m^{-2(1+\beta)}\right),\\
R_{\Delta,J}(\widehat\tau,\tau_0)
&=O_{\mathbb P_{\mathrm{nuis}}}\!\left(
m^2n_1^\delta\phi_{n_1}L\log^2 n_1
+m^{-2\beta}\right).
\end{aligned}
\]
%\xrcm{We can hide this rate, and directly use $m_{opt}$ to get the rate below (20)}
As in Theorem~\ref{prop:oracle_fixed_t}(ii),
conditional Markov's inequality gives the same
respective orders under general $\mathbb P$
for the realized average squared errors in levels
and time-normalized contrasts. Balancing the two terms in the temporal contrast bound gives
\begin{equation}
\label{eq:m_opt}
m_{\mathrm{opt}}(\delta)
\asymp
\left(n_1^\delta\phi_{n_1}L\log^2 n_1
\right)^{-1/\{2(1+\beta)\}}.
\end{equation}
We choose
$m\asymp m_{\mathrm{opt}}(\delta)$ as it minimizes the contrast bound in order and gives
\begin{equation} \label{junk}
\begin{aligned}
R_J(\widehat\tau,\tau_0)
&=O_{\mathbb P_{\mathrm{nuis}}}\!\left(
n_1^\delta\phi_{n_1}L\log^2 n_1\right),
\qquad R_{\Delta,J}(\widehat\tau,\tau_0)
=O_{\mathbb P_{\mathrm{nuis}}}\!\left(
(n_1^\delta\phi_{n_1}L\log^2 n_1)^{\beta/(1+\beta)}\right).
\end{aligned}
\end{equation}
\noindent
Suppose the fixed-time bounds in Theorem~\ref{prop:oracle_fixed_t} hold uniformly over $t_1<\cdots<t_J$. Let the superscript $\mathrm{sep}$ denote separate fits at the $J$
evaluation times, so $R_J^{\mathrm{sep}}$ is the average of their
fixed-time risks $R_1$. Applying Theorem~\ref{prop:oracle_fixed_t}
uniformly over the grid   gives
\[
\begin{aligned}
R_J^{\mathrm{sep}}
=O_{\mathbb P_{\mathrm{nuis}}}\!\left(n^{\delta}_{1}\phi_{n_1}L\log^2 n_1\right),\qquad
R_{\Delta,J}^{\mathrm{sep}}
=O_{\mathbb P_{\mathrm{nuis}}}\!\Big(
\frac{n^{\delta}_{1}\phi_{n_1}L\log^2 n_1}{J-1}
\sum_{j=1}^{J-1}\frac{1}{(t_{j+1}-t_j)^2}
\Big).
\end{aligned}
\]
 For an equally spaced grid over a fixed time window,
separate fitting yields
\(
R_{\Delta,J}^{\mathrm{sep}}
=O_{\mathbb P_{\mathrm{nuis}}}
\big(J^2n_1^\delta\phi_{n_1}L\log^2 n_1\big).
\)
The joint temporal contrast bound in \eqref{junk}
is asymptotically lower when
\(
J^{2(1+\beta)}n_1^\delta
\phi_{n_1}L\log^2 n_1\to\infty.
\)
For example, 
this may  hold when $J\asymp n_1^{1/2}$.
 The theoretical advantage is based on upper bounds,
as the separate-fitting bound ignores dependence
across time points. Simulations further demonstrate
improved temporal contrast accuracy.

%\xrcm{We will reconstruct this paragraph, make the logic clearer.

%%%%%%%%%%%%%%%%%%%%

To quantify the sampling variability of the ensemble
estimator, we use its first-order H\'ajek projection,
whose covariance depends on two subsample fits
sharing exactly one observation. Specifically,
consider two estimators, each fitted using $r$
training observations, with one observation
shared between them.  
Conditional on $\mathcal D_{\mathrm{nuis}}$,
both fits use the same $\widehat\eta$, with
all remaining observations and fitting randomness
independent across fits.
 Define the single-overlap covariance
\begin{equation}
\label{eq:zeta_r}
\zeta_r^{(1)}(\mathbf x_*,t;\mathbf x_*',t')
=
\operatorname{Cov}_{\mathrm{train}}\!\left\{
\widehat\psi_r(\mathbf x_*;t),
\widehat\psi_r'(\mathbf x_*';t')\right\}.
\end{equation}
Write $\zeta_r^{(1)}(\mathbf x_*;t)
=\zeta_r^{(1)}(\mathbf x_*,t;\mathbf x_*,t)$.
The first-order covariance and variance components
of the ensemble estimator are
\begin{equation}
\label{eq:CATE_var_true}
\Gamma_\psi(\mathbf x_*,t;\mathbf x_*',t')
=\frac{r^2}{n_1}
\zeta_r^{(1)}(\mathbf x_*,t;\mathbf x_*',t'),
\qquad
\sigma_\psi^2(\mathbf x_*;t)
=\Gamma_\psi(\mathbf x_*,t;\mathbf x_*,t).
\end{equation}

Since $\tanh'(\psi_0)=1-\tau_0^2$, the first-order contrast
variances corresponding to \eqref{eq:var_temporal} are
\begin{align}
\sigma_{\mathrm{pt}}^2(\mathbf x_*,\mathbf x_*';t)
={}&
\{1-\tau_0^2(\mathbf x_*;t)\}^2\sigma_\psi^2(\mathbf x_*;t)
+\{1-\tau_0^2(\mathbf x_*';t)\}^2\sigma_\psi^2(\mathbf x_*';t)
\nonumber\\
&-2\{1-\tau_0^2(\mathbf x_*;t)\}
    \{1-\tau_0^2(\mathbf x_*';t)\}
    \Gamma_\psi(\mathbf x_*,t;\mathbf x_*',t),
\nonumber\\[2pt]
\sigma_{\mathrm{tm}}^2(t,t';\mathbf x_*)
={}&
\{1-\tau_0^2(\mathbf x_*;t)\}^2\sigma_\psi^2(\mathbf x_*;t)
+\{1-\tau_0^2(\mathbf x_*;t')\}^2\sigma_\psi^2(\mathbf x_*;t')
\nonumber\\
&-2\{1-\tau_0^2(\mathbf x_*;t)\}
    \{1-\tau_0^2(\mathbf x_*;t')\}
    \Gamma_\psi(\mathbf x_*,t;\mathbf x_*,t').
\label{eq:CATE_contrast_true}
\end{align}
For the temporal contrast per unit time, the variance is
$\sigma_{\mathrm{tm}}^2(t,t';\mathbf x_*)/(t'-t)^2$.
For inference, $\psi_{m,\mathrm{apx}}$ in \eqref{eq:tau_m_apx}
uses the architecture at sample size $r$.

\begin{assumption}[Optimization and Nuisance Errors]
\label{assum:opt_nuisance_error}
For DNN fits based on $r$
training observations, assume
\(
\mathcal E_{\mathrm{opt},r,J}
+\widehat{\mathcal E}_n^2
=O_{\mathbb P_{\mathrm{nuis}}}\!\left(
r^\delta\phi_r L\log^2 r\right),
\)
where $\mathcal E_{\mathrm{opt},r,J}$ and
$\widehat{\mathcal E}_n$ are defined in
\eqref{eq:optimization_error} and
\eqref{nuis-error}, respectively.
\end{assumption}
 
This assumption bounds the optimization and squared nuisance
errors by the stochastic rate at subsample size $r$,
allowing the inference analysis to focus on ensemble
bias and variance.
For subsample size $r$, define
\[
M_r(\mathbf x)
=\mathbb E_{\mathrm{train}}\!\left[
\sup_{t\in\mathcal T}
|(\widehat\psi_r-\psi_0)(\mathbf x;t)|^2
+\sup_{t\in\mathcal T}
|\partial_t(\widehat\psi_r-\psi_0)(\mathbf x;t)|^2
\right].
\]
The following corollary converts the integrated risk
bounds into control of latent estimation errors and
temporal derivatives at covariate profiles outside
a set of vanishing probability.

\begin{corollary}
\label{lem:An}
Suppose the conditions of Theorem~\ref{thm:joint_oracle} and
Assumption~\ref{assum:opt_nuisance_error} hold at subsample size
$r$, with $0<\delta<c_{\mathrm{eff}}$. For any deterministic
sequence $\ell_n\to\infty$, there exists a deterministic set
$\mathcal A_n\subseteq\mathcal X$ with
$P_{\mathbf X}(\mathcal A_n^c)\le\ell_n^{-1}$ such that,
for every deterministic sequence $\mathbf x_{*,n}\in\mathcal A_n$,
\[
\mathbb P_{\mathrm{nuis}}\!\left\{
M_r(\mathbf x_{*,n})>
\ell_n^3\big(m^3r^\delta\phi_r L\log^2 r+m^{1-2\beta}\big)
\right\}\longrightarrow0.
\]
\end{corollary}

\begin{assumption}[Pointwise Nondegeneracy]
\label{assum:xi}
For a fixed evaluation point $(\mathbf x_*;t)\in\mathcal X\times\mathcal T$, with $\mathbf x_*\in\mathcal A_n$ in Corollary \ref{lem:An} for all sufficiently large $n$.
For some $\nu\ge1$ and $c_\zeta>0$, at each such point,
\[
\mathbb P_{\mathrm{nuis}}\!\left\{
\zeta_r^{(1)}(\mathbf x_*;t)\ge c_\zeta r^{-\nu}
\right\}\longrightarrow1.
\]
\end{assumption}

This condition prevents the single-overlap covariance at the
evaluation points from decaying faster than $r^{-\nu}$, as in
subsampled neural-network inference \citep{Meng2025-zh} and
honest random forests \citep{Wager&Athey}.

\begin{assumption}[Bias Domination]
\label{assum:bias_domination}
At each evaluation profile $\mathbf x_*$ in
Assumption~\ref{assum:xi}, for some $C>0$ and the same $\nu$,
\[
\mathbb P_{\mathrm{nuis}}\!\left\{
\sup_{t\in\mathcal T}
\left|
\mathbb E_{\mathrm{train}}[\widehat\psi_r(\mathbf x_*;t)]
-\psi_{m,\mathrm{apx}}(\mathbf x_*;t)
\right|^2
\le C\left(
\|\psi_{m,\mathrm{apx}}-\psi_0\|_\infty^2+r^{-\nu}
\right)
\right\}\longrightarrow1.
\]
\end{assumption}

Assumption~\ref{assum:bias_domination} bounds the squared
conditional fitting bias by the squared uniform approximation
error plus $r^{-\nu}$, allowing nonzero bias even under exact
approximation, unlike \citet[Proposition~3.6]{Meng2025-zh}.
For evaluation points with positive single-overlap variances,
we define
\[
\rho_r(\mathbf x_*,t;\mathbf x_*',t')
=
1-\frac{|\zeta_r^{(1)}(\mathbf x_*,t;\mathbf x_*',t')|}
{\{\zeta_r^{(1)}(\mathbf x_*;t)\,
\zeta_r^{(1)}(\mathbf x_*';t')\}^{1/2}},
\]
where smaller $\rho_r$ indicates stronger absolute first-order
correlation.

Let $1/2<\beta\le1$, $0<\delta<c_{\mathrm{eff}}$, and
$r=\lfloor n_1^\alpha\rfloor$ with $0<\alpha<1$.
For pointwise CATEs and covariate contrasts, use
\begin{equation}
\label{eq:inference_tuning}
\alpha_{\mathrm{lower,pt}}
=\frac{1}{2-\nu+c_{\mathrm{eff}}-\delta}
<\alpha<\frac1\nu,
\qquad
1<\nu<1+\frac{c_{\mathrm{eff}}-\delta}{2}.
\end{equation}
For temporal contrasts, take $1/2<\beta<1$ and use
\begin{equation}
\label{eq:inference_tuning_tm}
\alpha_{\mathrm{lower,tm}}
=\biggl\{2-\nu+
\frac{\beta(c_{\mathrm{eff}}-\delta)}{1+\beta}\biggr\}^{-1}
<\alpha<\frac1\nu,
\qquad
1<\nu<1+\frac{\beta(c_{\mathrm{eff}}-\delta)}{2(1+\beta)}.
\end{equation}
The bounds on $\nu$ ensure nonempty ranges for $\alpha$.
The temporal conditions suffice for all three targets.
The full tuning ranges are imposed for the subsequent CLTs.
Bias negligibility requires only the corresponding lower
bound on $\alpha$.
If $\partial_t\psi_0$ is Lipschitz, any working
$1/2<\beta<1$ may be used consistently in
\eqref{eq:m_opt} and \eqref{eq:inference_tuning_tm}.

\begin{proposition}[Negligible Bias and Contrast Centering]
\label{cor:asymptotic_bias}
Suppose the conditions of Corollary~\ref{lem:An} and
Assumptions~\ref{assum:xi} and~\ref{assum:bias_domination}
hold at the prespecified evaluation points, with covariate
profiles in $\mathcal A_n$ for all sufficiently large $n$.
Take an integer $m\asymp m_{\mathrm{opt}}(\delta)$ in
\eqref{eq:m_opt}, evaluated at subsample size $r$.
For parts~(i) and~(ii), require
$0<\alpha_{\mathrm{lower,pt}}<\alpha$.
For part~(iii), require $0<\alpha_{\mathrm{lower,tm}}<\alpha$.

\begin{enumerate}[(i)]
\item \emph{Pointwise bias.}
\[
\frac{
|\mathbb E_{\mathrm{train}}[\widehat\psi_r(\mathbf x_*,t)]
-\psi_0(\mathbf x_*,t)|^2}
{\sigma_\psi^2(\mathbf x_*,t)}
=o_{\mathbb P_{\mathrm{nuis}}}(1).
\]

\item \emph{Covariate contrast.}
If, for some $c_\rho>0$,
\(
\mathbb P_{\mathrm{nuis}}\{
\rho_r(\mathbf x_*,t,\mathbf x_*',t)\ge c_\rho
\}\to1,
\)
then
\[
\frac{
\big|
\tanh\{\mathbb E_{\mathrm{train}}
[\widehat\psi_r(\mathbf x_*,t)]\}
-\tanh\{\mathbb E_{\mathrm{train}}
[\widehat\psi_r(\mathbf x_*',t)]\}
-\theta^{\mathrm{pt}}(\mathbf x_*,\mathbf x_*',t)
\big|^2}
{\sigma_{\mathrm{pt}}^2(\mathbf x_*,\mathbf x_*',t)}
=o_{\mathbb P_{\mathrm{nuis}}}(1).
\]

\item \emph{Temporal contrast.}
Let $t<t'$, with $t'-t$ possibly tending to zero.
If, for some $c_\rho>0$,
\(
\mathbb P_{\mathrm{nuis}}\{
\rho_r(\mathbf x_*,t,\mathbf x_*,t')
\ge c_\rho(t'-t)^2
\}\to1,
\)
then
\[
\frac{
\big|
\tanh\{\mathbb E_{\mathrm{train}}
[\widehat\psi_r(\mathbf x_*,t')]\}
-\tanh\{\mathbb E_{\mathrm{train}}
[\widehat\psi_r(\mathbf x_*,t)]\}
-\theta^{\mathrm{tm}}(t,t',\mathbf x_*)
\big|^2}
{\sigma_{\mathrm{tm}}^2(t,t',\mathbf x_*)}
=o_{\mathbb P_{\mathrm{nuis}}}(1).
\]
\end{enumerate}
Here $\psi_0$ and $\sigma_\psi^2$ are as in
\eqref{eq:psi_true} and \eqref{eq:CATE_var_true}.
The contrasts and their variances are as in
\eqref{eq:cate_contrasts} and \eqref{eq:CATE_contrast_true}.
\end{proposition}

The correlation condition in Proposition~\ref{cor:asymptotic_bias}(iii) is motivated
by mean-square differentiability of the first-order projection
process, whose increments have variance of order $(t'-t)^2$
when the derivative variance is nondegenerate.
The proposition establishes negligible pointwise bias and
contrast centering errors relative to their standard errors.
The lower bounds on $\alpha$ control these errors.
Part~(iii) requires a larger lower bound on $\alpha$ because
slope approximation has squared error $m^{-2\beta}$,
compared with $m^{-2(1+\beta)}$ for levels.
Since the conditional fitting bias relative to the spline
approximant is itself a spline, these centering results
require neither a separate increment-bias assumption nor
a minimum time separation.

\begin{theorem}[Pointwise Asymptotic Normality]
\label{thm:pointwise_clt}
Suppose the conditions of Proposition~\ref{cor:asymptotic_bias}(i) 
hold at $(\mathbf x_*;t)$ and $B\gtrsim n_1$ and $m$ in \eqref{eq:m_opt}. Then, with \eqref{eq:inference_tuning} under
$\mathbb P$,
\[
\frac{\bar\psi_B(\mathbf x_*;t)-\psi_0(\mathbf x_*;t)}
     {\sigma_\psi(\mathbf x_*;t)}
\xrightarrow{d}\mathcal N(0,1),
\]
with $\bar\psi_B$, $\psi_0$, and $\sigma_\psi^2$ as in
Algorithm~\ref{algo:single_split_esm}, \eqref{eq:psi_true},
and \eqref{eq:CATE_var_true}, respectively.
\end{theorem}

Theorem~\ref{thm:pointwise_clt} establishes pointwise asymptotic normality on the $\psi$ scale, with the lower bound on $\alpha$ from Proposition~\ref{cor:asymptotic_bias}(i) making the bias negligible and the additional upper bound $\alpha<1/\nu$ ensuring that the first-order H\'ajek projection dominates.

\begin{theorem}[Asymptotic Normality of Contrasts]
\label{thm:contrast_clt}
Let $B\gtrsim n_1$. With the contrasts, estimators, and variances
in \eqref{eq:cate_contrasts}, \eqref{eq:cate_contrast_estimates},
and \eqref{eq:CATE_contrast_true}, respectively, the following
hold under $\mathbb P$.

\begin{enumerate}[(i)]
\item \emph{Covariate contrast.}
Under the conditions of Proposition~\ref{cor:asymptotic_bias}(ii) and \eqref{eq:inference_tuning},
\[
\frac{\widehat\theta_B^{\mathrm{pt}}(\mathbf x_*,\mathbf x_*';t)
      -\theta^{\mathrm{pt}}(\mathbf x_*,\mathbf x_*';t)}
     {\sigma_{\mathrm{pt}}(\mathbf x_*,\mathbf x_*';t)}
\xrightarrow{d}\mathcal N(0,1).
\]

\item \emph{Temporal contrast.}
 In Algorithm~\ref{algo:single_split_esm}, each subsample
estimator minimizes the loss, up to the specified
optimization error, over the network class
\eqref{dnn-class}, subject to
\[
\|\partial_t\psi_{\mathbf f}\|_\infty
=
\sup_{\mathbf x\in\mathcal X,\,t\in\mathcal T}
\left|
\sum_{\ell=1}^{m}
f_\ell(\mathbf x)\frac{d}{dt} B_{\ell,m}(t)
\right|
\le F_\Delta,
\]
where $F_\Delta>K_t$ is fixed.
Under the conditions of
Proposition~\ref{cor:asymptotic_bias}(iii)
and the inference tuning requirements
\eqref{eq:inference_tuning_tm}, the temporal
contrast estimator is asymptotically normal:
\[
\frac{\widehat\theta_B^{\mathrm{tm}}(t,t';\mathbf x_*)
      -\theta^{\mathrm{tm}}(t,t';\mathbf x_*)}
     {\sigma_{\mathrm{tm}}(t,t';\mathbf x_*)}
\xrightarrow{d}\mathcal N(0,1).
\]

\end{enumerate}
\end{theorem}
Here, the constraint $F_\Delta>K_t$ preserves the
approximation rate because the spline derivative
converges uniformly to the true derivative.

\begin{theorem}[Consistency of the Infinitesimal Jackknife]
\label{thm:ij_variance_consistency}
Under the conditions of Theorem~\ref{thm:pointwise_clt}, the
adjusted IJ estimator in \eqref{eq:ij_cov} satisfies
\[
\frac{\widehat\sigma_\psi^2(\mathbf x_*;t)}
     {\sigma_\psi^2(\mathbf x_*;t)}
\xrightarrow{\mathbb P}1,
\qquad
\widehat\sigma_\psi^2(\mathbf x_*;t)
=\widehat\Gamma_\psi(\mathbf x_*,t;\mathbf x_*,t).
\]
Under the respective conditions of
Theorem~\ref{thm:contrast_clt}(i) and~(ii), the variance
estimators in \eqref{eq:var_temporal} satisfy
\[
\frac{\widehat\sigma_{\mathrm{pt}}^2(\mathbf x_*,\mathbf x_*';t)}
     {\sigma_{\mathrm{pt}}^2(\mathbf x_*,\mathbf x_*';t)}
\xrightarrow{\mathbb P}1,
\qquad
\frac{\widehat\sigma_{\mathrm{tm}}^2(t,t';\mathbf x_*)}
     {\sigma_{\mathrm{tm}}^2(t,t';\mathbf x_*)}
\xrightarrow{\mathbb P}1,
\]
with variances as in \eqref{eq:CATE_var_true} and
\eqref{eq:CATE_contrast_true}.
\end{theorem}

Together with Theorems~\ref{thm:pointwise_clt}
and \ref{thm:contrast_clt}, ratio consistency
of the adjusted IJ variance estimator ensures
asymptotic coverage of $1-\varsigma$ for the
confidence intervals in \eqref{eq:CATE_CI}
and \eqref{eq:contrast_CI}.

\section{Simulation Study}
\label{sec:sim}

Our simulations address three objectives:
({\bf Aim 1}) we evaluate whether the shared temporal
representation improves estimation of time-varying
treatment effects, particularly temporal contrasts,
relative to separate time-specific models.
We also examine sensitivity to the working spline
dimension and basis misspecification. These
experiments use the true nuisance functions to
isolate the contribution of temporal modeling;
({\bf Aim 2}) we assess the finite-sample performance
of the proposed ensemble inference procedure,
focusing on confidence interval coverage,
standard error estimation, and robustness to
propensity score and survival model misspecification
when nuisance functions are estimated using
sample splitting;
({\bf Aim 3}) we benchmark the proposed estimator
against survival Bayesian additive regression
trees (survival BART)
\citep{sparapani2016nonparametric},
causal survival forests (CSF)
\citep{cui2023estimating},
doubly robust Bayesian additive regression
trees (DR-BART) \citep{chipman2010bart},
and doubly robust random forests (DR-RF)
\citep{athey2019generalized,Sun2026-wa},
comparing estimation accuracy and inferential
performance.

\medskip
\noindent
{\bf Setup} 
 
We generate independent covariates
$X_k\sim\operatorname{Unif}[-1,1]$, $k=1,\ldots,10$,
and treatment assignments
$W\mid\mathbf X=\mathbf x\sim
\operatorname{Bernoulli}\{e(\mathbf x)\}$, where
$e(\mathbf x)=\operatorname{expit}(\boldsymbol\alpha^\top\mathbf x)$
and $\alpha_k=\{1-2(k-1)/9\}/\sqrt{10}$.
The evaluation window is $\mathcal T=[0.3,1.2]$,
with $u=(t-0.3)/0.9$. Define
\begin{align*}
g_1(\mathbf x)
&=0.75\tanh\{1.5(x_1+x_2x_3)\}+0.25x_4x_5,\\
g_2(\mathbf x)
&=0.6\tanh\{1.5(x_6-x_7x_8)\}+0.4x_9x_{10},\\
g_3(\mathbf x)
&=-0.5g_1(\mathbf x)
+0.5\tanh\{1.5(x_5+x_6x_9)\}.
\end{align*}
We consider three CATE surfaces:
\begin{align*}
\tau_0^{\mathrm{bern}}(\mathbf x;t)
&=\tanh\!\left[\frac12\{
g_1(\mathbf x)(1-u)^2
+2g_2(\mathbf x)u(1-u)
+g_3(\mathbf x)u^2\}\right],\\
\tau_0^{\mathrm{exp}}(\mathbf x;t)
&=\tanh\!\left[\frac13\{
g_1(\mathbf x)(1-u)+g_2(\mathbf x)u
+4g_3(\mathbf x)u(1-u)e^{a(u-1/2)}\}\right],\\
\tau_0^{\mathrm{rat}}(\mathbf x;t)
&=\tanh\!\left[\frac13\{
g_1(\mathbf x)(1-u)+g_2(\mathbf x)u
+\frac{4g_3(\mathbf x)u(1-u)}
{1+a(u-1/2)}\}\right],
\end{align*}
where $a=0.25$. The latent predictors are
quadratic, exponential, and rational in time,
respectively. All three CATE surfaces have
uniformly bounded second time derivatives
and satisfy the temporal smoothness condition
with $\beta=1$.

To construct valid survival distributions, extend
each CATE trajectory to $d_{\mathbf x}(t)$ on
$[0,\infty)$ using $v(z)=3z^2-2z^3$:
\[
d_{\mathbf x}(t)=
\begin{cases}
\tau_0(\mathbf x;0.3)v(t/0.3),&t<0.3,\\
\tau_0(\mathbf x;t),&0.3\le t\le1.2,\\
\tau_0(\mathbf x;1.2)
\{1-v((t-1.2)/0.3)\},&1.2<t<1.5,\\
0,&t\ge1.5.
\end{cases}
\]
Let $V_{\mathbf x}(t)=\int_0^t|d'_{\mathbf x}(s)|\,ds$
and $V_{\mathbf x,\infty}=V_{\mathbf x}(\infty)$.
We set
\[
\begin{aligned}
M(\mathbf x,t)
&=\left(1-\frac{V_{\mathbf x,\infty}}2\right)e^{-t}
+\frac{V_{\mathbf x,\infty}-V_{\mathbf x}(t)}2,\\
S^w(\mathbf x;t)
&=M(\mathbf x,t)+\frac{2w-1}{2}d_{\mathbf x}(t).
\end{aligned}
\]
Since $V_{\mathbf x,\infty}\le2$ uniformly,
these are valid survival functions with
$S^1-S^0=\tau_0$ on $\mathcal T$.
Event times are generated by numerical inversion,
with independent censoring
$C\sim\operatorname{Exp}(0.1)$.

We use true nuisance functions in the joint
modelling and sensitivity experiments to isolate
estimation and approximation effects, while
estimating nuisance functions from a separate
sample in the inference experiments.
We compare separate time-specific DNNs with
the joint DNN on the equally spaced grid
$t_j=0.3+0.9(j-1)/(J-1)$.
The joint estimator uses clamped B-splines
of degree $\min(3,m-1)$.

To assess the predicted advantage of joint
fitting for time-normalized contrasts on finer
grids, we fix $n=3000$ and consider
$J\in\{5,7,9,13,17,25,33,49,97\}$,
with spline dimensions
$m\in\{2,3,4,5,6,8\}$ restricted to $m\le J$.

For each configuration, we conduct 200  Monte Carlo replications, using  80/20 training-validation splits
across methods. Networks have two hidden layers (128 and 64 units)
and are trained with AdamW (learning rate $10^{-3}$,
weight decay $0.01$) for up to 500 epochs, with early
stopping after 40 epochs. 
\medskip

\noindent
{\bf Aim 1} 
 
  We evaluate
$\mathrm{RMSE}_{\Delta,J}=\sqrt{R_{\Delta,J}}$
on 2000 independent test profiles shared across methods
and replications, reporting the ratio relative to
separate DNNs at the same $J$.

\begin{figure}[htbp]
\centering

{\small\textbf{Bernstein truth}\par}
\includegraphics[
  width=\textwidth,
  height=0.25\textheight,
  keepaspectratio
]{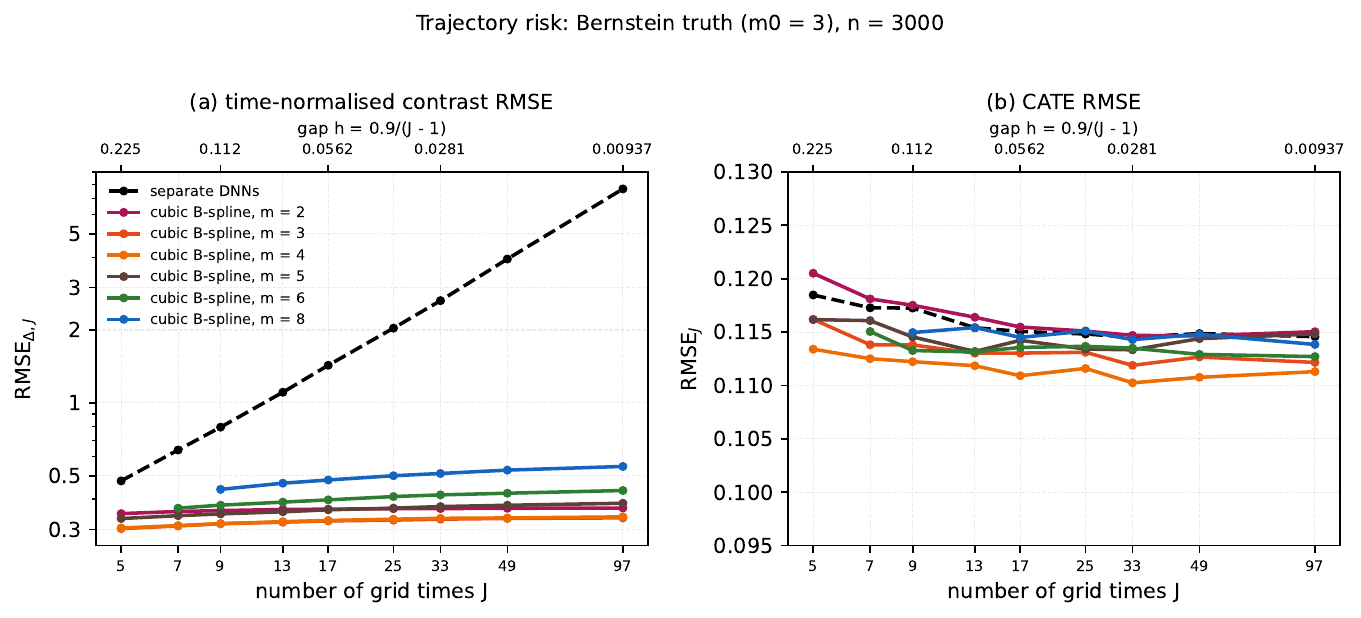}

\medskip
{\small\textbf{Exponential-hump truth}\par}
\includegraphics[
  width=\textwidth,
  height=0.25\textheight,
  keepaspectratio
]{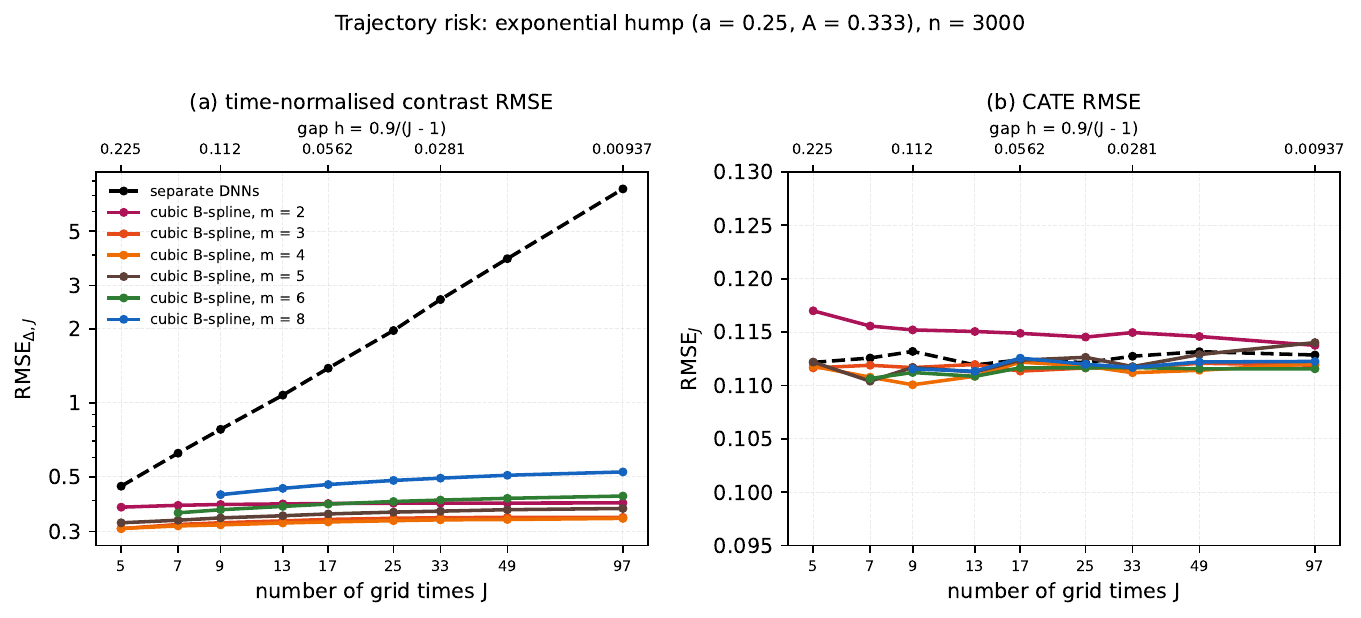}

\medskip
{\small\textbf{Rational-hump truth}\par}
\includegraphics[
  width=\textwidth,
  height=0.25\textheight,
  keepaspectratio
]{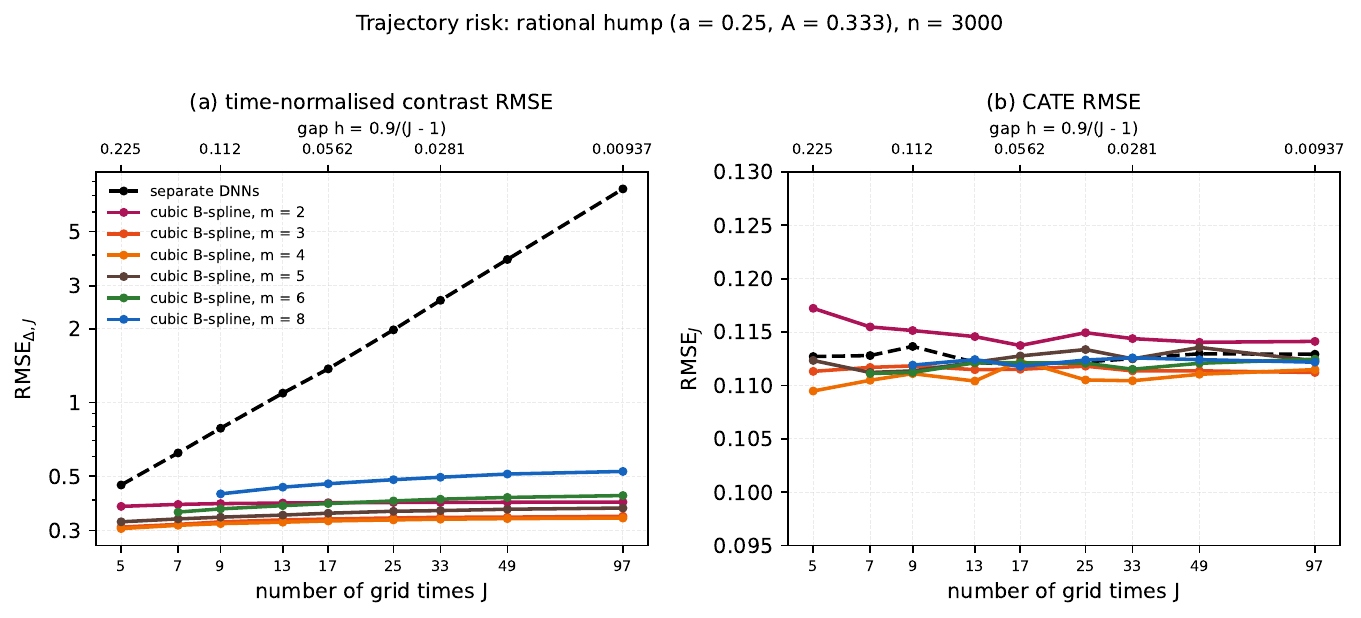}

\caption{ 
Time-normalized contrast RMSE ($\mathrm{RMSE}_{\Delta,J}$;
left, log scale) and CATE level RMSE ($\mathrm{RMSE}_J$;
right, common scale) across grid sizes $J$.
Rows correspond to the Bernstein, exponential-hump,
and rational-hump CATE surfaces, and $m$ denotes the
fitted spline dimension. Results are based on
200 replications with $n=3000$, known nuisance
functions, and $a=0.25$ for both hump scenarios.
}

\label{fig:trajectory_all}
\end{figure}

Figure~\ref{fig:trajectory_all} compares CATE level and
time-normalized contrast RMSE across grid sizes at $n=3000$.
Under the Bernstein truth, the contrast RMSE for $m=4$
remains nearly constant (0.31--0.34) as $J$ increases
from 5 to 97, whereas that of separate DNNs rises
from 0.46 to over 7. At $J=97$, however, both methods
achieve a level RMSE of approximately 0.11, showing
that similar level accuracy can conceal substantial
differences in temporal contrast accuracy.
The exponential and rational truths show similar patterns.
As $J$ increases from 25 to 97, the contrast RMSE
for $m=4$ remains near 0.34, while that of separate
DNNs increases from 1.93 to approximately 7.2.
The approximation-only contrast RMSEs are 0.0016
and 0.0033, respectively, compared with fitted RMSEs
of 0.343 and 0.340. Increasing $m$ to 8 raises
the fitted contrast RMSE to approximately 0.52
at $J=97$. These findings demonstrate that joint
fitting can substantially improve temporal contrast
estimation without improving CATE level accuracy.

\begin{figure}[htbp]
\centering

{\small\textbf{Bernstein truth}\par}
\includegraphics[width=\textwidth]{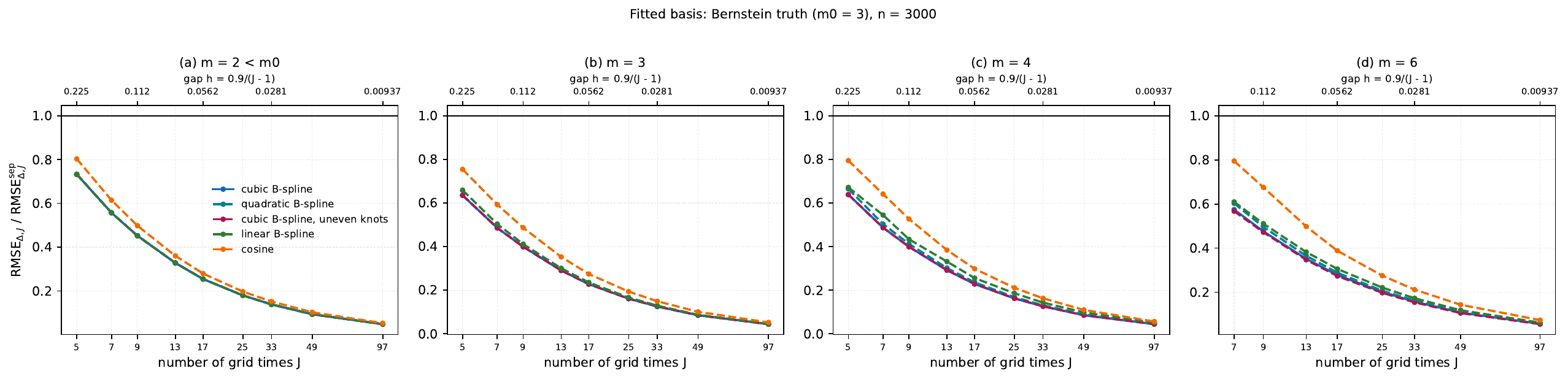}

\medskip
{\small\textbf{Exponential-hump truth}\par}
\includegraphics[width=\textwidth]{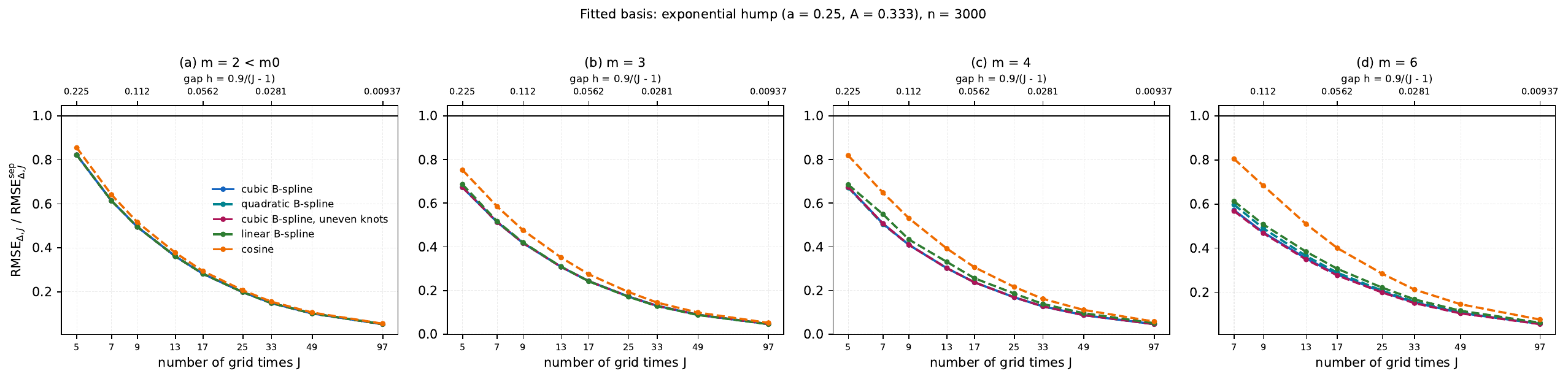}

\medskip
{\small\textbf{Rational-hump truth}\par}
\includegraphics[width=\textwidth]{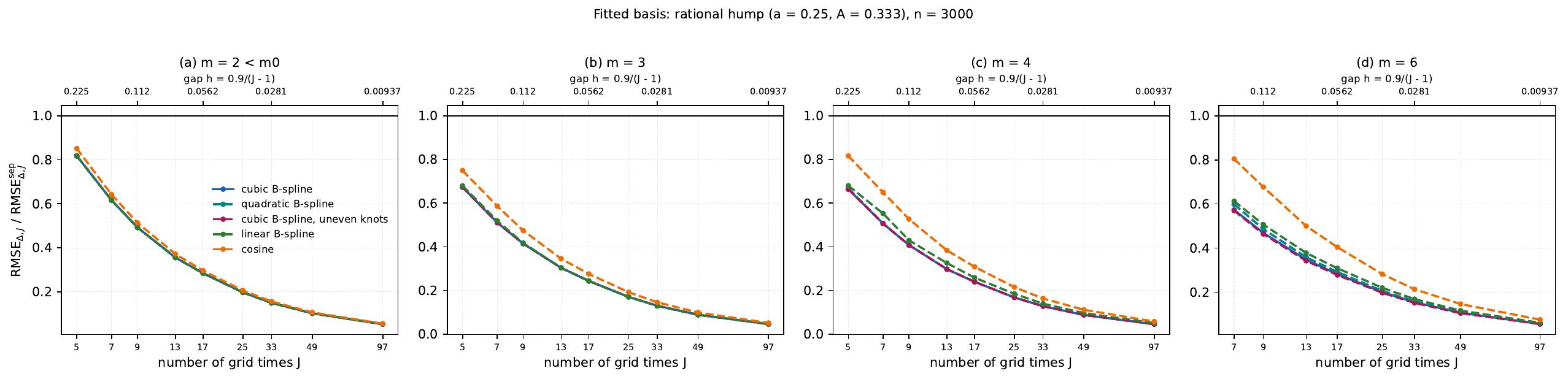}

\caption{Sensitivity of temporal contrast estimation to the choice of
temporal basis across three CATE surfaces (rows) and working
dimensions $m=2,3,4,6$ (columns).
The vertical axis shows the ratio of time-normalized contrast
RMSE to that of separately fitted DNNs on the same evaluation
grid, with values below one indicating improved accuracy.
Results are based on $n=3000$, 200 Monte Carlo replications,
and known nuisance functions.
}
\label{fig:sensitivity_all}
\end{figure}

We further examine sensitivity to the working
dimension and temporal basis using true nuisance
functions. Figure~\ref{fig:sensitivity_all} compares
the primary B-splines with quadratic and piecewise-linear
splines, nonuniform-knot B-splines, and the cosine basis
$\{1,\cos(\pi u),\ldots,\cos((m-1)\pi u)\}$.
Nonuniform knots are obtained by transforming each
interior knot $\xi$ to
$\xi+0.25\sin(2\pi\xi)/(2\pi)$, preserving
quasi-uniform spacing.
At $m=3$, the primary basis  represents
the Bernstein latent predictor, whereas the
nonpolynomial hump predictors are not exactly
represented by finite-dimensional spline spaces.
Nevertheless, their approximation-only contrast
errors remain small relative to fitted errors.
The uniform- and nonuniform-knot bases coincide
for $m\le4$, and knot perturbations have little
effect at $m=6$. The cosine basis also outperforms
separate DNNs: at $m=6$ and $J=7$, the relative
contrast RMSE is approximately 0.60 for cubic
splines and 0.80 for the cosine basis, with the
difference narrowing as $J$ increases.

\medskip
\noindent
{\bf Aim 2} 

We assess inference under the Bernstein truth with
$n=3000$ and estimated nuisance functions. The
propensity score is
\(
e(\mathbf{x})=\operatorname{expit}
\{-0.1+0.6x_1-0.6x_6+1.2x_2x_3
+1.4(x_5^2-1/3)\},
\)
and covariate-dependent exponential censoring
yields approximately $20\%$ censoring.

As a practical implementation, we use five-fold
cross-fitting to improve data utilization and
estimation stability, with a single-split
sensitivity analysis for Design~1. The correctly
specified propensity model includes
$(\mathbf{x},x_2x_3,x_5^2)$, while its misspecified
version uses only $\mathbf{x}$. The survival model
is either the correctly specified parametric family
with unknown amplitudes or a misspecified Cox model
linear in $\mathbf{x}$. Design~1 specifies both
models correctly, while Designs~2 and~3 misspecify
the propensity and survival models, respectively,
to assess double robustness. All designs use
identical data and subsamples within each replication.

The ensemble uses $B=1000$ subsamples of size
$r=\lfloor n^{0.90}\rfloor$. Each coefficient network
has two ReLU hidden layers (256 and 128 units)
and $m=4$ outputs, combined with cubic B-splines
on $J=25$ equally spaced times in $[0.3,1.2]$.
The choice $m=4$ exactly represents the Bernstein
latent predictor and lies within the range
performing well in the sensitivity experiments.
Networks are trained using full-batch Adam for
200 epochs (learning rate $10^{-3}$, weight
decay $10^{-4}$).
We evaluate pointwise coverage at 80 fixed profiles
from the central $80\%$ of CATE trajectories.
Temporal contrasts compare the first and third
quarter points of the evaluation window, while
covariate contrasts use 20 profile pairs differing
only in $x_1$ by 0.6.

\begin{table}[htbp]
\centering\small\setlength{\tabcolsep}{6pt}
\begin{tabular}{l l rrrrr}
\toprule
Estimand & Design & $|$Bias$|$ & EmpSD & SE & CP & AIL \\
\midrule
$\tau_0$ & 1 & 0.0454 & 0.3819 & 0.3926 & 0.947 & 1.2973 \\
 & 2 & 0.0473 & 0.3855 & 0.3972 & 0.947 & 1.3083 \\
 & 3 & 0.0432 & 0.3847 & 0.3950 & 0.948 & 1.3028 \\
\addlinespace
$\theta^{\mathrm{tm}}$ & 1 & 0.0408 & 0.1412 & 0.1530 & 0.947 & 0.5996 \\
 & 2 & 0.0509 & 0.1430 & 0.1565 & 0.942 & 0.6133 \\
 & 3 & 0.0491 & 0.1356 & 0.1461 & 0.939 & 0.5725 \\
\addlinespace
$\theta^{\mathrm{pt}}$ & 1 & 0.0254 & 0.2351 & 0.2485 & 0.957 & 0.9742 \\
 & 2 & 0.0261 & 0.2369 & 0.2516 & 0.958 & 0.9864 \\
 & 3 & 0.0255 & 0.2353 & 0.2501 & 0.958 & 0.9803 \\
\bottomrule
\end{tabular}  
\caption{Performance of the ESM estimator for pointwise
CATEs $\tau_0(\mathbf{x}_*;t_j)$, temporal contrasts
$\theta^{\mathrm{tm}}(t,t';\mathbf x_*)$, and
between-patient contrasts
$\theta^{\mathrm{pt}}(\mathbf{x}^{(1)}_*,\mathbf{x}^{(2)}_*;t_j)$
under Designs~1--3. Results are based on
$m=4$, $J=25$, $r=\lfloor n^{0.90}\rfloor$,
and 200 Monte Carlo replications, with a nominal
coverage level of $95\%$.}

\label{tab:esm}
\end{table}

\begin{figure}[htbp]
\centering
\includegraphics[width=\textwidth]{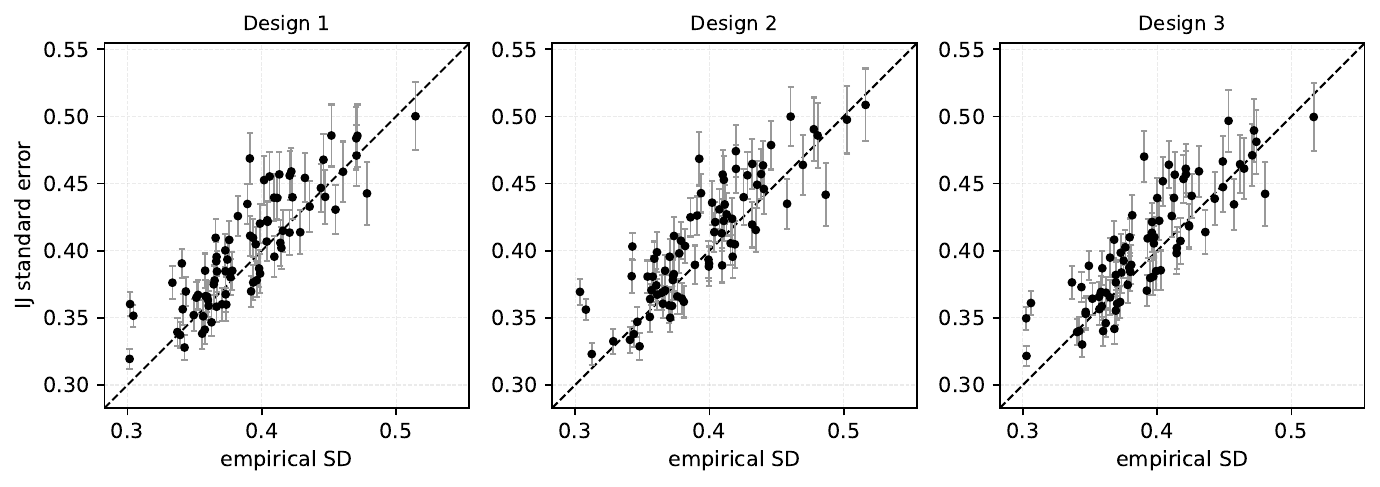}
 \caption{Comparison of empirical standard deviations
of CATE estimates and mean infinitesimal jackknife
(IJ) standard errors across 80 evaluation profiles
at $t=0.75$ under Designs~1--3.
Results are based on $\alpha=0.90$, $m=4$, $J=25$,
and 200 Monte Carlo replications.
Error bars represent $95\%$ Monte Carlo confidence
intervals for the mean IJ standard errors.}
\label{fig:sim_sd_se}
\end{figure}

Table~\ref{tab:esm} summarizes inference results over
200 replications. Empirical coverage is close to 95\%
for pointwise CATEs (0.947--0.948), temporal contrasts
(0.939--0.947), and covariate contrasts (0.957--0.958),
including when either nuisance model is misspecified.
IJ standard errors agree closely with empirical standard
deviations ($\mathrm{SE}/\mathrm{EmpSD}=1.03$--$1.09$;
Figure~\ref{fig:sim_sd_se}), and mean absolute bias
remains below 0.34 standard errors.

\begin{figure}[htbp]
\centering
\includegraphics[width=\textwidth]{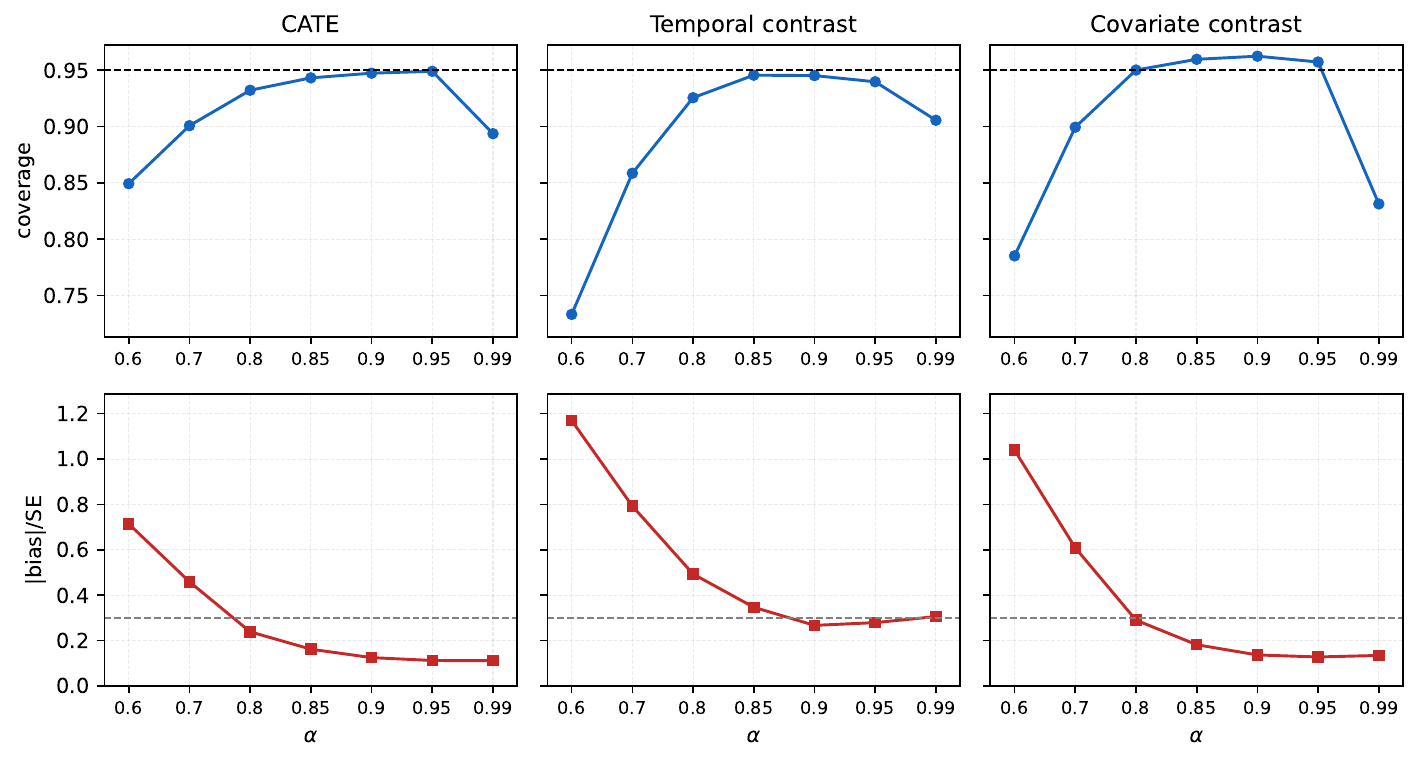}
 \caption{Sensitivity of empirical coverage (top row) and
absolute bias relative to standard error (bottom row)
to the subsample exponent $\alpha$, with
$r=\lfloor n^\alpha\rfloor$, for the CATE (left),
temporal contrast (middle), and between-patient
contrast (right). Results are based on Design~1,
$m=4$, $J=25$, and the same 100 Monte Carlo
replications across all values of $\alpha$.
The bias-to-SE ratio is defined as the absolute
Monte Carlo bias divided by the mean IJ standard
error. Dashed lines indicate the nominal
coverage level of $0.95$ (top) and a reference
bias-to-SE ratio of $0.3$ (bottom).}
\label{fig:sim_alpha}
\end{figure}

In a 50-replication sensitivity analysis, single splitting
with 20\% or 30\% reserved for nuisance estimation
yields coverage of 0.940--0.967, but pointwise intervals
are 5--7\% longer than with five-fold cross-fitting.
Figure~\ref{fig:sim_alpha} shows near-nominal coverage
for $\alpha=0.85$--$0.95$. The pointwise bias-to-SE
ratio decreases at smaller $\alpha$ than the temporal
contrast ratio, consistent with the sufficient bounds
in Proposition~\ref{cor:asymptotic_bias}.
At $\alpha=0.90$, the ratios are 0.12, 0.10, and 0.27
for pointwise CATEs, covariate contrasts, and temporal
contrasts, respectively. Coverage deteriorates at
$\alpha=0.99$ despite smaller bias-to-SE ratios,
indicating that bias reduction alone does not ensure
valid inference.

\medskip
\noindent
{\bf Aim 3}
 
We compare our estimator with survival BART
\citep{sparapani2016nonparametric}, causal survival
forests (CSF) \citep{cui2023estimating}, DR-BART
\citep{chipman2010bart}, and DR-RF
\citep{athey2019generalized,Sun2026-wa}
over 200 replications using identical evaluation
profiles and contrasts. Our method achieves
near-nominal coverage for pointwise CATEs
(0.947--0.948) and covariate contrasts
(0.957--0.958), comparable to competing methods.
Its main advantage is for temporal contrasts,
with coverage of 0.939--0.947 and interval
lengths of 0.57--0.61 across nuisance designs.
In comparison, DR-RF, CSF, and survival BART
substantially undercover (0.676--0.861), while
DR-BART produces overly conservative intervals
(coverage 1.000; lengths 1.84--2.33).
Thus, our method offers a better balance of
coverage and precision for temporal contrasts.
Further results appear in
Appendix~\ref{sec:supp_sim}.

\section{Application to the Boston Lung Cancer Study}
\label{sec:realdata}

We estimate the effect of perioperative adjuvant chemotherapy
on overall survival in the Boston Lung Cancer Study
\citep{alvarez2026effects}, with time zero defined as
the date of diagnosis and death as the event of interest.
From $7{,}755$ records, we retain patients with
surgically resected non-small-cell lung cancer who
underwent surgery within $180$ days of diagnosis
between $1993$ and $2016$, with administrative
censoring at ten years. We restrict the analysis
to stages II and III because stage I included
too few treated patients, while chemotherapy
for stage IV was largely palliative. The resulting
cohort comprises $936$ patients, with $34.2\%$
censored. Covariates include age, body mass
index, smoking pack-years, sex, and cancer stage,
with missing values handled using median
imputation and missingness indicators.
 
To reduce sensitivity to a particular sample split
and improve estimation stability, we use five-fold
cross-fitting as a practical extension of the
single-split procedure underlying our theoretical
analysis. Specifically, the nuisance functions
are estimated using arm-specific Cox models for
survival and censoring and a logistic propensity
model, with estimated propensity scores trimmed
to $[0.10,0.90]$, excluding two patients.
The ensemble comprises $B=1{,}500$ subsamples
of size $r=n^{0.90}=471$, using the network
architecture described in Section~\ref{sec:sim},
trained for $100$ epochs, with $m=4$ temporal
basis functions evaluated on a grid from nine
months to five years. At five years, $45.9\%$
of the cohort remains at risk. The reference
patient is a 65-year-old man with a body mass
index of $26$ and a smoking history of $35$
pack-years, evaluated separately at stages II
and III.

\begin{figure}[htbp]
\centering
\includegraphics[width=\textwidth]{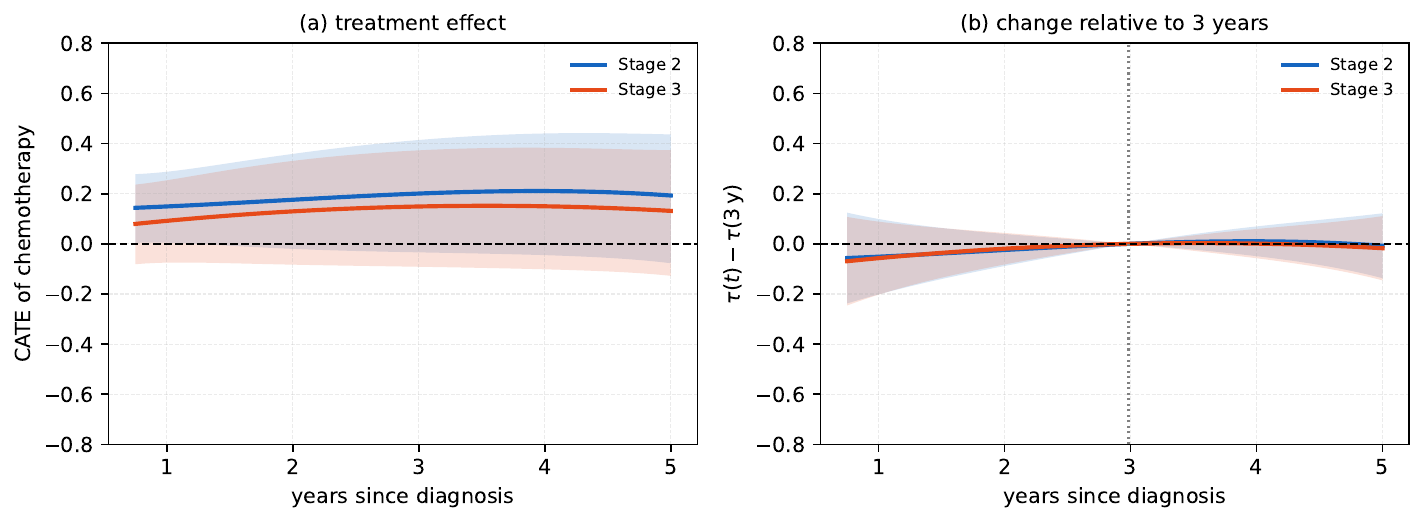}
\caption{Pointwise conditional average treatment effects (CATEs) for the unified anchor patient across disease stages (left) and their temporal contrasts relative to the three-year follow-up time (right), stratified by stage.}
\label{fig:blcs_stage}
\end{figure}

Figure~\ref{fig:blcs_stage} shows positive estimated
treatment benefits in both stages. For stage II,
the estimated benefit is $0.148$ (95\% CI:
$[0.005, 0.286]$) at one year, $0.200$
($[-0.034, 0.414]$) at three years, and $0.193$
at five years. The confidence intervals exclude
zero only between nine and fourteen months after
diagnosis. For stage III, the estimated benefit
is $0.149$ ($[-0.091, 0.373]$) at three years,
with confidence intervals including zero throughout
follow-up. In both stages, estimated benefits
increase during the first few years before leveling
off, reaching approximately $0.21$ near four years
in stage II and $0.15$ near three and a half years
in stage III. Confidence intervals widen over time
as the number of patients at risk decreases.
The confidence intervals for both the stage
differences and temporal contrasts include zero
throughout follow-up. At three years, the estimated
stage difference is $0.051$ (95\% CI:
$[-0.279, 0.382]$), providing limited evidence
of differences in treatment benefits or their
temporal patterns between stages.

\begin{figure}[htbp]\centering
\includegraphics[width=1.0\textwidth, height=6cm]{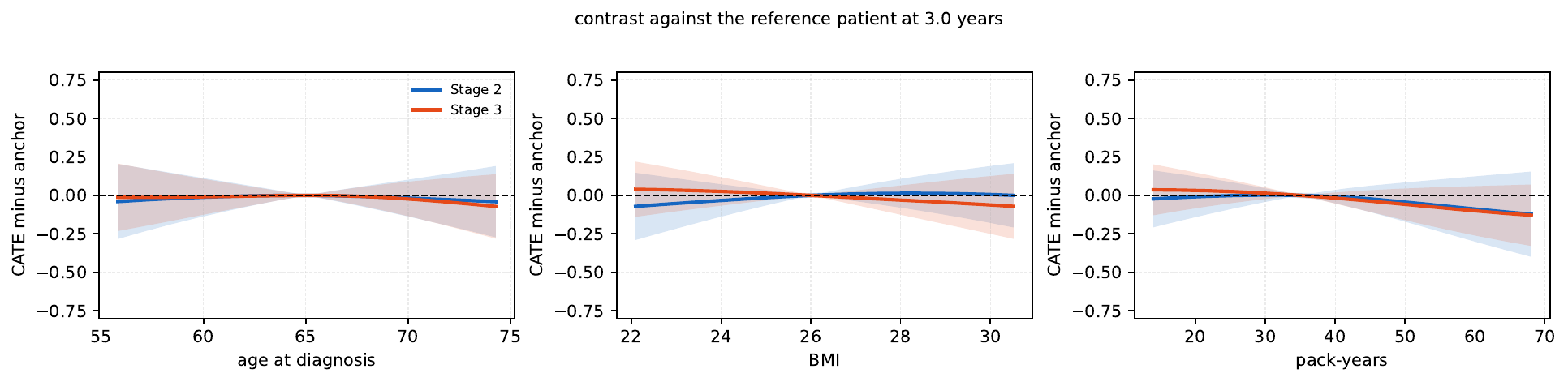}
\caption{Covariate contrasts in treatment effects at
three years relative to a common reference patient
(a man aged $65$ years, with BMI 26 and 35 pack-years).
Positive values indicate greater treatment benefits
than those estimated for the reference patient. Each curve is zero at the reference value by construction.}
\label{fig:blcs_cov_stage}
\end{figure}

Figure~\ref{fig:blcs_cov_stage} examines how estimated
three-year treatment benefits vary with age, body mass
index, and smoking history in stages II and III.
The most noticeable variation is associated with
smoking history, with estimated benefits declining
by approximately $0.12$ at $68$ pack-years in both
stages. Variation with age and body mass index is
less pronounced. As body mass index increases from
$22$ to $30.5$, the estimated benefit increases
slightly in stage II but decreases by $0.07$ in
stage III. In stage III, the estimated benefit
is also $0.07$ lower at age $74$ than at the
reference age. However, none of the corresponding
$95\%$ confidence intervals excludes zero,
suggesting that the observed patterns provide
limited evidence of treatment-effect heterogeneity
across these patient characteristics.
As this observational analysis may be subject to residual confounding
and treatment-timing bias, particularly because chemotherapy may
begin after diagnosis, causal interpretation requires caution.

\section{Discussion}
\label{sec:discussion}

 We develop a unified framework for estimation and inference
of heterogeneous treatment effects with censored survival
outcomes, combining neural networks for nonparametric covariate
effects \citep{schmidt2020deep} with a shared spline basis
for latent CATE trajectories.
By jointly modeling trajectories through coefficient functions
shared across time, rather than fitting separate models,
our approach yields sharper upper bounds for time-normalized
temporal contrast risks, avoiding inverse-gap amplification
on finer grids.
For inference, subsampled ensembles and incomplete
U-statistic theory \citep{mentch2016quantifying,Wager&Athey}
establish asymptotic normality and consistent variance
estimation for pointwise CATEs, temporal contrasts, and
covariate contrasts, including temporal contrasts over
shrinking intervals with potentially vanishing variances.
A single ensemble enables joint covariance estimation.

Several directions merit investigation.
Adaptive temporal bases may improve approximation of latent
CATE trajectories and derivatives, reducing temporal contrast
bias. Simultaneous confidence bands and data-adaptive
contrasts could broaden inference.
Although doubly robust pseudo-outcomes mitigate nuisance
model misspecification
\citep{robins1994estimation,chernozhukov2018double},
causal identification still requires unconfoundedness,
positivity, and conditional independent censoring.
Future work could address sensitivity analysis,
competing risks, informative censoring, time-varying
treatments, and more computationally efficient ensembles
that preserve inferential validity.

\bibliography{reference}
\bibliographystyle{ims}

\appendix
\setcounter{equation}{0}
\renewcommand{\theequation}{A\arabic{equation}}
\setcounter{table}{0}
\renewcommand{\thetable}{A\arabic{table}}

\section{Proofs of Main Results}
\label{sec:supp_main_proofs}

%We use the probability conventions of Section~\ref{sec:estimation}. Table~\ref{tab:proof_notation} groups the notation by the results that use it.
 
 \begingroup
\small
\begin{longtable}{@{}p{0.29\linewidth}@{\hspace{1em}}p{0.66\linewidth}@{}}
\caption{Notation dictionary.}
\label{tab:proof_notation}\\
\toprule
Notation & Meaning and reference \\
\midrule
\endfirsthead
\multicolumn{2}{@{}l}{\tablename~\thetable\ (continued)}\\
\toprule
Notation & Meaning and reference \\
\midrule
\endhead
\midrule
\multicolumn{2}{r@{}}{\emph{Continued on the next page}}\\
\endfoot
\bottomrule
\endlastfoot

\multicolumn{2}{@{}l}{\emph{Samples and probability conventions (throughout)}}\\*
$\mathcal D_n,\ \mathcal D_{\mathrm{nuis}},\
 \mathcal D_{\mathrm{train}}$
 & Full, nuisance, and training samples; $n_1=|\mathcal D_{\mathrm{train}}|$, indexed by $i=1,\ldots,n_1$. \\
$\mathbb P,\ \mathbb E$
 & Probability and expectation over data, splitting, subsampling, and fitting randomness. \\
$\mathcal S_{\mathrm{nuis}}$;\newline
$\mathbb P_{\mathrm{nuis}},\ \mathbb E_{\mathrm{nuis}}$
 & Nuisance-sample sigma-field; $\mathbb P_{\mathrm{nuis}}$ and $\mathbb E_{\mathrm{nuis}}$ are its probability and expectation. \\
$\mathbb P_{\mathrm{train}},\
  \mathbb E_{\mathrm{train}},\
  \operatorname{Cov}_{\mathrm{train}}$
 & Average over training data, subsampling, and fitting; $\mathbb E[U]=\mathbb E_{\mathrm{nuis}}[\mathbb E_{\mathrm{train}}[U]]$. \\

\addlinespace
\multicolumn{2}{@{}l}{\emph{Working model and oracle bounds
(Theorems~\ref{prop:oracle_fixed_t} and~\ref{thm:joint_oracle})}}\\*
$\eta,\ \widehat\eta$;\newline
$\varphi_i(t),\ \widehat\varphi_i(t)$
 & Nuisance functions and estimates; $\widehat\varphi_i(t)=\widehat\pi_i^1(t)-\widehat\pi_i^0(t)$, with $\varphi_i(t)$ using true nuisances (Section~\ref{sec:causal_setup}). \\
$\tau_0,\ \psi_0$;\newline
$\mathcal G,\ \mathcal G_{\mathrm{joint}}^*$
 & $\tau_0=S^1-S^0=\tanh(\psi_0)$; $\psi_0(\cdot;t)\in\mathcal G$. The joint class bounds levels and slopes by $K_t$ and imposes $\beta$-H\"older derivative continuity (Definitions~\ref{def:composite_holder} and~\ref{def:joint_composite}). \\
$\mathcal F_{\mathrm{joint}}(L,\mathbf p,\mathbf s,F)$
 & Coefficient networks: depth $L$, widths $\mathbf p$, output sparsities $\mathbf s=(s_1,\ldots,s_m)$, envelope $F$, and $p_{L+1}=m$ ($m=1$ for separate fits; \eqref{dnn-class}, Assumption~\ref{assum:dnn}). \\
$m,\ J,\ \mathcal B_m,\ \mathbf B$
 & Spline dimension, grid size, spline space, and evaluation matrix $\mathbf B_{j\ell}=B_{\ell,m}(t_j)$ (\eqref{eq:readout}). \\
$\psi_{\mathbf f},\ \tau_{\mathbf f},\ \mathcal T_m$
 & $\psi_{\mathbf f}(\mathbf x;t)=\sum_{\ell=1}^m f_\ell(\mathbf x)B_{\ell,m}(t)$, $\tau_{\mathbf f}=\tanh(\psi_{\mathbf f})$; $\mathcal T_m$ collects these surfaces (Section~\ref{sec:dnn_architecture}). \\
$\|g\|_{n_1},\ \|g\|_{L_2(P_{\mathbf X})}$
 & $\|g\|_{n_1}^2=n_1^{-1}\sum_i|g(\mathbf X_i)|^2$; $\|g\|_{L_2(P_{\mathbf X})}^2=\int|g(\mathbf x)|^2dP_{\mathbf X}(\mathbf x)$ with fitted $g$ fixed. Joint empirical norm: $J^{-1}\sum_j\|g(\cdot;t_j)\|_{n_1}^2$. \\
$\|g\|_\infty$
 & Supremum over $\mathcal X$ for slices, or $\mathcal X\times\mathcal T$ for surfaces. \\
$R_1,\ R_J,\ R_{\Delta,J}$
 & Expected squared population errors: fixed time, grid average, and adjacent time-normalized contrasts (\eqref{eq:joint_risk}, \eqref{eq:contrast_risk}); conditional on nuisance data. \\
$\tau_{\mathrm{apx}}(\cdot;t_j)$;\newline
$\tau_{m,\mathrm{apx}},\ \psi_{m,\mathrm{apx}}$
 & Fixed-time CATE approximation \eqref{eq:tau_fixed_apx}; joint latent approximation and $\tau_{m,\mathrm{apx}}=\tanh\circ\psi_{m,\mathrm{apx}}$ \eqref{eq:tau_m_apx}, also used at subsample size $r$. \\
$\mathcal E_{\mathrm{opt},n,J},\
  \widehat{\mathcal E}_n$
 & Conditional optimization gap \eqref{eq:optimization_error} and nuisance error \eqref{nuis-error}; use $\mathcal E_{\mathrm{opt},n,1}$ at one time and $\mathcal E_{\mathrm{opt},r,J}$ for subsamples. \\
$c_{\mathrm{eff}},\ \phi_n,\ \delta,\
 m_{\mathrm{opt}}(\delta)$
 & $\phi_n=n^{-c_{\mathrm{eff}}}$: squared covariate rate \eqref{eq:c_eff&phi_n}; $\delta$: approximation budget; $m_{\mathrm{opt}}(\delta)$ balances temporal contrast terms \eqref{eq:m_opt}. \\

\addlinespace
\multicolumn{2}{@{}l}{\emph{Subsampling, bias, and inference}}\\*

$r,\ \alpha,\ B$\newline
$\widehat\psi_r,\ \widehat\psi^b,\ \bar\psi_B,\ \bar\tau_B$
 & $r=\lfloor n_1^\alpha\rfloor$; $B$ fits $\widehat\psi^b$ of generic type $\widehat\psi_r$; $\bar\psi_B=B^{-1}\sum_b\widehat\psi^b$, $\bar\tau_B=\tanh(\bar\psi_B)$ (Algorithm~\ref{algo:single_split_esm}). \\

$M_r(\mathbf x),\ \mathcal A_n,\ \ell_n$
 & Conditional mean squared uniform level and slope errors; Corollary~\ref{lem:An} controls $M_r$ on $\mathcal A_n$, with $P_{\mathbf X}(\mathcal A_n^c)\le\ell_n^{-1}$, $\ell_n\to\infty$. \\

$\zeta_r^{(1)},\ \nu,\ \rho_r$
 & $\zeta_r^{(1)}$: covariance of two fits sharing one observation, otherwise independent \eqref{eq:zeta_r}; $\nu$: Assumption~\ref{assum:xi}; $\rho_r$: one minus absolute first-order correlation (Proposition~\ref{cor:asymptotic_bias}). \\

$\alpha_{\mathrm{lower,pt}},\
  \alpha_{\mathrm{lower,tm}}$
 & Lower $\alpha$ bounds for negligible centering bias (Proposition~\ref{cor:asymptotic_bias}); CLT ranges in \eqref{eq:inference_tuning} and \eqref{eq:inference_tuning_tm}. \\

$\Gamma_\psi,\ \widehat\Gamma_\psi$\newline
$\sigma_\psi^2,\ \widehat\sigma_\psi^2$
 & First-order latent covariance $\Gamma_\psi=(r^2/n_1)\zeta_r^{(1)}$ and adjusted IJ estimate $\widehat\Gamma_\psi$; diagonals $\sigma_\psi^2$, $\widehat\sigma_\psi^2$ (\eqref{eq:CATE_var_true}, \eqref{eq:ij_cov}). \\

$\check\Gamma_\psi,\ \check\sigma_\psi^2$
 & Infinite-ensemble IJ covariance and diagonal ($B\to\infty$, samples fixed); training-sample dependent (proof of Theorem~\ref{thm:ij_variance_consistency}). \\

$\theta^{\mathrm{pt}},\ \widehat\theta_B^{\mathrm{pt}}$\newline
$\theta^{\mathrm{tm}},\ \widehat\theta_B^{\mathrm{tm}}$
 & Ordered between-profile and later-minus-earlier temporal CATE contrasts; estimates substitute $\bar\tau_B$ for $\tau_0$ (\eqref{eq:cate_contrasts}, \eqref{eq:cate_contrast_estimates}). \\

$\sigma_{\mathrm{pt}},\ \widehat\sigma_{\mathrm{pt}}$\newline
$\sigma_{\mathrm{tm}},\ \widehat\sigma_{\mathrm{tm}}$
 & First-order and estimated contrast standard errors, with argument orders matching $\theta^{\mathrm{pt}}$ and $\theta^{\mathrm{tm}}$ (\eqref{eq:CATE_contrast_true}, \eqref{eq:var_temporal}). \\

$F_\Delta$
 & Subsample latent slope envelope $\|\partial_t\psi_{\mathbf f}\|_\infty\le F_\Delta$, $F_\Delta>K_t$ (Theorem~\ref{thm:contrast_clt}(ii)). \\

\addlinespace
\multicolumn{2}{@{}l}{\emph{Quantities used in the inference proofs}}\\*

$Y_b^{\mathrm{pt}}$
 & $Y_b^{\mathrm{pt}}=\{1-\tau_0(\mathbf x_*,t)^2\}\widehat\psi^b(\mathbf x_*,t)-\{1-\tau_0(\mathbf x_*',t)^2\}\widehat\psi^b(\mathbf x_*',t)$. Used in Theorems~\ref{thm:contrast_clt} and~\ref{thm:ij_variance_consistency}. \\

$Y_b^{\mathrm{tm}}$
 & $Y_b^{\mathrm{tm}}=\big[\{1-\tau_0(\mathbf x_*,t')^2\}\widehat\psi^b(\mathbf x_*,t')-\{1-\tau_0(\mathbf x_*,t)^2\}\widehat\psi^b(\mathbf x_*,t)\big]/(t'-t)$. Used in the same proofs. \\
  %Its ensemble average has first-order variance$\sigma_{\mathrm{tm}}^2(t,t',\mathbf x_*)/(t'-t)^2$. \\

$I_{b,i},\ \bar I_{\cdot,i}$
 & $I_{b,i}=\mathbb I(i\in\mathcal I^b)$; $\bar I_{\cdot,i}=B^{-1}\sum_b I_{b,i}$ (Algorithm~\ref{algo:single_split_esm}, \eqref{eq:ij_cov}). \\

$N_i,\ N_{i_1\ldots i_k},\ p_k$
 & Subsample counts containing $i$ or $\{i_1,\ldots,i_k\}$; $N_{i_1\ldots i_k}\sim\operatorname{Bin}(B,p_k)$, $p_k=\binom{n_1-k}{r-k}/\binom{n_1}{r}$ (proof of Theorem~\ref{thm:pointwise_clt}). \\

$h_{1,r},\ \bar h_{1,r},\
  \mathring T_{n_1,r}$
 & First-order kernel, its training mean, and H\'ajek projection $\mathring T_{n_1,r}=(r/n_1)\sum_i h_{1,r}(D_i,\mathbf x_*,t)$; variances $\zeta_r^{(1)}$ and $\sigma_\psi^2$ \eqref{eq:pointwise_hajek}. \\

$T_k,\ V_k$
 & Order-$k$ Hoeffding components and variances $V_k=\operatorname{Var}_{\mathrm{train}}\{T_k(D_1,\ldots,D_k)\}$ \eqref{eq:pointwise_hoeff}; proofs of Theorems~\ref{thm:pointwise_clt} and~\ref{thm:ij_variance_consistency}. \\

$\mathbb E_{\mathrm{train}}
  [\,\cdot\mid\mathcal D_{\mathrm{train}}];\newline \mathrm{Var}_{\mathrm{train}}
  [\,\cdot\mid\mathcal D_{\mathrm{train}}];$\newline$\mathrm{Cov}_{\mathrm{train}}
  [\,\cdot\mid\mathcal D_{\mathrm{train}}]$
 & Hold both samples fixed; average over subsampling and network fitting (proof of Theorem~\ref{thm:ij_variance_consistency}). \\

$M_i,\ \mathbf M$
 & $M_i=\operatorname{Cov}_{\mathrm{train}}\{I_{b,i},\widehat\psi^b(\mathbf x_*,t)\mid\mathcal D_{\mathrm{train}}\}$; $\mathbf M=(M_i)_{i=1}^{n_1}$ defines infinite-ensemble IJ estimates (proof of Theorem~\ref{thm:ij_variance_consistency}). \\

$Z_{b,i},\ Z_{b,i}^{\circ}$\newline
$\mathbf Z_b,\ \mathbf Z_b^\circ,\ \mathbf \Sigma_Z$
 & Centered inclusion--prediction products: $Z_{b,i}$ uses empirical means $\bar I_{\cdot,i},\bar\psi_B$; $Z_{b,i}^\circ$ uses $r/n_1$ and conditional prediction means. Vectors stack subjects; $\mathbf\Sigma_Z=\operatorname{Cov}_{\mathrm{train}}(\mathbf Z_b^\circ\mid\mathcal D_{\mathrm{train}})$ (proof of Theorem~\ref{thm:ij_variance_consistency}). \\

$\check{\mathbf\Sigma}_{\mathrm{pt}},\
  \widehat{\mathbf\Sigma}_{\mathrm{pt}}$
 & Infinite-ensemble and adjusted IJ covariance matrices for latent predictions at $(\mathbf x_*,t)$ and $(\mathbf x_*',t)$ (proof of Theorem~\ref{thm:ij_variance_consistency}). \\

$\check{\mathbf\Sigma}_{\mathrm{tm}},\
  \widehat{\mathbf\Sigma}_{\mathrm{tm}}$
 & Analogous matrices for latent level at $t$ and slope $\{\bar\psi_B(\mathbf x_*,t')-\bar\psi_B(\mathbf x_*,t)\}/(t'-t)$ (proof of Theorem~\ref{thm:ij_variance_consistency}). \\

$\|\cdot\|_{\mathrm{op}},\
  \preceq,\ \succeq$
 & Operator norm $\|\mathbf H\|_{\mathrm{op}}=\sup_{\|\mathbf v\|_2=1}\|\mathbf H\mathbf v\|_2$; $\mathbf H\preceq\mathbf K$ iff $\mathbf K-\mathbf H\succeq0$ (positive semidefinite). \\

\end{longtable}
\endgroup
%%%%%%%% Proof %%%%%% of %%%%%5 Theorem 1 %%%%%%%%%%%%%%%
 \noindent
\textbf{Proof of Theorem~\ref{prop:oracle_fixed_t}.}
Fix $t_j$ and define
\[
\Omega_n=\bigcap_{w\in\{0,1\}}
\left\{
\inf_{\mathbf x}\widehat e^w(\mathbf x)\ge c/2,\quad
\inf_{(\mathbf x,t)}\widehat G^w(\mathbf x;t)\ge c_G/2,\quad
\|\widehat S^w\|_\infty\le2
\right\},
\]
where the infima are over $\mathcal X$ and
$\mathcal X\times\mathcal T$, respectively.
By Assumptions~\ref{assum:causal}--\ref{assum:nuisance},
$\Omega_n$ is $\mathcal S_{\mathrm{nuis}}$-measurable
and $\mathbb P_{\mathrm{nuis}}(\Omega_n)\to1$.
The estimator and comparator belong to
$\mathcal H_{t_j}
=\{\tanh\circ f:
f\in\mathcal F_{\mathrm{joint}}(L,\mathbf p,s_1,F),
\ p_{L+1}=1\}$.
Let $N=\mathcal N(n_1^{-1},\mathcal H_{t_j},
\|\cdot\|_\infty)$, with net centers in
$\mathcal H_{t_j}$, and
$\|g\|_{n_1}^2=n_1^{-1}\sum_i g(\mathbf X_i)^2$.
We establish on $\Omega_n$
\begin{equation}
\label{eq:fixed_risk_comparison}
\begin{aligned}
R_1(\widehat\tau(\cdot;t_j),\tau_0(\cdot;t_j))
\le{}&\tfrac43\mathbb E_{\mathrm{train}}
\|\widehat\tau(\cdot;t_j)-\tau_0(\cdot;t_j)\|_{n_1}^2
+C\frac{1+\log N}{n_1},
\end{aligned}
\end{equation}
and
\begin{equation}
\label{eq:fixed_empirical_upper}
\begin{aligned}
&\tfrac23\mathbb E_{\mathrm{train}}
\|\widehat\tau(\cdot;t_j)-\tau_0(\cdot;t_j)\|_{n_1}^2\\
&\quad\le
\mathcal E_{\mathrm{opt},n,1}
+2\|\tau_{\mathrm{apx}}(\cdot;t_j)
-\tau_0(\cdot;t_j)\|_\infty^2
+C\left\{\frac{1+\log N}{n_1}
+\widehat{\mathcal E}_n^2\right\}.
\end{aligned}
\end{equation}

\medskip
\noindent\emph{Part~(i).}
Combining these inequalities gives
\[
\begin{aligned}
R_1(\widehat\tau(\cdot;t_j),\tau_0(\cdot;t_j))
\le{}&2\mathcal E_{\mathrm{opt},n,1}
+4\|\tau_{\mathrm{apx}}(\cdot;t_j)
-\tau_0(\cdot;t_j)\|_\infty^2+C\left\{\frac{1+\log N}{n_1}
+\widehat{\mathcal E}_n^2\right\}.
\end{aligned}
\]
Since $\tanh$ is $1$-Lipschitz,
\citet[Lemma~5]{schmidt2020deep} and
Assumption~\ref{assum:dnn} with $m=1$ yield
\[
\frac{1+\log N}{n_1}
\lesssim\frac{(s_1+1)L\log n_1}{n_1}
\lesssim n_1^\delta\phi_{n_1}L\log^2n_1.
\]
This proves (i).

\medskip
\noindent\emph{Part~(ii).}
Apply \citet[proof of Theorem~1, eq.~(26)]{schmidt2020deep}
with the architecture enlargement of
\citet{Ghosal2026} at
$n_\star=\lceil n_1^{1+\delta/(1-c_{\mathrm{eff}})}\rceil$.
Then $\log n_\star\asymp\log n_1$,
$n_\star\phi_{n_\star}\asymp
n_1^{1+\delta}\phi_{n_1}$, and
$n_\star\phi_{n_\star}\log n_\star
\asymp n_1^{1+\delta}\phi_{n_1}\log n_1$.
With sufficiently large capacity constants in
Assumption~\ref{assum:dnn},
\begin{equation}
\label{eq:fixed_scalar_approx}
\begin{aligned}
\|\tau_{\mathrm{apx}}(\cdot;t_j)
-\tau_0(\cdot;t_j)\|_\infty^2
&\le
\inf_{f\in\mathcal F_{\mathrm{joint}}
(L,\mathbf p,(s_1),F)}
\sup_{\mathbf x\in\mathcal X}
|f(\mathbf x)-\psi_0(\mathbf x;t_j)|^2\\
&\quad\lesssim\phi_{n_\star}
\asymp\phi_{n_1}^{\,1+\delta/(1-c_{\mathrm{eff}})}
=o(n_1^\delta\phi_{n_1}L\log^2n_1).
\end{aligned}
\end{equation}
The first inequality follows from
\eqref{eq:tau_fixed_apx} and the Lipschitz
property of $\tanh$; admissibility follows from
$F\ge\|\psi_0(\cdot;t_j)\|_\infty$.
Part~(i), \eqref{eq:fixed_scalar_approx},
the hypothesis of (ii), and
$\mathbb P_{\mathrm{nuis}}(\Omega_n^c)\to0$
give
$R_1(\widehat\tau(\cdot;t_j),\tau_0(\cdot;t_j))
=O_{\mathbb P_{\mathrm{nuis}}}
(n_1^\delta\phi_{n_1}L\log^2n_1)$.
For $u>0$ and $0<\varepsilon<1$, conditional
Markov's inequality and the tower property yield
\[
\begin{aligned}
&\mathbb P\left\{
\|\widehat\tau(\cdot;t_j)-\tau_0(\cdot;t_j)\|
_{L_2(P_{\mathbf X})}^2>u\right\}\le
\mathbb P_{\mathrm{nuis}}\left\{
R_1(\widehat\tau(\cdot;t_j),\tau_0(\cdot;t_j))
>\varepsilon u\right\}+\varepsilon.
\end{aligned}
\]
Taking $u=K n_1^\delta\phi_{n_1}L\log^2n_1$,
then increasing $K$ and decreasing $\varepsilon$,
proves (ii).

\medskip
\noindent\emph{Verification of \eqref{eq:fixed_risk_comparison}.}
Let $v_1,\ldots,v_N$ be the net centers and
$k_*$ the smallest index with
$\|\widehat\tau(\cdot;t_j)-v_{k_*}(\cdot;t_j)\|_\infty
\le n_1^{-1}$.
For independent ghost observations
$\mathbf X'_i\sim P_{\mathbf X}$, define
\[
\begin{aligned}
q_k&=\|v_k(\cdot;t_j)-\tau_0(\cdot;t_j)\|
_{L_2(P_{\mathbf X})}^2,\\
D_k&=\frac1{n_1}\sum_{i=1}^{n_1}
\Big[
\{v_k(\mathbf X'_i;t_j)-\tau_0(\mathbf X'_i;t_j)\}^2
-\{v_k(\mathbf X_i;t_j)-\tau_0(\mathbf X_i;t_j)\}^2
\Big].
\end{aligned}
\]
Each summand is centered, bounded by $4$, and
has variance at most $8q_k$.
Bernstein's inequality and a union bound
\citep[proof of Lemma~4]{schmidt2020deep}
give, for $v=\{u+\log(2N)\}/n_1$,
\[
\mathbb P_{\mathrm{train}}\left\{
|D_k|\le4\sqrt{q_kv}+\tfrac83v
\le q_k/4+19v,\ \forall k\le N
\right\}\ge1-e^{-u}.
\]
Tail integration yields
$\mathbb E_{\mathrm{train}}|D_{k_*}|
\le\tfrac14\mathbb E_{\mathrm{train}}q_{k_*}
+C(1+\log N)/n_1$.
Since net approximation changes squared errors
by at most $4/n_1$,
\[
\begin{aligned}
\mathbb E_{\mathrm{train}}q_{k_*}
\le R_1(\widehat\tau,\tau_0)+4/n_1,\quad
R_1(\widehat\tau,\tau_0)
\le\mathbb E_{\mathrm{train}}
\|\widehat\tau-\tau_0\|_{n_1}^2
+\mathbb E_{\mathrm{train}}|D_{k_*}|+8/n_1,
\end{aligned}
\]
where all functions are evaluated at $t_j$.
Rearrangement proves \eqref{eq:fixed_risk_comparison}.

\medskip
\noindent\emph{Verification of \eqref{eq:fixed_empirical_upper}.}
On $\Omega_n$, the nuisance bounds and
$0\le Y_i(t_j)\le1$ give
$|\widehat\varphi_i(t_j)|
\le4+\frac2c(\frac2{c_G}+2)=M/2$.
Set
$\xi_i(t_j)=\widehat\varphi_i(t_j)
-\mathbb E_{\mathrm{train}}
[\widehat\varphi_i(t_j)\mid\mathbf X_i]$
and
$b(\mathbf x;t_j)=
\mathbb E_{\mathrm{train}}
[\widehat\varphi_i(t_j)\mid\mathbf X_i=\mathbf x]
-\tau_0(\mathbf x;t_j)$.
Then
$\widehat\varphi_i(t_j)-\tau_0(\mathbf X_i;t_j)
=\xi_i(t_j)+b(\mathbf X_i;t_j)$,
with $|\xi_i(t_j)|\le M$.
Conditional on the covariates, the $\xi_i$ are
independent and centered.

Independent censoring gives
\[
\begin{aligned}
&\mathbb E_{\mathrm{train}}
[\widehat\pi_i^w(t_j)\mid\mathbf X_i=\mathbf x]
-S^w(\mathbf x;t_j)=\frac{(\widehat e^w-e^w)
(\widehat S^w-S^w)}
{\widehat e^w}
-\frac{e^wS^w(\widehat G^w-G^w)}
{\widehat e^w\widehat G^w},
\end{aligned}
\]
where the functions on the right are evaluated
at $(\mathbf x,t_j)$ as appropriate.
Thus \eqref{nuis-error} gives
$\|b(\cdot;t_j)\|_\infty
\lesssim\widehat{\mathcal E}_n$.
For
$L_{n_1}(\tau)=n_1^{-1}\sum_i
\{\widehat\varphi_i(t_j)-\tau(\mathbf X_i;t_j)\}^2$,
expansion yields
\begin{equation}
\label{eq:fixed_loss_identity}
\begin{aligned}
L_{n_1}(\tau)-L_{n_1}(\tau_0)
&=\|\tau-\tau_0\|_{n_1}^2
-\frac2{n_1}\sum_i
\{\widehat\varphi_i(t_j)-\tau_0(\mathbf X_i;t_j)\}
\{\tau(\mathbf X_i;t_j)-\tau_0(\mathbf X_i;t_j)\}\\
&=\|\tau-\tau_0\|_{n_1}^2
-\frac2{n_1}\sum_i
\{\xi_i(t_j)+b(\mathbf X_i;t_j)\}
\{\tau(\mathbf X_i;t_j)-\tau_0(\mathbf X_i;t_j)\}.
\end{aligned}
\end{equation}
Here and below, $\tau$ and $\tau_0$ in
empirical norms denote their $t_j$ slices.
By \eqref{eq:optimization_error},
$\mathbb E_{\mathrm{train}}
[L_{n_1}(\widehat\tau)-L_{n_1}(\tau_{\mathrm{apx}})]
\le\mathcal E_{\mathrm{opt},n,1}$.
The deterministic comparator has zero expected
$\xi_i$ term. Applying $2|uv|\le u^2+v^2$
to its bias term gives
\begin{equation}
\label{eq:fixed_loss_comparison}
\begin{aligned}
\mathbb E_{\mathrm{train}}
\|\widehat\tau-\tau_0\|_{n_1}^2&\le\mathcal E_{\mathrm{opt},n,1}
+2\|\tau_{\mathrm{apx}}-\tau_0\|_\infty^2
+\|b\|_\infty^2\\
&+
\mathbb E_{\mathrm{train}}\left[
\frac2{n_1}\sum_i
\{\xi_i(t_j)+b(\mathbf X_i;t_j)\}
\{\widehat\tau(\mathbf X_i;t_j)-\tau_0(\mathbf X_i;t_j)\}
\right].
\end{aligned}
\end{equation}
\noindent
For each net center, define
\[
Z_k=
\frac{\sum_i\xi_i(t_j)
\{v_k(\mathbf X_i;t_j)-\tau_0(\mathbf X_i;t_j)\}}
{\sqrt{n_1}\,
\|v_k(\cdot;t_j)-\tau_0(\cdot;t_j)\|_{n_1}},
\]
setting $Z_k=0$ when the denominator vanishes.
Conditional Hoeffding and a union bound give
\[
\mathbb P_{\mathrm{train}}\left\{
\max_{k\le N}|Z_k|>u\mid\mathbf X_1,\ldots,\mathbf X_{n_1}
\right\}
\le2Ne^{-u^2/(2M^2)}.
\]
Hence
$\mathbb E_{\mathrm{train}}\max_{k\le N}Z_k^2
\lesssim1+\log N$.
Using $v_{k_*}$ as in
\citet[proof of Lemma~4]{schmidt2020deep},
Cauchy--Schwarz and Young's inequality yield
\[
\begin{aligned}
&\mathbb E_{\mathrm{train}}\bigg|
\frac2{n_1}\sum_i\xi_i(t_j)
\{\widehat\tau(\mathbf X_i;t_j)-\tau_0(\mathbf X_i;t_j)\}
\bigg|\\
&\le\frac{2M}{n_1}
+\frac2{\sqrt{n_1}}\mathbb E_{\mathrm{train}}
\big[
(\|\widehat\tau-\tau_0\|_{n_1}
+n_1^{-1})\max_{k\le N}|Z_k|
\big]\\
&\le\tfrac16\mathbb E_{\mathrm{train}}
\|\widehat\tau-\tau_0\|_{n_1}^2
+C\frac{1+\log N}{n_1}.
\end{aligned}
\]
Similarly,
\[
\begin{aligned}
&\mathbb E_{\mathrm{train}}\left|
\frac2{n_1}\sum_i b(\mathbf X_i;t_j)
\{\widehat\tau(\mathbf X_i;t_j)-\tau_0(\mathbf X_i;t_j)\}
\right|\le\tfrac16\mathbb E_{\mathrm{train}}
\|\widehat\tau-\tau_0\|_{n_1}^2
+6\|b(\cdot;t_j)\|_\infty^2.
\end{aligned}
\]
Substituting into \eqref{eq:fixed_loss_comparison}
and using $\|b(\cdot;t_j)\|_\infty
\lesssim\widehat{\mathcal E}_n$
proves \eqref{eq:fixed_empirical_upper}.

%%%%%%%%% Proof of Theorem 2 %%%%%%%%%%%%%% %%%%%%%%%%%%%%

%%%%%%%%%Proof of Theorem 2%%%%%%%%%%%%%%%%%%%%%%%
 \medskip
\noindent
\textbf{Proof of Theorem~\ref{thm:joint_oracle}.}
Work throughout on the event $\Omega_n$ of
Theorem~\ref{prop:oracle_fixed_t}, whose bounds hold
uniformly on $\mathcal X\times\mathcal T$.
All conditional inequalities have constants independent
of $n_1,m,J$, and the grid gaps.

\medskip
\noindent
\emph{Part~(i).}
As in the proof of Theorem~\ref{prop:oracle_fixed_t},
define
$\xi_{ij}=\widehat\varphi_i(t_j)
-\mathbb E_{\mathrm{train}}[
\widehat\varphi_i(t_j)\mid\mathbf X_i]$
and
$b_j(\mathbf x)=\mathbb E_{\mathrm{train}}[
\widehat\varphi_i(t_j)\mid\mathbf X_i=\mathbf x]
-\tau_0(\mathbf x,t_j)$.
Then
$\widehat\varphi_i(t_j)
=\tau_0(\mathbf X_i,t_j)+b_j(\mathbf X_i)+\xi_{ij}$,
where
$\mathbb E_{\mathrm{train}}[\xi_{ij}\mid\mathbf X_i]=0$.
The calculation in \eqref{eq:fixed_empirical_upper} gives
$\max_{j\le J}\|b_j\|_\infty
\lesssim\widehat{\mathcal E}_n$
and
$\max_{i\le n_1,j\le J}|\xi_{ij}|\le M$.
Conditional on the training covariates, the vectors
$(\xi_{i1},\ldots,\xi_{iJ})$ are independent across $i$.

Let
$N_j=\mathcal N(n_1^{-1},
\{\tau(\cdot,t_j):\tau\in\mathcal T_m\},
\|\cdot\|_\infty)$.
For networks $\mathbf f,\mathbf g$ in \eqref{dnn-class},
the nonnegative B-splines in \eqref{eq:readout}
sum to one. Since $\tanh$ is $1$-Lipschitz,
\[
\begin{aligned}
&\|\tau_{\mathbf f}(\cdot,t_j)
-\tau_{\mathbf g}(\cdot,t_j)\|_\infty\le
\sum_{\ell=1}^m B_{\ell,m}(t_j)
\|f_\ell-g_\ell\|_\infty
\le
\max_{\ell:B_{\ell,m}(t_j)>0}
\|f_\ell-g_\ell\|_\infty.
\end{aligned}
\]
At most a fixed number of basis functions are nonzero
at each $t_j$ \citep{deboor2001practical}.
Cover the corresponding coefficient classes at radius
$(2n_1)^{-1}$ and select a network from each nonempty
combination of covering balls. The selected slices
form an $n_1^{-1}$-cover.
Since output $\ell$ has at most $s_\ell$ nonzero
parameters, \citet[Lemma~5 and Remark~1]{schmidt2020deep}
and Assumption~\ref{assum:dnn} yield
\begin{equation}
\label{eq:joint_entropy}
\log N_j
\lesssim
\sum_{\ell:B_{\ell,m}(t_j)>0}
(s_\ell+1)L\log n_1,
\qquad
\max_{1\le j\le J}
\frac{1+\log N_j}{n_1}
\lesssim n_1^\delta\phi_{n_1}L\log^2 n_1 .
\end{equation}

For deterministic $\tau_{\mathbf g}\in\mathcal T_m$,
let
$L_{n_1}^{(j)}(\tau)
=n_1^{-1}\sum_{i=1}^{n_1}
\{\widehat\varphi_i(t_j)-\tau(\mathbf X_i,t_j)\}^2$
and
$L_{n_1,J}(\tau)
=J^{-1}\sum_{j=1}^J L_{n_1}^{(j)}(\tau)$.
The arguments for \eqref{eq:fixed_risk_comparison}
and \eqref{eq:fixed_loss_identity} apply uniformly
to each slice, including the jointly fitted estimator.
Thus,
\[
\begin{aligned}
\mathbb E_{\mathrm{train}}\!\left[
\|\widehat\tau(\cdot,t_j)-\tau_0(\cdot,t_j)\|
_{L_2(P_{\mathbf X})}^2\right]
&\le
2\mathbb E_{\mathrm{train}}\!\left[
L_{n_1}^{(j)}(\widehat\tau)
-L_{n_1}^{(j)}(\tau_{\mathbf g})\right]
+4\|\tau_{\mathbf g}(\cdot,t_j)
-\tau_0(\cdot,t_j)\|_\infty^2\\
&\qquad+
C\left\{\frac{1+\log N_j}{n_1}
+\widehat{\mathcal E}_n^2\right\}.
\end{aligned}
\]
Averaging over $j$, using
$\mathbb E_{\mathrm{train}}[
L_{n_1,J}(\widehat\tau)
-L_{n_1,J}(\tau_{\mathbf g})]
\le\mathcal E_{\mathrm{opt},n,J}$
from \eqref{eq:optimization_error},
and applying \eqref{eq:joint_entropy} give
\begin{equation}
\label{eq:joint_level_upper}
\begin{aligned}
R_J(\widehat\tau,\tau_0)
\le{}&2\mathcal E_{\mathrm{opt},n,J}
+4\|\tau_{\mathbf g}-\tau_0\|_\infty^2+C\{n_1^\delta\phi_{n_1}L\log^2 n_1
+\widehat{\mathcal E}_n^2\}.
\end{aligned}
\end{equation}
Taking $\tau_{\mathbf g}=\tau_{m,\mathrm{apx}}$
proves (i).

\medskip
\noindent
\emph{Part~(ii).}
Fix $\mathbf x\in\mathcal X$ and
$\tau_{\mathbf f},\tau_{\mathbf g}\in\mathcal T_m$.
Write $v_\ell=f_\ell(\mathbf x)-g_\ell(\mathbf x)$,
$S_\ell=\{t\in\mathcal T:B_{\ell,m}(t)>0\}$,
and let $\overline S_\ell$ be its closure.
The B-spline partition of unity and bounded
coefficients imply
$\|\psi_{\mathbf f}\|_\infty,
\|\psi_{\mathbf g}\|_\infty\le F$.
Since $\inf_{|z|\le F}\tanh'(z)>0$,
the lower Gram bound gives
\begin{equation}
\label{eq:joint_gram_control}
\frac1m\sum_{\ell=1}^m v_\ell^2
\lesssim
\frac1J\sum_{j=1}^J
\left|\sum_{\ell=1}^m
v_\ell B_{\ell,m}(t_j)\right|^2
\lesssim
\frac1J\sum_{j=1}^J
|\tau_{\mathbf f}(\mathbf x,t_j)
-\tau_{\mathbf g}(\mathbf x,t_j)|^2 .
\end{equation}

Let $\mathbf u_\ell$ have $k$th entry
$\mathbb I(\overline S_k\cap\overline S_\ell
\ne\varnothing)$.
Fixed spline order implies
$\|\mathbf u_\ell\|_2^2\lesssim1$ and
$(\mathbf B\mathbf u_\ell)_j=1$
whenever $t_j\in\overline S_\ell$.
By the upper Gram bound,
\[
\frac1J\sum_{j=1}^J
\mathbb I(t_j\in\overline S_\ell)
\le
\mathbf u_\ell^\top
\frac{\mathbf B^\top\mathbf B}{J}
\mathbf u_\ell
\lesssim\frac1m.
\]
Since $\overline S_\ell$ is an interval and $m\le J$,
the number of adjacent grid intervals intersecting it
satisfies
\[
\begin{aligned}
&\sum_{j=1}^{J-1}
\mathbb I\big(
\overline S_\ell\cap[t_j,t_{j+1}]
\ne\varnothing\big)\le
2+\sum_{j=1}^J
\mathbb I(t_j\in\overline S_\ell)
\lesssim J/m.
\end{aligned}
\]

By the chain rule,
\[
\begin{aligned}
\partial_t(\tau_{\mathbf f}
-\tau_{\mathbf g})(\mathbf x,t)
&=\tanh'\{\psi_{\mathbf f}(\mathbf x,t)\}
\sum_{\ell=1}^m v_\ell\partial_tB_{\ell,m}(t)\\
&\quad+
\big[\tanh'\{\psi_{\mathbf f}(\mathbf x,t)\}
-\tanh'\{\psi_{\mathbf g}(\mathbf x,t)\}\big]
\sum_{\ell=1}^m
g_\ell(\mathbf x)\frac{d}{dt}B_{\ell,m}(t).
\end{aligned}
\]
The B-spline differentiation formula
\citep[Ch.~X, eq.~(12)]{deboor2001practical}
and quasi-uniform knots imply
$\max_\ell\|\partial_tB_{\ell,m}\|_\infty
\lesssim m$.
Using bounded coefficients and derivatives of
$\tanh$ gives
\[
|\partial_t(\tau_{\mathbf f}
-\tau_{\mathbf g})(\mathbf x,t)|^2
\lesssim
m^2\sum_{\ell:t\in\overline S_\ell}v_\ell^2.
\]
The fundamental theorem of calculus and
Cauchy--Schwarz yield
\[
\begin{aligned}
\left|
\frac{(\tau_{\mathbf f}-\tau_{\mathbf g})
(\mathbf x,t_{j+1})
-(\tau_{\mathbf f}-\tau_{\mathbf g})
(\mathbf x,t_j)}
{t_{j+1}-t_j}
\right|^2
&\le
\frac1{t_{j+1}-t_j}
\int_{t_j}^{t_{j+1}}
|\partial_t(\tau_{\mathbf f}
-\tau_{\mathbf g})(\mathbf x,u)|^2\,du\\
&\quad\lesssim
m^2\sum_{\ell=1}^m v_\ell^2
\mathbb I\big(
\overline S_\ell\cap[t_j,t_{j+1}]
\ne\varnothing\big).
\end{aligned}
\]
Averaging over $j$ and using the support count,
\[
\begin{aligned}
&\frac1{J-1}\sum_{j=1}^{J-1}
\left|
\frac{(\tau_{\mathbf f}-\tau_{\mathbf g})
(\mathbf x,t_{j+1})
-(\tau_{\mathbf f}-\tau_{\mathbf g})
(\mathbf x,t_j)}
{t_{j+1}-t_j}
\right|^2\\
&\quad\lesssim
\frac{m^2}{J-1}
\sum_{\ell=1}^m v_\ell^2
\sum_{j=1}^{J-1}
\mathbb I\big(
\overline S_\ell\cap[t_j,t_{j+1}]
\ne\varnothing\big)
\lesssim m\sum_{\ell=1}^m v_\ell^2.
\end{aligned}
\]
Combining this with \eqref{eq:joint_gram_control},
integrating over $P_{\mathbf X}$,
and taking $\mathbb E_{\mathrm{train}}$ gives
\begin{equation}
\label{eq:joint_contrast_transfer}
R_{\Delta,J}(\tau_{\mathbf f},\tau_{\mathbf g})
\le C m^2 R_J(\tau_{\mathbf f},\tau_{\mathbf g}).
\end{equation}

The comparison holds for each realization of the
coefficients, including $\tau_{\mathbf f}=\widehat\tau$.
Take $\tau_{\mathbf g}=\tau_{m,\mathrm{apx}}
=\tanh\circ\psi_{m,\mathrm{apx}}\in\mathcal T_m$
from \eqref{eq:tau_m_apx}.
The fundamental theorem of calculus and
Cauchy--Schwarz give
\[
\begin{aligned}
&\left|
\frac{(\tau_{m,\mathrm{apx}}-\tau_0)
(\mathbf x,t_{j+1})
-(\tau_{m,\mathrm{apx}}-\tau_0)
(\mathbf x,t_j)}
{t_{j+1}-t_j}
\right|^2\\
&\quad\le
\frac1{t_{j+1}-t_j}
\int_{t_j}^{t_{j+1}}
|\partial_u(\tau_{m,\mathrm{apx}}
-\tau_0)(\mathbf x,u)|^2\,du\le
\|\partial_t(\tau_{m,\mathrm{apx}}
-\tau_0)\|_\infty^2.
\end{aligned}
\]
Hence
$R_{\Delta,J}(\tau_{m,\mathrm{apx}},\tau_0)
\le\|\partial_t(\tau_{m,\mathrm{apx}}
-\tau_0)\|_\infty^2$.
Using \eqref{eq:joint_contrast_transfer}
and the squared triangle inequality,
\[
\begin{aligned}
R_{\Delta,J}(\widehat\tau,\tau_0)
&\le
2Cm^2R_J(\widehat\tau,\tau_{m,\mathrm{apx}})
+2\|\partial_t(\tau_{m,\mathrm{apx}}
-\tau_0)\|_\infty^2\\
&\le
4Cm^2\bigl\{
R_J(\widehat\tau,\tau_0)
+\|\tau_{m,\mathrm{apx}}-\tau_0\|_\infty^2
\bigr\}+
2\|\partial_t(\tau_{m,\mathrm{apx}}
-\tau_0)\|_\infty^2.
\end{aligned}
\]
Applying part~(i) proves (ii).

\medskip
\noindent
Before proving Corollary~\ref{lem:An}, we establish the following
lemma and defer its proof to the end of all the proofs.

%%%%%%%%%%%%%%%Proof of Corollary 1%%%%%%%%%%%%%%%%%%%%%%

\begin{lemma}[Approximation Lemma]
\label{lem:apprx}
Suppose $\tau_0\in
\mathcal G_{\mathrm{joint}}^*(\mathcal G,K_t,\beta)$,
$0<\beta\le1$, and Assumption~\ref{assum:dnn} holds at
sample size $r$, with $0\le\delta<c_{\mathrm{eff}}$.
For the B-spline basis in \eqref{eq:readout}, the approximations in \eqref{eq:tau_m_apx},
with architecture evaluated at sample size $r$, satisfy
\begin{align*}
\|\psi_{m,\mathrm{apx}}-\psi_0\|_\infty^2
+\|\tau_{m,\mathrm{apx}}-\tau_0\|_\infty^2
&\lesssim m^{-2(1+\beta)}
+\phi_r^{\,1+\delta/(1-c_{\mathrm{eff}})},\\
\|\partial_t(\psi_{m,\mathrm{apx}}-\psi_0)\|_\infty^2
+\|\partial_t(\tau_{m,\mathrm{apx}}-\tau_0)\|_\infty^2
&\lesssim m^{-2\beta}
+m^2\phi_r^{\,1+\delta/(1-c_{\mathrm{eff}})}.
\end{align*}
The constants are independent of $r$, $m$, and $J$.
\end{lemma}

\medskip
\noindent
\textbf{Proof of Corollary~\ref{lem:An}.}
Let $\Omega_n\in\mathcal S_{\mathrm{nuis}}$ be the event
in Theorem~\ref{thm:joint_oracle} at sample size $r$, with
$\mathbb P_{\mathrm{nuis}}(\Omega_n)\to1$.
On $\Omega_n$, apply part~(i) of Theorem \ref{thm:joint_oracle} with map
$\tanh(\psi_{m,\mathrm{apx}})$. Lemma~\ref{lem:apprx} gives
\[
\begin{aligned}
R_J(\widehat\tau_r,\tau_0)
\lesssim{}&
\mathcal E_{\mathrm{opt},r,J}
+r^\delta\phi_r L\log^2r+\widehat{\mathcal E}_n^2+m^{-2(1+\beta)}
+\phi_r^{\,1+\delta/(1-c_{\mathrm{eff}})}.
\end{aligned}
\]
Since
$\phi_r^{\,1+\delta/(1-c_{\mathrm{eff}})}
\lesssim r^\delta\phi_r L\log^2r$,
Assumption~\ref{assum:opt_nuisance_error} and
$\mathbb P_{\mathrm{nuis}}(\Omega_n)\to1$ imply
\begin{equation}
\label{eq:lem1_R_J_bound}
R_J(\widehat\tau_r,\tau_0)
=O_{\mathbb P_{\mathrm{nuis}}}
\bigl(r^\delta\phi_r L\log^2r+m^{-2(1+\beta)}\bigr).
\end{equation}
\noindent
For $q=\sum_{\ell=1}^m v_\ell B_{\ell,m}\in\mathcal B_m$,
the basis bounds and the lower Gram bound give
\begin{equation}
\label{eq:spline_level_slope}
\|q\|_\infty^2+\|\partial_tq\|_\infty^2
\lesssim m^2\sum_{\ell=1}^m v_\ell^2
\lesssim\frac{m^3}{J}\sum_{j=1}^J|q(t_j)|^2.
\end{equation}
At each $\mathbf x$, apply \eqref{eq:spline_level_slope}
to $q(t)=(\widehat\psi_r-\psi_{m,\mathrm{apx}})(\mathbf x,t)$.
The bounded latent ranges give
\(
|\widehat\psi_r(\mathbf x,t)-\psi_0(\mathbf x,t)|
\lesssim|\widehat\tau_r(\mathbf x,t)-\tau_0(\mathbf x,t)|.
\)

\[
\begin{aligned}
M_r(\mathbf x)
\lesssim{}&
\frac{m^3}{J}\sum_{j=1}^J
\mathbb E_{\mathrm{train}}
|\widehat\tau_r(\mathbf x,t_j)-\tau_0(\mathbf x,t_j)|^2+m^3\|\psi_{m,\mathrm{apx}}-\psi_0\|_\infty^2
+\|\partial_t(\psi_{m,\mathrm{apx}}-\psi_0)\|_\infty^2.
\end{aligned}
\]
Integrating and using Lemma~\ref{lem:apprx} and
\eqref{eq:lem1_R_J_bound},
\begin{align}
\int_{\mathcal X}M_r(\mathbf x)\,dP_{\mathbf X}(\mathbf x)
&\lesssim
m^3R_J(\widehat\tau_r,\tau_0)
+m^{1-2\beta}
+m^3\phi_r^{\,1+\delta/(1-c_{\mathrm{eff}})}
\nonumber\\
&=O_{\mathbb P_{\mathrm{nuis}}}
\bigl(m^3r^\delta\phi_r L\log^2r+m^{1-2\beta}\bigr).
\label{eq:integrated_M_r}
\end{align}
\noindent
Define
\[
\begin{aligned}
\Omega'_n
&=\Omega_n\cap
\biggl\{
\int_{\mathcal X}M_r(\mathbf x)\,dP_{\mathbf X}(\mathbf x)
\le\ell_n
\bigl(m^3r^\delta\phi_r L\log^2r+m^{1-2\beta}\bigr)
\biggr\},\\
\mathcal A_n
&=\Bigl\{\mathbf x\in\mathcal X
\ \big|\
\mathbb E_{\mathrm{nuis}}
[\mathbb I(\Omega'_n)M_r(\mathbf x)]
\le\ell_n^2
(m^3r^\delta\phi_r L\log^2r+m^{1-2\beta})
\Bigr\}.
\end{aligned}
\]
By \eqref{eq:integrated_M_r},
$\mathbb P_{\mathrm{nuis}}(\Omega'_n)\to1$.
The set $\mathcal A_n$ is deterministic, and Tonelli's
theorem and Markov's inequality give
\[
P_{\mathbf X}(\mathcal A_n^c)
\le
\frac{
\mathbb E_{\mathrm{nuis}}\bigl[
\mathbb I(\Omega'_n)
\int_{\mathcal X}M_r(\mathbf x)\,dP_{\mathbf X}(\mathbf x)
\bigr]}
{\ell_n^2
\bigl(m^3r^\delta\phi_r L\log^2r+m^{1-2\beta}\bigr)}
\le\ell_n^{-1}.
\]
For every deterministic $\mathbf x_{*,n}\in\mathcal A_n$,
\[
\begin{aligned}
&\mathbb P_{\mathrm{nuis}}\bigl\{
M_r(\mathbf x_{*,n})>
\ell_n^3
\bigl(m^3r^\delta\phi_r L\log^2r+m^{1-2\beta}\bigr)
\bigr\}\\
&\quad\le
\mathbb P_{\mathrm{nuis}}\bigl((\Omega'_n)^c\bigr)
+
\frac{
\mathbb E_{\mathrm{nuis}}
[\mathbb I(\Omega'_n)M_r(\mathbf x_{*,n})]}
{\ell_n^3
\bigl(m^3r^\delta\phi_r L\log^2r+m^{1-2\beta}\bigr)}
\le o(1)+\ell_n^{-1}\longrightarrow0.
\end{aligned}
\]

%%%%%%%%%%%%%%%%%%%%%%%%%%%%%%%%%

%%%%%%%%%%%%%%%%%%%Proof of Proposition 1%%%%%%%%%%%%%%%%%%%

\medskip
\noindent
\textbf{Proof of Proposition~\ref{cor:asymptotic_bias}.}
In each part, work on the intersection of the events in
Assumptions~\ref{assum:xi} and~\ref{assum:bias_domination}
and the stated correlation condition when applicable.
Its $\mathbb P_{\mathrm{nuis}}$-probability tends to one.
We first prove part~(i) under
$0<\alpha_{\mathrm{lower,pt}}<\alpha<1$.
Since $m\asymp m_{\mathrm{opt}}(\delta)$ in \eqref{eq:m_opt},
we have
\[
m^{-2(1+\beta)}
\asymp r^\delta\phi_r L\log^2r
=r^{-(c_{\mathrm{eff}}-\delta)}L\log^2r.
\]
Since $0<\delta<c_{\mathrm{eff}}<1$ and
$\nu\ge1$ in Assumption~\ref{assum:xi},
$\nu>c_{\mathrm{eff}}-\delta$, and
\[
\frac{r^{-\nu}}{m^{-2(1+\beta)}}
\asymp
\frac{r^{c_{\mathrm{eff}}-\delta-\nu}}{L\log^2r}
\longrightarrow0,
\qquad
\frac{\phi_r^{\,1+\delta/(1-c_{\mathrm{eff}})}}
     {m^{-2(1+\beta)}}
\asymp
\frac{r^{-\delta/(1-c_{\mathrm{eff}})}}{L\log^2r}
\longrightarrow0.
\]
This implies
\(
\phi_r^{\,1+\delta/(1-c_{\mathrm{eff}})}+r^{-\nu}
=o\bigl(m^{-2(1+\beta)}\bigr).
\)
Lemma~\ref{lem:apprx} therefore gives
\[
\|\psi_{m,\mathrm{apx}}-\psi_0\|_\infty^2
\lesssim m^{-2(1+\beta)},
\qquad
\|\partial_t(\psi_{m,\mathrm{apx}}-\psi_0)\|_\infty^2
\lesssim m^{-2\beta}.
\]
By Assumption~\ref{assum:bias_domination} and the squared
triangle inequality,
\begin{equation}
\label{eq:cor_bias_level}
\begin{aligned}
&\sup_{s\in\mathcal T}
\big|\mathbb E_{\mathrm{train}}
[\widehat\psi_r(\mathbf x_*,s)]-\psi_0(\mathbf x_*,s)\big|^2
\lesssim
\|\psi_{m,\mathrm{apx}}-\psi_0\|_\infty^2+r^{-\nu}
\lesssim m^{-2(1+\beta)}.
\end{aligned}
\end{equation}
Assumption~\ref{assum:xi} gives
\(
\sigma_\psi^2(\mathbf x_*,t)
=(r^2/n_1)\zeta_r^{(1)}(\mathbf x_*,t)
\gtrsim r^{2-\nu}/n_1.
\)
Consequently,
\begin{equation}
\label{eq:cor_bias_ratio}
\begin{aligned}
\frac{
|\mathbb E_{\mathrm{train}}[\widehat\psi_r(\mathbf x_*,t)]
-\psi_0(\mathbf x_*,t)|^2}
{\sigma_\psi^2(\mathbf x_*,t)}
&\lesssim n_1r^{\nu-2}m^{-2(1+\beta)}
\asymp
n_1^{1-\alpha(2-\nu+c_{\mathrm{eff}}-\delta)}
L\log^2r
\longrightarrow0,
\end{aligned}
\end{equation}
since $\alpha>\alpha_{\mathrm{lower,pt}}$ and
$L\lesssim(\log r)^\mu$. This proves part~(i).

To compare with Corollary~\ref{lem:An}, the conditional
projection inequality gives
\[
c_\zeta r^{1-\nu}
\le r\zeta_r^{(1)}(\mathbf x_*,t)
\le \operatorname{Var}_{\mathrm{train}}
[\widehat\psi_r(\mathbf x_*,t)]
\le M_r(\mathbf x_*).
\]
At the chosen $m$,
\[
m^3r^\delta\phi_r L\log^2r+m^{1-2\beta}
\asymp
(r^\delta\phi_r L\log^2r)^{(2\beta-1)/\{2(1+\beta)\}}.
\]
Thus the lower variance bound is of smaller order than
this upper error rate whenever
\(
\nu>1+(2\beta-1)(c_{\mathrm{eff}}-\delta)/\{2(1+\beta)\}.
\)
the compatibility lower bound is
smaller than the upper bound in \eqref{eq:inference_tuning},
\[
1+\frac{(2\beta-1)(c_{\mathrm{eff}}-\delta)}{2(1+\beta)}
<
1+\frac{c_{\mathrm{eff}}-\delta}{2},
\]
for every $0<\delta<c_{\mathrm{eff}}$ and $1/2<\beta\le1$.

We next prove part~(ii). Boundedness of $\psi_0$ implies
$\inf_{\mathbf x,s}\{1-\tau_0(\mathbf x,s)^2\}>0$.
The definitions of $\rho_r$ and
$\sigma_{\mathrm{pt}}^2$ give
\[
\begin{aligned}
\sigma_{\mathrm{pt}}^2(\mathbf x_*,\mathbf x_*',t)
\ge{}&
\rho_r(\mathbf x_*,t,\mathbf x_*',t)
\bigl[
\{1-\tau_0(\mathbf x_*,t)^2\}^2
\sigma_\psi^2(\mathbf x_*,t)\\
&\hspace{15mm}
+\{1-\tau_0(\mathbf x_*',t)^2\}^2
\sigma_\psi^2(\mathbf x_*',t)
\bigr]
\gtrsim \frac{r^{2-\nu}}{n_1}.
\end{aligned}
\]
Since $\tanh$ is $1$-Lipschitz, \eqref{eq:cor_bias_level}
at the two profiles yields
\[
\begin{aligned}
&\big|
\tanh\{\mathbb E_{\mathrm{train}}
[\widehat\psi_r(\mathbf x_*,t)]\}
-\tanh\{\mathbb E_{\mathrm{train}}
[\widehat\psi_r(\mathbf x_*',t)]\}
-\theta^{\mathrm{pt}}(\mathbf x_*,\mathbf x_*',t)
\big|^2\\
&\le
2|\mathbb E_{\mathrm{train}}[\widehat\psi_r(\mathbf x_*,t)]
-\psi_0(\mathbf x_*,t)|^2
+2|\mathbb E_{\mathrm{train}}[\widehat\psi_r(\mathbf x_*',t)]
-\psi_0(\mathbf x_*',t)|^2\\
&\lesssim m^{-2(1+\beta)}.
\end{aligned}
\]
Dividing by the variance bound and applying
\eqref{eq:cor_bias_ratio} proves part~(ii).
The compatibility comparison in part~(i) applies at
both profiles.

Finally, we prove part~(iii) under
$0<\alpha_{\mathrm{lower,tm}}<\alpha<1$. Since
\(
\mathbb E_{\mathrm{train}}[\widehat\psi_r(\mathbf x_*,\cdot)]
-\psi_{m,\mathrm{apx}}(\mathbf x_*,\cdot)\in\mathcal B_m,
\)
the spline inverse inequality and
Assumption~\ref{assum:bias_domination} give
\[
\begin{aligned}
\sup_{s\in\mathcal T}
\big|
\partial_s\{\mathbb E_{\mathrm{train}}
[\widehat\psi_r(\mathbf x_*,s)]
-\psi_{m,\mathrm{apx}}(\mathbf x_*,s)\}
\big|^2
&\lesssim
m^2\sup_{s\in\mathcal T}
|\mathbb E_{\mathrm{train}}[\widehat\psi_r(\mathbf x_*,s)]
-\psi_{m,\mathrm{apx}}(\mathbf x_*,s)|^2\\
&\quad\lesssim
m^2\bigl(\|\psi_{m,\mathrm{apx}}-\psi_0\|_\infty^2
+r^{-\nu}\bigr)
\lesssim m^{-2\beta}.
\end{aligned}
\]
Adding the derivative approximation bound from
Lemma~\ref{lem:apprx} yields
\begin{equation}
\label{eq:cor_bias_slope}
\sup_{s\in\mathcal T}
\big|
\partial_s\{\mathbb E_{\mathrm{train}}
[\widehat\psi_r(\mathbf x_*,s)]-\psi_0(\mathbf x_*,s)\}
\big|^2
\lesssim m^{-2\beta}.
\end{equation}
By the chain rule,
\eqref{eq:cor_bias_level}, \eqref{eq:cor_bias_slope},
and $\|\partial_t\psi_0\|_\infty\le K_t$,
\[
\begin{aligned}
&\sup_{s\in\mathcal T}
\big|
\partial_s\{\tanh(\mathbb E_{\mathrm{train}}
[\widehat\psi_r(\mathbf x_*,s)])
-\tau_0(\mathbf x_*,s)\}
\big|^2
\lesssim
\sup_{s\in\mathcal T}
\big|\partial_s\{\mathbb E_{\mathrm{train}}
[\widehat\psi_r(\mathbf x_*,s)]-\psi_0(\mathbf x_*,s)\}\big|^2\\
&\hspace{60mm}
+K_t^2\sup_{s\in\mathcal T}
|\mathbb E_{\mathrm{train}}[\widehat\psi_r(\mathbf x_*,s)]
-\psi_0(\mathbf x_*,s)|^2
\lesssim m^{-2\beta}.
\end{aligned}
\]
The fundamental theorem of calculus therefore gives
\[
\begin{aligned}
&\big|
\tanh\{\mathbb E_{\mathrm{train}}
[\widehat\psi_r(\mathbf x_*,t')]\}
-\tanh\{\mathbb E_{\mathrm{train}}
[\widehat\psi_r(\mathbf x_*,t)]\}
-\theta^{\mathrm{tm}}(t,t',\mathbf x_*)
\big|^2
\lesssim (t'-t)^2m^{-2\beta}.
\end{aligned}
\]
The covariance inequality used in part~(ii) and the temporal
correlation condition imply
\[
\sigma_{\mathrm{tm}}^2(t,t',\mathbf x_*)
\gtrsim
\rho_r(\mathbf x_*,t,\mathbf x_*,t')
\frac{r^{2-\nu}}{n_1}
\gtrsim (t'-t)^2\frac{r^{2-\nu}}{n_1}.
\]
Thus the ratio in part~(iii) is bounded by
\[
\begin{aligned}
n_1r^{\nu-2}m^{-2\beta}
&\asymp
n_1^{1-\alpha\{2-\nu+
\beta(c_{\mathrm{eff}}-\delta)/(1+\beta)\}}
(L\log^2r)^{\beta/(1+\beta)}
\longrightarrow0,
\end{aligned}
\]
since $\alpha>\alpha_{\mathrm{lower,tm}}$, including when
$t'-t\to0$. For the temporal compatibility comparison,
$\alpha_{\mathrm{lower,tm}}<1$ requires
$\nu<1+\beta(c_{\mathrm{eff}}-\delta)/(1+\beta)$. the corresponding comparison with upper bound of $\nu$ in
\eqref{eq:inference_tuning_tm} is
\[
1+\frac{(2\beta-1)(c_{\mathrm{eff}}-\delta)}{2(1+\beta)}
<
1+\frac{\beta(c_{\mathrm{eff}}-\delta)}{2(1+\beta)},
\]
which holds for every $0<\delta<c_{\mathrm{eff}}$ and
$1/2<\beta<1$. At $\beta=1$, the two bounds coincide.
Since the events used in each part have
$\mathbb P_{\mathrm{nuis}}$-probability tending to one,
all three ratios are $o_{\mathbb P_{\mathrm{nuis}}}(1)$.

%%%%%%%%%%%%%%%%%%%%Proof of Theorem 3%%%%%%%%%%%%%%%%%%%%%%%%%%%%
\medskip
\noindent
\textbf{Proof of Theorem~\ref{thm:pointwise_clt}.}
Fix $(\mathbf x_*,t)$ with $\mathbf x_*\in\mathcal A_n$
for all sufficiently large $n$.
By Assumption~\ref{assum:xi} and
Proposition~\ref{cor:asymptotic_bias}(i), there exist
$\epsilon_n\downarrow0$ and $E_n\in\mathcal S_{\mathrm{nuis}}$
with $\mathbb P_{\mathrm{nuis}}(E_n)\to1$ such that
\[
\zeta_r^{(1)}(\mathbf x_*,t)\ge c_\zeta r^{-\nu},
\qquad
\frac{
|\mathbb E_{\mathrm{train}}[\widehat\psi_r(\mathbf x_*,t)]
-\psi_0(\mathbf x_*,t)|}
{\sigma_\psi(\mathbf x_*,t)}
\le\epsilon_n
\quad\text{on }E_n.
\]
All conditional calculations below are on $E_n$.
Each selected subsample has the same marginal distribution,
so
\(
\mathbb E_{\mathrm{train}}[\bar\psi_B(\mathbf x_*,t)]
=
\mathbb E_{\mathrm{train}}[\widehat\psi_r(\mathbf x_*,t)].
\)
Define
\[
h_{1,r}(D_i,\mathbf x_*,t)
=
\mathbb E_{\mathrm{train}}
[\widehat\psi^b(\mathbf x_*,t)\mid D_i,I_{b,i}=1]
-\mathbb E_{\mathrm{train}}[\widehat\psi_r(\mathbf x_*,t)].
\]
This does not depend on $b$ by symmetry.
An estimator excluding $i$ is independent of $D_i$.
Moreover,
\(
N_i=\sum_{b=1}^B I_{b,i}
\sim\operatorname{Bin}(B,r/n_1).
\)
Hence the H\'ajek projection is
\begin{equation}
\label{eq:pointwise_hajek}
\begin{aligned}
\mathring T_{n_1,r}(\mathbf x_*,t)
&=
\sum_{i=1}^{n_1}
\bigl\{
\mathbb E_{\mathrm{train}}[\bar\psi_B(\mathbf x_*,t)\mid D_i]
-\mathbb E_{\mathrm{train}}[\widehat\psi_r(\mathbf x_*,t)]
\bigr\}\\
&=
\sum_{i=1}^{n_1}
\frac{\mathbb E_{\mathrm{train}}[N_i]}B
h_{1,r}(D_i,\mathbf x_*,t)
=
\frac r{n_1}\sum_{i=1}^{n_1}h_{1,r}(D_i,\mathbf x_*,t).
\end{aligned}
\end{equation}

The two estimators in \eqref{eq:zeta_r} share one observation
and are otherwise conditionally independent. Therefore,
\[
\begin{aligned}
&\mathbb E_{\mathrm{train}}[h_{1,r}(D_i,\mathbf x_*,t)]
=0,\qquad
\operatorname{Var}_{\mathrm{train}}
[h_{1,r}(D_i,\mathbf x_*,t)]=\zeta_r^{(1)}(\mathbf x_*,t),\\
&\hspace{20mm}\operatorname{Var}_{\mathrm{train}}
[\mathring T_{n_1,r}(\mathbf x_*,t)]
=\frac{r^2}{n_1}\zeta_r^{(1)}(\mathbf x_*,t)
=\sigma_\psi^2(\mathbf x_*,t).
\end{aligned}
\]
By \eqref{dnn-class}, $B_{\ell,m}(t)\ge0$, and
$\sum_{\ell=1}^mB_{\ell,m}(t)=1$,
\(
|\widehat\psi^b(\mathbf x_*,t)|
\le\sum_{\ell=1}^m
|\widehat f_\ell^b(\mathbf x_*)|B_{\ell,m}(t)
\le F.
\)
Thus
\[
|h_{1,r}(D_i,\mathbf x_*,t)|\le2F,
\qquad
\mathbb E_{\mathrm{train}}
|h_{1,r}(D_i,\mathbf x_*,t)|^3
\le2F\zeta_r^{(1)}(\mathbf x_*,t).
\]
The Berry--Esseen inequality gives
\[
\begin{aligned}
\sup_{z\in\mathbb R}
\biggl|
\mathbb P_{\mathrm{train}}\biggl\{
\frac{\mathring T_{n_1,r}(\mathbf x_*,t)}
     {\sigma_\psi(\mathbf x_*,t)}
\le z
\biggr\}-\Phi(z)
\biggr|
\lesssim
\frac1{\sqrt{n_1\zeta_r^{(1)}(\mathbf x_*,t)}}
\lesssim\sqrt{\frac{r^\nu}{n_1}}
\longrightarrow0,
\end{aligned}
\]
where $\Phi$ is the standard normal distribution function
and $\alpha<1/\nu$ by \eqref{eq:inference_tuning}.

We next establish H\'ajek projection dominance.
Let $\mathcal I^b=\{k_1,\ldots,k_r\}$.
After averaging over fitting randomness, the symmetric
subsample kernel has the Hoeffding decomposition
\begin{equation}
\label{eq:pointwise_hoeff}
\begin{aligned}
&\mathbb E_{\mathrm{train}}
[\widehat\psi^b(\mathbf x_*,t)
 \mid\mathcal D_{\mathrm{train}},\mathcal I^b]
-\mathbb E_{\mathrm{train}}[\widehat\psi_r(\mathbf x_*,t)]\\
&\qquad=
\sum_{i=1}^r T_1(D_{k_i})
+\sum_{1\le i_1<i_2\le r}
T_2(D_{k_{i_1}},D_{k_{i_2}})
+\cdots+T_r(D_{k_1},\ldots,D_{k_r}),
\end{aligned}
\end{equation}
where $T_1(D_i)=h_{1,r}(D_i,\mathbf x_*,t)$ and
\(
\mathbb E_{\mathrm{train}}
[T_k(D_1,\ldots,D_k)\mid D_1,\ldots,D_{k-1}]=0
\)
for $k\ge2$.
Each fixed set of $k$ observations appears in
$\binom{n_1-k}{r-k}$ of the $\binom{n_1}{r}$ subsamples.
Consequently,
\[
\begin{aligned}
\mathbb E_{\mathrm{train}}
[\bar\psi_B(\mathbf x_*,t)\mid\mathcal D_{\mathrm{train}}]
-\mathbb E_{\mathrm{train}}[\widehat\psi_r(\mathbf x_*,t)]
=
\sum_{k=1}^r
\frac{\binom{n_1-k}{r-k}}{\binom{n_1}{r}}
\sum_{i_1<\cdots<i_k}
T_k(D_{i_1},\ldots,D_{i_k}).
\end{aligned}
\]
The $k=1$ term is $\mathring T_{n_1,r}(\mathbf x_*,t)$.
Set
\[
R(\mathbf x_*,t)
=
\bar\psi_B(\mathbf x_*,t)
-\mathbb E_{\mathrm{train}}[\widehat\psi_r(\mathbf x_*,t)]
-\mathring T_{n_1,r}(\mathbf x_*,t),
\qquad
V_k=\operatorname{Var}_{\mathrm{train}}
[T_k(D_1,\ldots,D_k)].
\]
Conditional variance decomposition gives
\[
\begin{aligned}
\mathbb E_{\mathrm{train}}[R(\mathbf x_*,t)^2]
=&
\mathbb E_{\mathrm{train}}\bigl[
\operatorname{Var}_{\mathrm{train}}
\{R(\mathbf x_*,t)\mid\mathcal D_{\mathrm{train}}\}
\bigr]+
\mathbb E_{\mathrm{train}}\bigl[
\{\mathbb E_{\mathrm{train}}
[R(\mathbf x_*,t)\mid\mathcal D_{\mathrm{train}}]\}^2
\bigr]\\
=&\mathcal I_1+\mathcal I_2.
\end{aligned}
\]

By Hoeffding orthogonality,
\[
\begin{aligned}
\mathcal I_2
&=
\sum_{k=2}^r
\frac{\binom{n_1-k}{r-k}^2\binom{n_1}{k}}
     {\binom{n_1}{r}^2}V_k
=
\sum_{k=2}^r
\frac{(n_1-k)!\,r!}{n_1!\,(r-k)!}
\binom rk V_k\\
&\le
\frac{r(r-1)}{n_1(n_1-1)}
\sum_{k=2}^r\binom rk V_k
\le
\frac{r(r-1)}{n_1(n_1-1)}
\operatorname{Var}_{\mathrm{train}}
[\widehat\psi_r(\mathbf x_*,t)],
\end{aligned}
\]
since the factorial ratio decreases with $k$. For $\mathcal I_1$, define
\[
N_{i_1\ldots i_k}
=
\sum_{b=1}^B
\mathbb I\bigl(\{i_1,\ldots,i_k\}\subseteq\mathcal I^b\bigr)
\sim\operatorname{Bin}(B,p_k),
\qquad
p_k=\frac{\binom{n_1-k}{r-k}}{\binom{n_1}{r}}.
\]
Averaging \eqref{eq:pointwise_hoeff} over the selected
subsamples gives
\[
\begin{aligned}
&\mathbb E_{\mathrm{train}}
[R(\mathbf x_*,t)
 \mid\mathcal D_{\mathrm{train}},\mathcal I^1,\ldots,\mathcal I^B]\\
&\qquad=
\frac1B\sum_{i=1}^{n_1}(N_i-Br/n_1)T_1(D_i)
+\frac1B\sum_{k=2}^r\sum_{i_1<\cdots<i_k}
N_{i_1\ldots i_k}T_k(D_{i_1},\ldots,D_{i_k}).
\end{aligned}
\]
Subtracting the conditional mean given
$\mathcal D_{\mathrm{train}}$ yields
\[
\begin{aligned}
&\mathbb E_{\mathrm{train}}
[R(\mathbf x_*,t)
 \mid\mathcal D_{\mathrm{train}},\mathcal I^1,\ldots,\mathcal I^B]
-\mathbb E_{\mathrm{train}}
[R(\mathbf x_*,t)\mid\mathcal D_{\mathrm{train}}]\\
&\qquad=
\frac1B\sum_{k=1}^r\sum_{i_1<\cdots<i_k}
(N_{i_1\ldots i_k}-Bp_k)
T_k(D_{i_1},\ldots,D_{i_k}).
\end{aligned}
\]
The inclusion counts are independent of the training observations,
so Hoeffding orthogonality eliminates cross terms after taking
$\mathbb E_{\mathrm{train}}$.
Conditional independence of fitting randomness then gives
\[
\begin{aligned}
\mathcal I_1
={}&
\frac1{B^2}\sum_{k=1}^r
\operatorname{Var}_{\mathrm{train}}(N_{1\ldots k})
\binom{n_1}{k}V_k+
\frac1B\mathbb E_{\mathrm{train}}\bigl[
\operatorname{Var}_{\mathrm{train}}
\{\widehat\psi^b(\mathbf x_*,t)
 \mid\mathcal D_{\mathrm{train}},\mathcal I^b\}
\bigr]\\
\le{}&
\frac1B\biggl\{
\sum_{k=1}^r\binom rk V_k
+
\mathbb E_{\mathrm{train}}\bigl[
\operatorname{Var}_{\mathrm{train}}
\{\widehat\psi^b(\mathbf x_*,t)
 \mid\mathcal D_{\mathrm{train}},\mathcal I^b\}
\bigr]
\biggr\}\\
={}&
\frac1B\operatorname{Var}_{\mathrm{train}}
[\widehat\psi_r(\mathbf x_*,t)].
\end{aligned}
\]
Here
\(
\operatorname{Var}_{\mathrm{train}}(N_{1\ldots k})
=Bp_k(1-p_k)\le Bp_k,\) and \(
p_k\binom{n_1}{k}=\binom rk,
\) and the last equality follows from
\eqref{eq:pointwise_hoeff} and total variance. Combining the bounds,
\[
\begin{aligned}
\frac{\mathbb E_{\mathrm{train}}[R(\mathbf x_*,t)^2]}
     {\sigma_\psi^2(\mathbf x_*,t)}
&\le
\biggl\{
\frac{n_1}{r^2B\zeta_r^{(1)}(\mathbf x_*,t)}
+\frac{r-1}{r(n_1-1)\zeta_r^{(1)}(\mathbf x_*,t)}
\biggr\}
\operatorname{Var}_{\mathrm{train}}
[\widehat\psi_r(\mathbf x_*,t)]\\
&\lesssim
\frac{n_1r^{\nu-2}}B+\frac{r^\nu}{n_1}
\lesssim r^{\nu-2}+n_1^{\alpha\nu-1}
\longrightarrow0.
\end{aligned}
\]
Indeed, $\operatorname{Var}_{\mathrm{train}}
[\widehat\psi_r(\mathbf x_*,t)]\le F^2$,
$B\gtrsim n_1$, and \eqref{eq:inference_tuning} gives
\(
\nu<1+\frac{c_{\mathrm{eff}}-\delta}{2}<2,\) and
\(\alpha\nu<1.\) Thus, for every $\varepsilon>0$,
\[
\mathbb P_{\mathrm{train}}\bigl\{
|R(\mathbf x_*,t)|>
\varepsilon\sigma_\psi(\mathbf x_*,t)
\bigr\}
\le
\frac{\mathbb E_{\mathrm{train}}[R(\mathbf x_*,t)^2]}
     {\varepsilon^2\sigma_\psi^2(\mathbf x_*,t)}
\longrightarrow0.
\]

Finally,
\[
\begin{aligned}
\frac{\bar\psi_B(\mathbf x_*,t)-\psi_0(\mathbf x_*,t)}
     {\sigma_\psi(\mathbf x_*,t)}
=
\frac{\mathring T_{n_1,r}(\mathbf x_*,t)}
     {\sigma_\psi(\mathbf x_*,t)}
+\frac{R(\mathbf x_*,t)}{\sigma_\psi(\mathbf x_*,t)}
+
\frac{\mathbb E_{\mathrm{train}}[\widehat\psi_r(\mathbf x_*,t)]
-\psi_0(\mathbf x_*,t)}
{\sigma_\psi(\mathbf x_*,t)}.
\end{aligned}
\]
The normal approximation, remainder bound, and bias bound
hold uniformly on $E_n$. Since
$\mathbb P_{\mathrm{nuis}}(E_n)\to1$, for every $z\in\mathbb R$,
\[
\mathbb P_{\mathrm{train}}\biggl\{
\frac{\bar\psi_B(\mathbf x_*,t)-\psi_0(\mathbf x_*,t)}
     {\sigma_\psi(\mathbf x_*,t)}
\le z
\biggr\}
=\Phi(z)+o_{\mathbb P_{\mathrm{nuis}}}(1).
\]
Taking $\mathbb E_{\mathrm{nuis}}$ and using boundedness
of conditional probabilities proves the convergence under
$\mathbb P$.

%%%%%%%%%Proof of Theorem 4%%%%%%%%%%%%%%%%
 \medskip
\noindent
\textbf{Proof of Theorem~\ref{thm:contrast_clt}.}
Work on the intersection of the events in
Assumptions~\ref{assum:xi} and~\ref{assum:bias_domination}
and the corresponding correlation condition, whose
$\mathbb P_{\mathrm{nuis}}$-probability tends to one.

\medskip
\noindent
\emph{Part~(i).}
For $\mathbf x\in\{\mathbf x_*,\mathbf x_*'\}$,
Theorem~\ref{thm:pointwise_clt} and
Proposition~\ref{cor:asymptotic_bias}(i) give
\[
\begin{aligned}
\bar\psi_B(\mathbf x,t)-\psi_0(\mathbf x,t)
=O_{\mathbb P}\{\sigma_\psi(\mathbf x,t)\},\qquad
\mathbb E_{\mathrm{train}}[\widehat\psi_r(\mathbf x,t)]
-\psi_0(\mathbf x,t)
=o_{\mathbb P_{\mathrm{nuis}}}
\{\sigma_\psi(\mathbf x,t)\}.
\end{aligned}
\]
Since $\tanh''$ is bounded, Taylor expansion yields
\[
\begin{aligned}
&\tanh\{\bar\psi_B(\mathbf x,t)\}
-\tanh\{\mathbb E_{\mathrm{train}}
[\widehat\psi_r(\mathbf x,t)]\}\\
&\quad=
\{1-\tau_0(\mathbf x,t)^2\}
\{\bar\psi_B(\mathbf x,t)
-\mathbb E_{\mathrm{train}}[\widehat\psi_r(\mathbf x,t)]\}
+O_{\mathbb P}\{\sigma_\psi^2(\mathbf x,t)\}.
\end{aligned}
\]
The correlation condition and boundedness of $\psi_0$
imply
\[
\sigma_{\mathrm{pt}}^2(\mathbf x_*,\mathbf x_*',t)
\gtrsim
\sigma_\psi^2(\mathbf x_*,t)
+\sigma_\psi^2(\mathbf x_*',t)
\gtrsim r^{2-\nu}/n_1.
\]
The Hoeffding variance bound gives
$\sigma_\psi^2(\mathbf x,t)
\le (r/n_1)\operatorname{Var}_{\mathrm{train}}
[\widehat\psi_r(\mathbf x,t)]
\le F^2r/n_1$.
Consequently,
\[
\frac{\sigma_\psi^2(\mathbf x_*,t)
+\sigma_\psi^2(\mathbf x_*',t)}
{\sigma_{\mathrm{pt}}(\mathbf x_*,\mathbf x_*',t)}
\lesssim\sqrt{r/n_1}\longrightarrow0.
\]
\noindent
Define
\[
\begin{aligned}
Y_b^{\mathrm{pt}}
=&\{1-\tau_0(\mathbf x_*,t)^2\}
\widehat\psi^b(\mathbf x_*,t)-\{1-\tau_0(\mathbf x_*',t)^2\}
\widehat\psi^b(\mathbf x_*',t).
\end{aligned}
\]
The expansion above and
Proposition~\ref{cor:asymptotic_bias}(ii) give
\begin{equation}
\label{eq:pt_linear_clt}
\begin{aligned}
&\widehat\theta_B^{\mathrm{pt}}
(\mathbf x_*,\mathbf x_*',t)
-\theta^{\mathrm{pt}}(\mathbf x_*,\mathbf x_*',t)=
\frac1B\sum_{b=1}^B
\{Y_b^{\mathrm{pt}}
-\mathbb E_{\mathrm{train}}[Y_b^{\mathrm{pt}}]\}
+o_{\mathbb P}\{
\sigma_{\mathrm{pt}}(\mathbf x_*,\mathbf x_*',t)\}.
\end{aligned}
\end{equation}
Since $|Y_b^{\mathrm{pt}}|\le2F$ and its
single-overlap variance satisfies
$n_1\sigma_{\mathrm{pt}}^2/r^2\gtrsim r^{-\nu}$,
the projection and inclusion-count arguments of
Theorem~\ref{thm:pointwise_clt} yield
\[
\frac{B^{-1}\sum_{b=1}^B
\{Y_b^{\mathrm{pt}}
-\mathbb E_{\mathrm{train}}[Y_b^{\mathrm{pt}}]\}}
{\sigma_{\mathrm{pt}}(\mathbf x_*,\mathbf x_*',t)}
\xrightarrow{d}\mathcal N(0,1).
\]
Together with \eqref{eq:pt_linear_clt}, this proves (i).

\medskip
\noindent
\emph{Part~(ii).}
We first verify that the slope restriction preserves
the approximation bounds of Corollary~\ref{lem:An}
and Proposition~\ref{cor:asymptotic_bias}.
Recall $\mathcal T_m$ and $\psi_{m,\mathrm{apx}}$
from \eqref{eq:tau_m_apx}, and define
\[
\begin{aligned}
\mathcal T_m^\Delta
&=\{\tanh\circ\psi\in\mathcal T_m:
\|\partial_t\psi\|_\infty\le F_\Delta\},\qquad
\psi_{m,\mathrm{apx}}^\Delta
\in\operatorname*{arg\,min}_{
\tanh\circ\psi\in\mathcal T_m^\Delta}
\|\psi-\psi_0\|_\infty.
\end{aligned}
\]
At $m$ in \eqref{eq:m_opt}, $m\to\infty$ and
\[
m^2\phi_r^{\,1+\delta/(1-c_{\mathrm{eff}})}
\asymp
\frac{m^{-2\beta}
r^{-\delta/(1-c_{\mathrm{eff}})}}
{L\log^2r}
\longrightarrow0.
\]
The inverse inequality in Lemma~\ref{lem:apprx}
therefore gives, for every original minimizer,
\[
\begin{aligned}
\|\partial_t\psi_{m,\mathrm{apx}}\|_\infty
\le K_t+
\|\partial_t(\psi_{m,\mathrm{apx}}-\psi_0)\|_\infty
&\le K_t+C\{
m^{-\beta}
+m\phi_r^{\{1+\delta/(1-c_{\mathrm{eff}})\}/2}
\}\\
&=K_t+o(1)<F_\Delta.
\end{aligned}
\]
Thus every original minimizer belongs to
$\mathcal T_m^\Delta$ for sufficiently large $r$.
Since $\mathcal T_m^\Delta\subseteq\mathcal T_m$,
the restricted and unrestricted minimum errors coincide,
and their sets of minimizers are equal.
We may take
$\psi_{m,\mathrm{apx}}^\Delta=\psi_{m,\mathrm{apx}}$,
with unchanged $L,\mathbf p,\mathbf s,F$.

Defining $\mathcal E_{\mathrm{opt},r,J}$ relative
to the empirical infimum over $\mathcal T_m^\Delta$,
we retain
$\mathbb E_{\mathrm{train}}[
L_{r,J}(\widehat\tau_r)
-L_{r,J}(\tau_{m,\mathrm{apx}})]
\le\mathcal E_{\mathrm{opt},r,J}$.
Theorem~\ref{thm:joint_oracle} therefore gives
the same estimation bounds for restricted estimators.
Corollary~\ref{lem:An} applies with $M_r$ and
$\mathcal A_n$ defined accordingly, as does
Proposition~\ref{cor:asymptotic_bias}.

Fix $t<t'$ and define
\[
Y_b^{\mathrm{tm}}
=
\frac{
\{1-\tau_0(\mathbf x_*,t')^2\}
\widehat\psi^b(\mathbf x_*,t')
-\{1-\tau_0(\mathbf x_*,t)^2\}
\widehat\psi^b(\mathbf x_*,t)}
{t'-t}.
\]
The slope restriction and
$\|\partial_t\psi_0\|_\infty\le K_t$ imply
\[
\begin{aligned}
|Y_b^{\mathrm{tm}}|
&\le\sup_{s\in\mathcal T}
\left|\partial_s[
\{1-\tau_0(\mathbf x_*,s)^2\}
\widehat\psi^b(\mathbf x_*,s)]\right|\le F_\Delta+2FK_t.
\end{aligned}
\]
By the temporal correlation condition,
\[
\frac{\sigma_{\mathrm{tm}}^2(t,t',\mathbf x_*)}
{(t'-t)^2}
\gtrsim
\sigma_\psi^2(\mathbf x_*,t)
+\sigma_\psi^2(\mathbf x_*,t')
\gtrsim r^{2-\nu}/n_1.
\]
Hence the single-overlap variance of
$Y_b^{\mathrm{tm}}$ is at least of order $r^{-\nu}$.
The projection and inclusion-count arguments of
Theorem~\ref{thm:pointwise_clt}, under
\eqref{eq:inference_tuning_tm}, yield
\[
\frac{t'-t}{B\sigma_{\mathrm{tm}}(t,t',\mathbf x_*)}
\sum_{b=1}^B
\{Y_b^{\mathrm{tm}}
-\mathbb E_{\mathrm{train}}[Y_b^{\mathrm{tm}}]\}
\xrightarrow{d}\mathcal N(0,1).
\]

It remains to bound the nonlinear remainder.
The slope restriction gives
$|\widehat\psi_r(\mathbf x_*,t)|\le F$
and
$|\widehat\psi_r(\mathbf x_*,t')
-\widehat\psi_r(\mathbf x_*,t)|
\le F_\Delta(t'-t)$.
The Hoeffding and finite-ensemble variance bounds give
\begin{equation}
\label{eq:tm_centered_moments}
\begin{aligned}
&\mathbb E_{\mathrm{train}}
\left|
\bar\psi_B(\mathbf x_*,t)
-\mathbb E_{\mathrm{train}}
[\widehat\psi_r(\mathbf x_*,t)]
\right|^2\\
&\quad+
\mathbb E_{\mathrm{train}}
\left|
\frac{
\bar\psi_B(\mathbf x_*,t')
-\bar\psi_B(\mathbf x_*,t)
-\mathbb E_{\mathrm{train}}
[\widehat\psi_r(\mathbf x_*,t')
-\widehat\psi_r(\mathbf x_*,t)]}
{t'-t}
\right|^2\lesssim
\frac r{n_1}+\frac1B
\lesssim\frac r{n_1}.
\end{aligned}
\end{equation}
By \eqref{eq:cor_bias_level} and
\eqref{eq:cor_bias_slope}, the sum of the squared
uniform level and derivative biases is
$O_{\mathbb P_{\mathrm{nuis}}}(m^{-2\beta})
=o_{\mathbb P_{\mathrm{nuis}}}(r/n_1)$,
since \eqref{eq:inference_tuning_tm} implies
\[
\frac{n_1m^{-2\beta}}r
=
(n_1r^{\nu-2}m^{-2\beta})r^{1-\nu}
\longrightarrow0.
\]
Thus the ensemble level and difference-quotient
errors are $O_{\mathbb P}(\sqrt{r/n_1})$.

Define
$G(u,v)=
[\tanh\{u+(t'-t)v\}-\tanh(u)]/(t'-t)$.
Its second derivatives satisfy
$|\partial_{uu}G(u,v)|
\le |v|\|\tanh'''\|_\infty$,
$|\partial_{uv}G(u,v)|
\le\|\tanh''\|_\infty$, and
$|\partial_{vv}G(u,v)|
\le(t_{\max}-t_{\min})\|\tanh''\|_\infty$.
By Definition~\ref{def:joint_composite} and the
slope restriction, the difference quotients of
$\psi_0$, $\mathbb E_{\mathrm{train}}[\widehat\psi_r]$,
and $\bar\psi_B$ are bounded by
$\max(K_t,F_\Delta)$.

Taylor expansion of $G$ at the true level and
difference quotient, using
\eqref{eq:tm_centered_moments},
\eqref{eq:cor_bias_level},
\eqref{eq:cor_bias_slope}, and
\eqref{eq:inference_tuning_tm}, gives
\begin{equation}
\label{eq:taylor_contrast}
\begin{aligned}
&\widehat\theta_B^{\mathrm{tm}}(t,t',\mathbf x_*)
-\bigl[
\tanh\{\mathbb E_{\mathrm{train}}
[\widehat\psi_r(\mathbf x_*,t')]\}
-\tanh\{\mathbb E_{\mathrm{train}}
[\widehat\psi_r(\mathbf x_*,t)]\}
\bigr]\\
&\quad=
\frac{t'-t}{B}\sum_{b=1}^B
\{Y_b^{\mathrm{tm}}
-\mathbb E_{\mathrm{train}}[Y_b^{\mathrm{tm}}]\}
+O_{\mathbb P}\bigl((t'-t)r/n_1\bigr).
\end{aligned}
\end{equation}
The remainder is negligible because
\[
\frac{(t'-t)r/n_1}
{\sigma_{\mathrm{tm}}(t,t',\mathbf x_*)}
\lesssim\sqrt{r^\nu/n_1}
\longrightarrow0.
\]
Combining \eqref{eq:taylor_contrast} with
Proposition~\ref{cor:asymptotic_bias}(iii) gives
\[
\begin{aligned}
&\widehat\theta_B^{\mathrm{tm}}(t,t',\mathbf x_*)
-\theta^{\mathrm{tm}}(t,t',\mathbf x_*)=
\frac{t'-t}{B}\sum_{b=1}^B
\{Y_b^{\mathrm{tm}}
-\mathbb E_{\mathrm{train}}[Y_b^{\mathrm{tm}}]\}
+o_{\mathbb P}\{
\sigma_{\mathrm{tm}}(t,t',\mathbf x_*)\}.
\end{aligned}
\]
The normal limit and Slutsky's theorem prove (ii),
including when $t'-t\to0$.
 
%%%%%%%%%%%%%%%%%%%%%%%%%%%%%%%%%%%%%%%%%%%%%%%%%%

 \medskip
\noindent
\textbf{Proof of Theorem~\ref{thm:ij_variance_consistency}.}
Fix $(\mathbf x_*,t)$ and work on the event in
Assumption~\ref{assum:xi}, whose
$\mathbb P_{\mathrm{nuis}}$-probability tends to one.
Define
$M_i=\operatorname{Cov}_{\mathrm{train}}
\{I_{b,i},\widehat\psi^b(\mathbf x_*,t)
\mid\mathcal D_{\mathrm{train}}\}$ and
$\check\sigma_\psi^2
=n_1(n_1-1)(n_1-r)^{-2}\sum_iM_i^2$.
The conditional covariance averages over subsampling
and DNN fitting with both samples fixed
(Table~\ref{tab:proof_notation}).

By \eqref{eq:pointwise_hoeff}, letting
$V_k=\operatorname{Var}_{\mathrm{train}}
[T_k(D_1,\ldots,D_k)]$, we have
$V_1=\zeta_r^{(1)}(\mathbf x_*,t)$ and
$\sum_{k=1}^r\binom rkV_k
\le\operatorname{Var}_{\mathrm{train}}
[\widehat\psi_r(\mathbf x_*,t)]\le F^2$.
The coefficient of an order-$k$ term indexed by
$i_1,\ldots,i_k$ in $M_i$ is
\[
\frac{\binom rk}{\binom{n_1}k}
\left(1-\frac r{n_1}\right)
\begin{cases}
1,&i\in\{i_1,\ldots,i_k\},\\
-k/(n_1-k),&\text{otherwise}.
\end{cases}
\]
Consequently,
\[
M_i^{(1)}
=\frac{r(n_1-r)}{n_1(n_1-1)}
\{h_{1,r}(D_i,\mathbf x_*,t)-\bar h_{1,r}\},
\]
where
$\bar h_{1,r}=n_1^{-1}\sum_i
h_{1,r}(D_i,\mathbf x_*,t)$.
By \eqref{eq:pointwise_hajek},
$\mathbb E_{\mathrm{train}}[h_{1,r}^4]
\le4F^2\zeta_r^{(1)}$ and
$n_1\zeta_r^{(1)}\gtrsim n_1r^{-\nu}\to\infty$.
Conditional Chebyshev's inequality yields
\begin{equation}
\label{eq:ij_first_order}
\begin{aligned}
&\frac{n_1(n_1-1)}{(n_1-r)^2}
\frac{\sum_i(M_i^{(1)})^2}{\sigma_\psi^2(\mathbf x_*,t)}
=
\frac{\sum_i
\{h_{1,r}(D_i,\mathbf x_*,t)-\bar h_{1,r}\}^2}
{(n_1-1)\zeta_r^{(1)}(\mathbf x_*,t)}
\xrightarrow{\mathbb P}1.
\end{aligned}
\end{equation}

Hoeffding orthogonality gives
\[
\begin{aligned}
&\frac{n_1(n_1-1)}{(n_1-r)^2}
\mathbb E_{\mathrm{train}}
\sum_i(M_i-M_i^{(1)})^2=
\sum_{k=2}^r
\frac{k(n_1-1)}{n_1-k}
\frac{\binom rk}{\binom{n_1}k}
\binom rkV_k
\lesssim(r/n_1)^2,
\end{aligned}
\]
using
$\binom rk/\binom{n_1}k\le(r/n_1)^k$
and $r/n_1\to0$.
Assumption~\ref{assum:xi} and conditional Markov's
inequality imply
\begin{equation}
\label{eq:ij_higher_order}
\begin{aligned}
&\frac{n_1(n_1-1)}{(n_1-r)^2}
\frac{\sum_i(M_i-M_i^{(1)})^2}
{\sigma_\psi^2(\mathbf x_*,t)}=
O_{\mathbb P}\left(
\frac1{n_1\zeta_r^{(1)}(\mathbf x_*,t)}
\right)=o_{\mathbb P}(1).
\end{aligned}
\end{equation}
The reverse triangle inequality and
\eqref{eq:ij_first_order}--\eqref{eq:ij_higher_order}
therefore give
\begin{equation}
\label{eq:ij_scalar_control}
\frac{\check\sigma_\psi^2(\mathbf x_*,t)}
{\sigma_\psi^2(\mathbf x_*,t)}
\xrightarrow{\mathbb P}1.
\end{equation}

For finite $B$, write
\[
\begin{aligned}
Z_{b,i}^{\circ}
&=\left(I_{b,i}-\frac r{n_1}\right)
\{\widehat\psi^b(\mathbf x_*,t)-\mu_D\},\qquad
Z_{b,i}=(I_{b,i}-\bar I_{\cdot,i})
\{\widehat\psi^b(\mathbf x_*,t)-\bar\psi_B(\mathbf x_*,t)\},
\end{aligned}
\]
where
$\mu_D=\mathbb E_{\mathrm{train}}
[\widehat\psi^b(\mathbf x_*,t)
\mid\mathcal D_{\mathrm{train}}]$.
Let
$\mathbf Z_b^\circ=(Z_{b,i}^\circ)_i$,
$\mathbf Z_b=(Z_{b,i})_i$,
$\mathbf M=(M_i)_i$,
$\mathbf I_b=(I_{b,i})_i$,
$\bar{\mathbf Z}^\circ=B^{-1}\sum_b\mathbf Z_b^\circ$,
$\bar{\mathbf Z}=B^{-1}\sum_b\mathbf Z_b$, and
$\mathbf\Sigma_Z=
\operatorname{Cov}_{\mathrm{train}}
(\mathbf Z_b^\circ\mid\mathcal D_{\mathrm{train}})$.
We use $\|\cdot\|_{\mathrm{op}}$ for the operator norm
and $\preceq$ for positive-semidefinite ordering.

By \eqref{eq:ij_cov},
\[
\widehat\sigma_\psi^2(\mathbf x_*,t)
=\frac{n_1(n_1-1)}{(n_1-r)^2}
\frac{\sum_{b\ne b'}\mathbf Z_b^\top\mathbf Z_{b'}}
{B(B-1)}.
\]
Uniform subsampling gives
\[
\operatorname{Cov}_{\mathrm{train}}
(\mathbf I_b\mid\mathcal D_{\mathrm{train}})
=\frac{r(n_1-r)}{n_1(n_1-1)}
\left(I_{n_1}-\frac{\mathbf1\mathbf1^\top}{n_1}\right).
\]
Since $|\widehat\psi^b|\le F$,
$\mathbb E_{\mathrm{train}}
[\mathbf Z_b^\circ\mid\mathcal D_{\mathrm{train}}]
=\mathbf M$ and
\[
\begin{aligned}
\mathbf\Sigma_Z+\mathbf M\mathbf M^\top
&=\mathbb E_{\mathrm{train}}
[\mathbf Z_b^\circ(\mathbf Z_b^\circ)^\top
\mid\mathcal D_{\mathrm{train}}]
\preceq4F^2
\operatorname{Cov}_{\mathrm{train}}
(\mathbf I_b\mid\mathcal D_{\mathrm{train}}).
\end{aligned}
\]
Thus
$\|\mathbf\Sigma_Z\|_{\mathrm{op}}\lesssim r/n_1$,
$\operatorname{tr}(\mathbf\Sigma_Z)\lesssim r$,
and $\|\mathbf M\|_2^2\lesssim r/n_1$.
Conditional independence across $b$ gives
\[
\mathbb E_{\mathrm{train}}\left[
\frac{\sum_{b\ne b'}
(\mathbf Z_b^\circ)^\top\mathbf Z_{b'}^\circ}
{B(B-1)}
\,\biggm|\mathcal D_{\mathrm{train}}\right]
=\|\mathbf M\|_2^2
\]
and
\begin{equation}
\label{eq:ij_cross_variance}
\begin{aligned}
&\operatorname{Var}_{\mathrm{train}}\left(
\frac{\sum_{b\ne b'}
(\mathbf Z_b^\circ)^\top\mathbf Z_{b'}^\circ}
{B(B-1)}
\,\biggm|\mathcal D_{\mathrm{train}}\right)=
\frac4B\mathbf M^\top\mathbf\Sigma_Z\mathbf M
+\frac2{B(B-1)}
\operatorname{tr}(\mathbf\Sigma_Z^2)\lesssim
\frac{r^2}{n_1^2B}+\frac{r^2}{n_1B^2}.
\end{aligned}
\end{equation}

The centering identity is
\[
\bar{\mathbf Z}-\bar{\mathbf Z}^\circ
=-\left(\bar I_{\cdot,i}-\frac r{n_1}\right)_{i=1}^{n_1}
\{\bar\psi_B(\mathbf x_*,t)-\mu_D\}.
\]
Conditional second moments and $B\gtrsim n_1$
give
\[
\begin{gathered}
\|\bar{\mathbf Z}-\bar{\mathbf Z}^\circ\|_2
=O_{\mathbb P}(\sqrt r/B),\qquad
B^{-1}\sum_b\|\mathbf Z_b-\mathbf Z_b^\circ\|_2^2
=O_{\mathbb P}(r/B),\\
\|\bar{\mathbf Z}^\circ\|_2
=O_{\mathbb P}(\sqrt{r/n_1}),\qquad
B^{-1}\sum_b\|\mathbf Z_b^\circ\|_2^2
=O_{\mathbb P}(r).
\end{gathered}
\]
Using
$\sum_{b\ne b'}\mathbf Z_b^\top\mathbf Z_{b'}
=B^2\|\bar{\mathbf Z}\|_2^2
-\sum_b\|\mathbf Z_b\|_2^2$,
we obtain
\begin{equation}
\label{eq:ij_centering_error}
\begin{aligned}
&\frac{\sum_{b\ne b'}
\{\mathbf Z_b^\top\mathbf Z_{b'}
-(\mathbf Z_b^\circ)^\top\mathbf Z_{b'}^\circ\}}
{B(B-1)}=
O_{\mathbb P}\left(
\frac r{B\sqrt{n_1}}+\frac r{B^{3/2}}\right).
\end{aligned}
\end{equation}
Equations~\eqref{eq:ij_cross_variance} and
\eqref{eq:ij_centering_error}, with conditional
Chebyshev's inequality, yield
\begin{equation}
\label{eq:ij_finite_B_error}
\begin{aligned}
\widehat\sigma_\psi^2(\mathbf x_*,t)
-\check\sigma_\psi^2(\mathbf x_*,t)
=O_{\mathbb P}(r/n_1^{3/2})=o_{\mathbb P}(r^{2-\nu}/n_1),
\end{aligned}
\end{equation}
since \eqref{eq:inference_tuning} implies
$r^{\nu-1}/\sqrt{n_1}
=\sqrt{r^\nu/n_1}\,r^{\nu/2-1}\to0$
for $\alpha\nu<1$ and $\nu<2$.
Together with \eqref{eq:ij_scalar_control}
and Assumption~\ref{assum:xi}, this proves
$\widehat\sigma_\psi^2/\sigma_\psi^2
\xrightarrow{\mathbb P}1$.

For contrasts, Theorem~\ref{thm:contrast_clt}
gives $|Y_b^{\mathrm{pt}}|\le2F$ and
$|Y_b^{\mathrm{tm}}|\le F_\Delta+2FK_t$, with
\begin{equation}
\label{eq:ij_contrast_scales}
\sigma_{\mathrm{pt}}^2(\mathbf x_*,\mathbf x_*',t)
\gtrsim\frac{r^{2-\nu}}{n_1},
\qquad
\frac{\sigma_{\mathrm{tm}}^2(t,t',\mathbf x_*)}
{(t'-t)^2}
\gtrsim\frac{r^{2-\nu}}{n_1}.
\end{equation}
Thus \eqref{eq:ij_scalar_control} and
\eqref{eq:ij_finite_B_error} also apply to
both contrast statistics.

Write $M_i(\mathbf x,t)$ for the corresponding
inclusion covariance and define
\[
\check\Gamma_\psi(\mathbf x,t,\mathbf x',t')
=\frac{n_1(n_1-1)}{(n_1-r)^2}
\sum_iM_i(\mathbf x,t)M_i(\mathbf x',t').
\]
The estimated covariance matrices are
\[
\widehat{\mathbf\Sigma}_{\mathrm{pt}}=
\begin{pmatrix}
\widehat\sigma_\psi^2(\mathbf x_*,t)&
\widehat\Gamma_\psi(\mathbf x_*,t,\mathbf x_*',t)\\
\widehat\Gamma_\psi(\mathbf x_*,t,\mathbf x_*',t)&
\widehat\sigma_\psi^2(\mathbf x_*',t)
\end{pmatrix}
\]
and
\[
\widehat{\mathbf\Sigma}_{\mathrm{tm}}
=
\begin{pmatrix}
1&0\\
-\frac1{t'-t}&\frac1{t'-t}
\end{pmatrix}
\begin{pmatrix}
\widehat\sigma_\psi^2(\mathbf x_*,t)&
\widehat\Gamma_\psi(\mathbf x_*,t,\mathbf x_*,t')\\
\widehat\Gamma_\psi(\mathbf x_*,t,\mathbf x_*,t')&
\widehat\sigma_\psi^2(\mathbf x_*,t')
\end{pmatrix}
\begin{pmatrix}
1&-\frac1{t'-t}\\
0&\frac1{t'-t}
\end{pmatrix}.
\]
Replacing $\widehat\sigma_\psi^2,\widehat\Gamma_\psi$
by $\check\sigma_\psi^2,\check\Gamma_\psi$
defines $\check{\mathbf\Sigma}_{\mathrm{pt}}$
and $\check{\mathbf\Sigma}_{\mathrm{tm}}$.
The bound on $\|\mathbf M\|_2^2$ and
\eqref{eq:ij_finite_B_error}, applied to each
coordinate and their sum, imply that both checked
matrices are positive semidefinite and
\[
\begin{aligned}
&\max\{
\|\check{\mathbf\Sigma}_{\mathrm{pt}}\|_{\mathrm{op}},
\|\check{\mathbf\Sigma}_{\mathrm{tm}}\|_{\mathrm{op}}
\}\lesssim r/n_1,\quad
&\max\{
\|\widehat{\mathbf\Sigma}_{\mathrm{pt}}
-\check{\mathbf\Sigma}_{\mathrm{pt}}\|_{\mathrm{op}},
\|\widehat{\mathbf\Sigma}_{\mathrm{tm}}
-\check{\mathbf\Sigma}_{\mathrm{tm}}\|_{\mathrm{op}}
\}
=O_{\mathbb P}(r/n_1^{3/2}).
\end{aligned}
\]

Consider the temporal contrast first.
Bounded derivatives of $\tanh^2$ and the
prescribed-pair bounds of
Theorem~\ref{thm:contrast_clt}(ii) give
\begin{equation}
\label{eq:ij_temporal_weight_error}
\begin{aligned}
&\left|
\frac{
(\bar\tau_B^2-\tau_0^2)(\mathbf x_*,t')
-(\bar\tau_B^2-\tau_0^2)(\mathbf x_*,t)}
{t'-t}\right|\\
&\quad\lesssim
|(\bar\psi_B-\psi_0)(\mathbf x_*,t)|
+\left|
\frac{
(\bar\psi_B-\psi_0)(\mathbf x_*,t')
-(\bar\psi_B-\psi_0)(\mathbf x_*,t)}
{t'-t}\right|=O_{\mathbb P}(\sqrt{r/n_1}),\\
&|(\bar\tau_B^2-\tau_0^2)(\mathbf x_*,t')|
=O_{\mathbb P}(\sqrt{r/n_1}).
\end{aligned}
\end{equation}
Set
\[
\mathbf w_B=
\begin{pmatrix}
\{\bar\tau_B(\mathbf x_*,t)^2
-\bar\tau_B(\mathbf x_*,t')^2\}/(t'-t)\\
1-\bar\tau_B(\mathbf x_*,t')^2
\end{pmatrix},
\]
and let $\mathbf w_0$ replace $\bar\tau_B$
by $\tau_0$.
Equation~\eqref{eq:ij_temporal_weight_error}
implies
$\|\mathbf w_B-\mathbf w_0\|_2
=O_{\mathbb P}(\sqrt{r/n_1})$.
The reverse triangle inequality and
$\|\check{\mathbf\Sigma}_{\mathrm{tm}}\|_{\mathrm{op}}
\lesssim r/n_1$ give
\[
\begin{aligned}
&\left|
\|\check{\mathbf\Sigma}_{\mathrm{tm}}^{1/2}
\mathbf w_B\|_2
-\|\check{\mathbf\Sigma}_{\mathrm{tm}}^{1/2}
\mathbf w_0\|_2\right|=O_{\mathbb P}(r/n_1)
=o_{\mathbb P}\left(
\sqrt{r^{2-\nu}/n_1}\right),
\end{aligned}
\]
because $\sqrt{r^\nu/n_1}\to0$. The slope restriction gives
$\|\mathbf w_B\|_2^2\le1+4F_\Delta^2$.
Hence \eqref{eq:var_temporal} and
\eqref{eq:ij_finite_B_error} imply
\[
\begin{aligned}
&\left|
\frac{\widehat\sigma_{\mathrm{tm}}^2
(t,t',\mathbf x_*)}{(t'-t)^2}
-\|\check{\mathbf\Sigma}_{\mathrm{tm}}^{1/2}
\mathbf w_B\|_2^2\right|
\le(1+4F_\Delta^2)
\|\widehat{\mathbf\Sigma}_{\mathrm{tm}}
-\check{\mathbf\Sigma}_{\mathrm{tm}}\|_{\mathrm{op}}=o_{\mathbb P}(r^{2-\nu}/n_1).
\end{aligned}
\]
Together with \eqref{eq:ij_scalar_control}
for $Y_b^{\mathrm{tm}}$ and
\eqref{eq:ij_contrast_scales}, this proves
\[
\frac{\widehat\sigma_{\mathrm{tm}}^2(t,t',\mathbf x_*)}
{\sigma_{\mathrm{tm}}^2(t,t',\mathbf x_*)}
\xrightarrow{\mathbb P}1.
\]

For the covariate contrast,
Theorem~\ref{thm:pointwise_clt} gives
$|(\bar\tau_B^2-\tau_0^2)(\mathbf x_*,t)|
+|(\bar\tau_B^2-\tau_0^2)(\mathbf x_*',t)|
=O_{\mathbb P}(\sqrt{r/n_1})$.
The same argument with
$\check{\mathbf\Sigma}_{\mathrm{pt}}$,
$\widehat{\mathbf\Sigma}_{\mathrm{pt}}$, and
$Y_b^{\mathrm{pt}}$ yields
\[
\frac{\widehat\sigma_{\mathrm{pt}}^2
(\mathbf x_*,\mathbf x_*',t)}
{\sigma_{\mathrm{pt}}^2
(\mathbf x_*,\mathbf x_*',t)}
\xrightarrow{\mathbb P}1.
\]
 
%%%%%%%%%%%%%%%%%%%%%%%%%%%%%%%%%%%%%%%%

 \medskip
\noindent
\textbf{Proof of Lemma~\ref{lem:apprx}.}
Let $\xi_\ell$ be the Greville knot average of
$B_{\ell,m}$. The B-spline identities
\citep[Chapter~IX]{deboor2001practical} give
\begin{equation}
\label{eq:lem1_identity}
\sum_{\ell=1}^m B_{\ell,m}(t)=1,\qquad
\sum_{\ell=1}^m\xi_\ell B_{\ell,m}(t)=t,\qquad
|\xi_\ell-t|\lesssim m^{-1}
\quad(t\in\overline S_\ell),
\end{equation}
where $S_\ell=\{t:B_{\ell,m}(t)>0\}$, as in
Theorem~\ref{thm:joint_oracle}.
Set
$\widetilde\psi_m(\mathbf x,t)
=\sum_{\ell=1}^m\psi_0(\mathbf x,\xi_\ell)B_{\ell,m}(t)$.
Equation~\eqref{eq:lem1_identity} implies
\begin{equation}
\label{eq:lem1_derivative_identity}
\sum_\ell(\xi_\ell-t)B_{\ell,m}(t)=0,\qquad
\sum_\ell B_{\ell,m}'(t)=0,\qquad
\sum_\ell(\xi_\ell-t)B_{\ell,m}'(t)=1.
\end{equation}
By Definition~\ref{def:joint_composite} and the
fundamental theorem of calculus, the remainder
$R_\ell(\mathbf x,t)
=\psi_0(\mathbf x,\xi_\ell)-\psi_0(\mathbf x,t)
-\partial_t\psi_0(\mathbf x,t)(\xi_\ell-t)$
satisfies $|R_\ell(\mathbf x,t)|
\lesssim|\xi_\ell-t|^{1+\beta}$.
The identities above yield
\[
\widetilde\psi_m-\psi_0
=\sum_\ell R_\ell B_{\ell,m},
\qquad
\partial_t(\widetilde\psi_m-\psi_0)
=\sum_\ell R_\ell B_{\ell,m}'.
\]
Since at most a fixed number of splines are active
at each $t$ and
$\|B_{\ell,m}'\|_\infty\lesssim m$,
\[
\|\widetilde\psi_m-\psi_0\|_\infty
\lesssim m^{-(1+\beta)},\qquad
\|\partial_t(\widetilde\psi_m-\psi_0)\|_\infty
\lesssim m^{-\beta}.
\]

Centering $\psi_0(\cdot,\xi_\ell)$ changes only
the final component in
Definition~\ref{def:composite_holder}, with a
uniformly enlarged H\"older bound and unchanged
$c_{\mathrm{eff}}$ in \eqref{eq:c_eff&phi_n}.
Assumption~\ref{assum:dnn} gives
\[
\begin{aligned}
r^{1+\delta}\phi_r
&\lesssim\min_l\left\lfloor\frac{p_l}{4m}\right\rfloor
\le\max_l\left\lfloor\frac{p_l}{4m}\right\rfloor
\lesssim r,\qquad
\lfloor s_\ell/4\rfloor
\asymp r^{1+\delta}\phi_r\log r,
\qquad \ell=1,\ldots,m.
\end{aligned}
\]
Thus \citet[Condition~2 and Theorem~1]{Ghosal2026},
with depth $L$, envelope $2F$, these widths and
sparsities, and unit parameter bounds, provides
deterministic networks $\widetilde f_\ell$ satisfying
\[
\max_{\ell\le m}
\|\widetilde f_\ell-\widetilde f_\ell(\mathbf0)
-\psi_0(\cdot,\xi_\ell)
+\psi_0(\mathbf0,\xi_\ell)\|_\infty^2
\lesssim\phi_r^{\,1+\delta/(1-c_{\mathrm{eff}})}.
\]
The constants are uniform in $\ell$.
Set $f_\ell=\widetilde f_\ell
-\widetilde f_\ell(\mathbf0)+\psi_0(\mathbf0,\xi_\ell)$.
For large $r$ and $F\ge\|\psi_0\|_\infty+1$,
\[
\max_{\ell\le m}
\|f_\ell-\psi_0(\cdot,\xi_\ell)\|_\infty^2
\lesssim\phi_r^{\,1+\delta/(1-c_{\mathrm{eff}})},
\qquad
\max_{\ell\le m}\|f_\ell\|_\infty\le F.
\]

Represent $f_\ell$ by two copies of the base
network, zeroing the second copy's input weights
and negating its output weights. Add
$\lceil F\rceil$ constant units in the last
hidden layer, each with output weight
$\psi_0(\mathbf0,\xi_\ell)/\lceil F\rceil$.
The resulting depth-$L$ scalar network has
\[
\begin{aligned}
p_l^{(\ell)}
&\le2\left\lfloor\frac{p_l}{4m}\right\rfloor
+\lceil F\rceil\mathbb I(l=L),\qquad
\sum_{l=0}^L\|\mathcal W_l^{(\ell)}\|_0
+\sum_{l=1}^L\|\mathbf v_l^{(\ell)}\|_0
\le2\lfloor s_\ell/4\rfloor
+2\lceil F\rceil\le s_\ell,
\end{aligned}
\]
with all parameters bounded by one.
Since $\min_l p_l/m\gtrsim
r^{1+\delta}\phi_r\to\infty$,
$\sum_{\ell=1}^mp_l^{(\ell)}
\le p_l/2+m\lceil F\rceil\le p_l$
for large $r$.

Following the parallel construction in
\citet[proof of Theorem~1]{schmidt2020deep},
stack the input matrices, use block-diagonal
hidden matrices
$\mathcal W_l=\operatorname{diag}
(\mathcal W_l^{(1)},\ldots,\mathcal W_l^{(m)})$,
and concatenate the shift vectors.
The resulting network retains input dimension $d$
and depth $L$, with $m$ outputs and at most
$s_\ell$ nonzero parameters per output.
Zero padding achieves the prescribed widths.
Hence
$\mathbf f=(f_1,\ldots,f_m)^\top
\in\mathcal F_{\mathrm{joint}}(L,\mathbf p,\mathbf s,F)$.

By \eqref{eq:lem1_identity} and the spline
derivative bound,
\[
\begin{aligned}
\|\psi_{\mathbf f}-\widetilde\psi_m\|_\infty^2
\lesssim\phi_r^{\,1+\delta/(1-c_{\mathrm{eff}})},
\qquad\|\partial_t(\psi_{\mathbf f}
-\widetilde\psi_m)\|_\infty^2
 \lesssim m^2\phi_r^{\,1+\delta/(1-c_{\mathrm{eff}})}.
\end{aligned}
\]
Combining these with the temporal spline errors,
\begin{equation}
\label{eq:apprx_constructed_latent}
\begin{aligned}
\|\psi_{\mathbf f}-\psi_0\|_\infty
\lesssim m^{-(1+\beta)}
+\phi_r^{\{1+\delta/(1-c_{\mathrm{eff}})\}/2},\quad
\|\partial_t(\psi_{\mathbf f}-\psi_0)\|_\infty
\lesssim m^{-\beta}
+m\phi_r^{\{1+\delta/(1-c_{\mathrm{eff}})\}/2}.
\end{aligned}
\end{equation}

By \eqref{eq:tau_m_apx}, optimality gives
$\|\psi_{m,\mathrm{apx}}-\psi_0\|_\infty
\le\|\psi_{\mathbf f}-\psi_0\|_\infty$
and
$\|\psi_{m,\mathrm{apx}}-\psi_{\mathbf f}\|_\infty
\le2\|\psi_{\mathbf f}-\psi_0\|_\infty$.
For each $\mathbf x$,
$\psi_{m,\mathrm{apx}}(\mathbf x,\cdot)
-\psi_{\mathbf f}(\mathbf x,\cdot)\in\mathcal B_m$.
On each knot interval $I$, this difference is
a polynomial of fixed degree with
$|I|\gtrsim m^{-1}$.
The rescaled polynomial Markov inequality yields
\[
\|\partial_t(\psi_{m,\mathrm{apx}}
-\psi_{\mathbf f})\|_\infty
\lesssim m\|\psi_{m,\mathrm{apx}}
-\psi_{\mathbf f}\|_\infty
\le2m\|\psi_{\mathbf f}-\psi_0\|_\infty.
\]
The triangle inequality and
\eqref{eq:apprx_constructed_latent} therefore give
\begin{equation}
\label{eq:apprx_latent_minimizer}
\begin{aligned}
\|\psi_{m,\mathrm{apx}}-\psi_0\|_\infty^2
&\lesssim m^{-2(1+\beta)}
+\phi_r^{\,1+\delta/(1-c_{\mathrm{eff}})},\\
\|\partial_t(\psi_{m,\mathrm{apx}}
-\psi_0)\|_\infty^2
&\lesssim m^{-2\beta}
+m^2\phi_r^{\,1+\delta/(1-c_{\mathrm{eff}})}.
\end{aligned}
\end{equation}

Finally, $\tau_{m,\mathrm{apx}}
=\tanh\circ\psi_{m,\mathrm{apx}}$
by \eqref{eq:tau_m_apx}.
Since $\|\tanh'\|_\infty\le1$,
$\|\tau_{m,\mathrm{apx}}-\tau_0\|_\infty
\le\|\psi_{m,\mathrm{apx}}-\psi_0\|_\infty$.
The chain rule and
$\|\partial_t\psi_0\|_\infty\le K_t$ give
\[
\begin{aligned}
\|\partial_t(\tau_{m,\mathrm{apx}}-\tau_0)\|_\infty
&\le\|\partial_t(\psi_{m,\mathrm{apx}}
-\psi_0)\|_\infty+\|\tanh''\|_\infty K_t
\|\psi_{m,\mathrm{apx}}-\psi_0\|_\infty\\
&\lesssim m^{-\beta}
+m\phi_r^{\{1+\delta/(1-c_{\mathrm{eff}})\}/2}.
\end{aligned}
\]
Thus \eqref{eq:apprx_latent_minimizer} yields
\[
\begin{aligned}
&\|\psi_{m,\mathrm{apx}}-\psi_0\|_\infty^2
+\|\tau_{m,\mathrm{apx}}-\tau_0\|_\infty^2\lesssim
m^{-2(1+\beta)}
+\phi_r^{\,1+\delta/(1-c_{\mathrm{eff}})},\\
&\|\partial_t(\psi_{m,\mathrm{apx}}-\psi_0)\|_\infty^2
+\|\partial_t(\tau_{m,\mathrm{apx}}-\tau_0)\|_\infty^2\lesssim
m^{-2\beta}
+m^2\phi_r^{\,1+\delta/(1-c_{\mathrm{eff}})}.
\end{aligned}
\]

\setcounter{table}{0}
\renewcommand{\thetable}{B\arabic{table}}
\section{Additional Simulation Results}
\label{sec:supp_sim}

This section provides implementation details and
additional simulation results for the comparison with the following competing methods. Briefly,
Survival BART \citep{sparapani2016nonparametric}
fits Bayesian discrete-time survival models
separately within each treatment arm and estimates
CATEs by differencing the resulting posterior
survival curves; Causal survival forests (CSF)
\citep{cui2023estimating} employ forest-based
weights in doubly robust estimating equations
with internally estimated nuisance functions.
DR-BART \citep{chipman2010bart} and DR-RF
\citep{athey2019generalized,Sun2026-wa}
regress the same cross-fitted pseudo-outcomes
used by our estimator on the covariates,
fitting separate models at each evaluation time.
Thus, the nuisance specifications in
Designs~1--3 affect DR-BART, DR-RF, and our
estimator, whereas CSF and survival BART
are fitted directly to the observed survival
data and yield identical results across designs.

All methods are evaluated using the same
200 Monte Carlo replications, prespecified
covariate profiles, evaluation times, and
profile pairs described in Section~\ref{sec:sim}.
Tables~\ref{tab:compare_cate}--\ref{tab:compare_tm}
report absolute bias, empirical standard deviation
(EmpSD), average estimated standard error (SE),
empirical coverage probability (CP), and average
interval length (AIL) for pointwise CATEs,
between-patient contrasts, and temporal contrasts,
respectively. All confidence intervals have
a nominal coverage level of $95\%$.

% --- TABLE 1: Pointwise CATE ---
\begin{table}[htbp]
\centering\small\setlength{\tabcolsep}{5pt}
\begin{tabular}{l l rrrrr}
\toprule
Design & Method & $|$Bias$|$ & EmpSD & SE & CP & AIL \\
\midrule
1 & Ours ($m=4$) & 0.0454 & 0.3819 & 0.3926 & 0.947 & 1.2973 \\
 & DR-RF & 0.0658 & 0.0684 & 0.1171 & 0.935 & 0.4591 \\
 & DR-BART & 0.0376 & 0.2782 & 0.4136 & 0.996 & 1.6215 \\
 & CSF & 0.0702 & 0.0450 & 0.0551 & 0.700 & 0.2158 \\
 & Survival BART & 0.0673 & 0.0949 & 0.1523 & 0.967 & 0.5969 \\
\addlinespace
2 & Ours ($m=4$) & 0.0473 & 0.3855 & 0.3972 & 0.947 & 1.3083 \\
 & DR-RF & 0.0649 & 0.0648 & 0.1104 & 0.927 & 0.4326 \\
 & DR-BART & 0.0400 & 0.2083 & 0.3306 & 0.997 & 1.2960 \\
 & CSF & 0.0702 & 0.0450 & 0.0551 & 0.700 & 0.2158 \\
 & Survival BART & 0.0673 & 0.0949 & 0.1523 & 0.967 & 0.5969 \\
\addlinespace
3 & Ours ($m=4$) & 0.0432 & 0.3847 & 0.3950 & 0.948 & 1.3028 \\
 & DR-RF & 0.0658 & 0.0687 & 0.1180 & 0.935 & 0.4626 \\
 & DR-BART & 0.0375 & 0.2818 & 0.4167 & 0.995 & 1.6336 \\
 & CSF & 0.0702 & 0.0450 & 0.0551 & 0.700 & 0.2158 \\
 & Survival BART & 0.0673 & 0.0949 & 0.1523 & 0.967 & 0.5969 \\
\bottomrule
\end{tabular} 
 
\caption{Comparison of methods for estimating
the pointwise CATE $\tau_0(\mathbf{x}_*;t_j)$
based on 200 Monte Carlo replications, $J=25$,
and a nominal coverage level of $95\%$.
Because CSF and survival BART are fitted directly
to the observed survival data without relying
on estimated nuisance functions, their results
are identical across nuisance designs.}

\label{tab:compare_cate}
\end{table}

For pointwise CATE estimation,
Table~\ref{tab:compare_cate} shows that our
estimator achieves stable coverage across
the three nuisance designs, with CP ranging
from $0.947$ to $0.948$ and absolute bias
from $0.0432$ to $0.0473$. The average estimated
standard errors ($0.393$--$0.397$) are close
to the empirical standard deviations
($0.382$--$0.386$). DR-RF achieves coverage
of $0.927$--$0.935$ with shorter intervals
($0.433$--$0.463$), whereas survival BART
attains coverage of $0.967$ with an average
interval length of $0.597$.
DR-BART produces coverage of $0.995$--$0.997$
with interval lengths of $1.296$--$1.634$.
Although CSF yields the shortest intervals
($0.216$), its coverage is only $0.700$.
Overall, our estimator maintains near-nominal
coverage with consistent performance across
the nuisance designs.

% --- TABLE 2: Temporal Contrast ---
\begin{table}[htbp]
\centering\small\setlength{\tabcolsep}{5pt}
\begin{tabular}{l l rrrrr}
\toprule
Design & Method & $|$Bias$|$ & EmpSD & SE & CP & AIL \\
\midrule
1 & Ours ($m=4$) & 0.0254 & 0.2351 & 0.2485 & 0.957 & 0.9742 \\
 & DR-RF & 0.0405 & 0.0502 & 0.1659 & 0.999 & 0.6503 \\
 & DR-BART & 0.0210 & 0.1350 & 0.2084 & 0.997 & 0.8168 \\
 & CSF & 0.0436 & 0.0323 & 0.0779 & 0.959 & 0.3054 \\
 & Survival BART & 0.0334 & 0.0639 & 0.1026 & 0.983 & 0.4023 \\
\addlinespace
2 & Ours ($m=4$) & 0.0261 & 0.2369 & 0.2516 & 0.958 & 0.9864 \\
 & DR-RF & 0.0385 & 0.0497 & 0.1566 & 0.998 & 0.6138 \\
 & DR-BART & 0.0211 & 0.1016 & 0.1685 & 0.998 & 0.6606 \\
 & CSF & 0.0436 & 0.0323 & 0.0779 & 0.959 & 0.3054 \\
 & Survival BART & 0.0334 & 0.0639 & 0.1026 & 0.983 & 0.4023 \\
\addlinespace
3 & Ours ($m=4$) & 0.0255 & 0.2353 & 0.2501 & 0.958 & 0.9803 \\
 & DR-RF & 0.0407 & 0.0498 & 0.1672 & 0.999 & 0.6555 \\
 & DR-BART & 0.0204 & 0.1358 & 0.2096 & 0.997 & 0.8215 \\
 & CSF & 0.0436 & 0.0323 & 0.0779 & 0.959 & 0.3054 \\
 & Survival BART & 0.0334 & 0.0639 & 0.1026 & 0.983 & 0.4023 \\
\bottomrule
\end{tabular}

\caption{Comparison of methods for estimating
the between-patient CATE contrast
$\theta_{\mathrm{pt}}(\mathbf{x}^{(1)}_*,
\mathbf{x}^{(2)}_*;t_j)$
based on 200 Monte Carlo replications, $J=25$,
and a nominal coverage level of $95\%$.}

 %CSF and Survival BART use the raw survival data, not the fitted nuisances, so their rows are the same in every design.}
\label{tab:compare_pt}
\end{table}

% --- TABLE 3: Between-Patient Contrast ---

\begin{table}[htbp]
\centering\small\setlength{\tabcolsep}{5pt}
\begin{tabular}{l l rrrrr}
\toprule
Design & Method & $|$Bias$|$ & EmpSD & SE & CP & AIL \\
\midrule
1 & Ours ($m=4$) & 0.0408 & 0.1412 & 0.1530 & 0.947 & 0.5996 \\
 & DR-RF & 0.0874 & 0.0655 & 0.1025 & 0.858 & 0.4019 \\
 & DR-BART & 0.0519 & 0.2664 & 0.5919 & 1.000 & 2.3203 \\
 & CSF & 0.1071 & 0.0436 & 0.0796 & 0.718 & 0.3120 \\
 & Survival BART & 0.0931 & 0.0353 & 0.0610 & 0.676 & 0.2391 \\
\addlinespace
2 & Ours ($m=4$) & 0.0509 & 0.1430 & 0.1565 & 0.942 & 0.6133 \\
 & DR-RF & 0.0884 & 0.0636 & 0.0993 & 0.845 & 0.3894 \\
 & DR-BART & 0.0581 & 0.1946 & 0.4706 & 1.000 & 1.8447 \\
 & CSF & 0.1071 & 0.0436 & 0.0796 & 0.718 & 0.3120 \\
 & Survival BART & 0.0931 & 0.0353 & 0.0610 & 0.676 & 0.2391 \\
\addlinespace
3 & Ours ($m=4$) & 0.0491 & 0.1356 & 0.1461 & 0.939 & 0.5725 \\
 & DR-RF & 0.0872 & 0.0653 & 0.1029 & 0.861 & 0.4034 \\
 & DR-BART & 0.0535 & 0.2676 & 0.5941 & 1.000 & 2.3289 \\
 & CSF & 0.1071 & 0.0436 & 0.0796 & 0.718 & 0.3120 \\
 & Survival BART & 0.0931 & 0.0353 & 0.0610 & 0.676 & 0.2391 \\
\bottomrule
\end{tabular}

\caption{Comparison of methods for estimating
the temporal CATE contrast
$\theta_{\mathrm{tm}}(\mathbf{x}_*;t,t')$
based on 200 Monte Carlo replications, $J=25$,
and a nominal coverage level of $95\%$.}

%CSF and Survival BART use the raw survival data, not the fitted nuisances, so their rows are the same in every design.}
\label{tab:compare_tm}
\end{table}

For between-patient contrasts,
Table~\ref{tab:compare_pt} shows that
our method achieves absolute bias of
$0.0254$--$0.0261$, empirical coverage
of $0.957$--$0.958$, and average interval
lengths of $0.974$--$0.986$.
The estimated standard errors
($0.249$--$0.252$) are close to the empirical
standard deviations ($0.235$--$0.237$).
CSF achieves comparable coverage ($0.959$)
with shorter intervals ($0.305$), while
survival BART achieves coverage of $0.983$
with interval length $0.402$.
DR-RF and DR-BART exhibit conservative
coverage ($0.998$--$0.999$ and
$0.997$--$0.998$, respectively), with
interval lengths of $0.614$--$0.656$
and $0.661$--$0.822$, respectively.
These results indicate that our procedure
provides stable inference for between-patient
contrasts across the nuisance designs.

The clearest advantage of the proposed method
emerges for temporal contrasts in
Table~\ref{tab:compare_tm}.
Across the three nuisance designs, our
estimator achieves coverage of
$0.939$--$0.947$, with absolute bias of
$0.0408$--$0.0509$ and average interval
lengths of $0.573$--$0.613$.
The estimated standard errors
($0.146$--$0.157$) are close to the
empirical standard deviations
($0.136$--$0.143$), providing evidence
of reliable variance estimation.
In contrast, DR-RF achieves coverage of
$0.845$--$0.861$, while CSF and survival
BART exhibit substantially lower coverage
of $0.718$ and $0.676$, respectively.
Although these methods produce shorter
intervals, their reduced coverage suggests
that the resulting intervals do not adequately
account for uncertainty and estimation bias
in temporal contrasts.
DR-BART achieves coverage of $1.000$
but produces  wider intervals
($1.845$--$2.329$), approximately three
to four times those of our method.

%Fitting one network per evaluation time is a natural alternative for the pointwise effect, and Theorem~\ref{thm:joint_oracle} at $m=J$ assigns it the same rate, but it cannot support either contrast: separately resampled ensembles carry no covariance between evaluation points, so $\psi_{\mathrm{tm}}$ and $\psi_{\mathrm{pt}}$ have no standard error.  A single fitted ensemble supplies all three estimands from the same influence contributions.

\end{document}